\documentclass[sigconf]{acmart}

\usepackage{acmart-taps}

\usepackage{multirow}
\usepackage{multicol}
\usepackage{booktabs}

\usepackage{colortbl}
\usepackage{siunitx}
\usepackage{soul}              

\newcommand{\edit}[1]{\textcolor{black}{#1}}
\newcommand{\revision}[1]{{\color{black} #1}}

\usepackage{tabularx}

\AtBeginDocument{%
  }

\setcopyright{acmlicensed}
\copyrightyear{2026}
\acmYear{2026}
\setcopyright{cc}
\setcctype{by-nc-nd}
\acmConference[CHI '26]{Proceedings of the 2026 CHI Conference on Human Factors in Computing Systems}{April 13--17, 2026}{Barcelona, Spain}
\acmBooktitle{Proceedings of the 2026 CHI Conference on Human Factors in Computing Systems (CHI '26), April 13--17, 2026, Barcelona, Spain}
\acmPrice{}
\acmDOI{10.1145/3772318.3790949}
\acmISBN{979-8-4007-2278-3/2026/04}

\begin{document}

\title[Examining How Teens and Parents Respond to Cybergrooming Advances]{From Vulnerable to Resilient: Examining Parent and Teen Perceptions on How to Respond to Unwanted Cybergrooming Advances}

\author{Xinyi Zhang}
\affiliation{%
  \institution{Virginia Tech}
  \city{Blacksburg}
  \state{Virginia}
  \country{United States}}
\email{xinyizhang@vt.edu}

\author{Mamtaj Akter}
\affiliation{%
 \institution{New York Institute of Technology}
 \city{New York}
 \state{New York}
 \country{United States}}
\email{mamtaj.akter@nyit.edu}

\author{Heajun An}
\affiliation{%
 \institution{Virginia Tech}
 \city{Blacksburg}
 \state{Virginia}
 \country{United States}}
\email{heajun@vt.edu}

\author{Minqian Liu}
\affiliation{%
 \institution{Virginia Tech}
 \city{Blacksburg}
 \state{Virginia}
 \country{United States}}
\email{minqianliu@vt.edu}

\author{Qi Zhang}
\affiliation{%
 \institution{Virginia Tech}
 \city{Falls Church}
 \state{Virginia}
 \country{United States}}
 \email{qiz21@vt.edu}

\author{Lifu Huang}
\affiliation{%
 \institution{University of California, Davis}
 \city{Davis}
 \state{California}
 \country{United States}}
\email{lfuhuang@ucdavis.edu}

\author{Jin-Hee Cho}
\affiliation{%
 \institution{Virginia Tech}
 \city{Falls Church}
 \state{Virginia}
 \country{United States}}
\email{jicho@vt.edu}

\author{Pamela J. Wisniewski}
\authornote{Co-corresponding authors: Pamela J. Wisniewski and Sang Won Lee.}
\affiliation{%
 \institution{International Computer Science Institute}
 \city{Berkeley}
 \state{California}
 \country{United States}}
\email{pwisniewski@icsi.berkeley.edu}

\author{Sang Won Lee}
\authornotemark[1]
\affiliation{%
  \institution{Virginia Tech}
  \city{Blacksburg}
  \state{Virginia}
  \country{United States}}
\email{sangwonlee@vt.edu}

\renewcommand{\shortauthors}{Zhang et al.}

\begin{abstract}

Cybergrooming is a form of online abuse that threatens teens' mental health and physical safety. Yet, most prior work has focused on detecting perpetrators’ behaviors, leaving a limited understanding of how teens might respond to such unwanted advances. To address this gap, we conducted an online survey with 74 participants---51 parents and 23 teens---who responded to simulated cybergrooming scenarios in two ways: responses that they think would make teens more vulnerable or resilient to unwanted sexual advances. Through a mixed-methods analysis, we identified four types of vulnerable responses (encouraging escalation, accepting an advance, displaying vulnerability, and negating risk concern) and four types of protective strategies (setting boundaries, directly declining, signaling risk awareness, and leveraging avoidance techniques). As the cybergrooming risk escalated, both vulnerable responses and protective strategies showed a corresponding progression. This study contributes a teen-centered understanding of cybergrooming, a labeled dataset, and a stage-based taxonomy of perceived protective strategies, while offering implications for educational programs and sociotechnical interventions.

\end{abstract}

\ccsdesc[500]{Human-centered computing~Empirical studies in HCI}

\keywords{Adolescent Online Safety, Cybergrooming, Resilience}

\maketitle

\section{Introduction}

Today, 95\% of U.S. teens have access to smartphones~\cite{Park_2025} and over 93\% use social media~\cite{DeAngelis_2024}, spending an average of 4.8 hours per day~\cite{DeAngelis_2024}, exposing them to risks such as cybergrooming~\cite{whittle2014their}.
These online risks have increased for minors; research has shown that \revision{over half of teens in the U.S. meet new people online~\cite{Lenhart_2015},} and 20\% of minors have encountered online sexual solicitations from adults~\cite{Thorn_2024}. Compared to more direct forms of online sexual exploitation, such as sextortion or sexual harassment, cybergrooming is a complex process of psychological manipulation that can extend over periods ranging from a few days to several years~\cite{de2017estrategias}. It involves a perpetrator who gradually befriended a teenager, strategically building trust and emotional bonds, and eventually sexually soliciting the teen either online or offline~\cite{mladenovic2021cyber, choo2009online, marchenko2017web}. Yet, research has \revision{also} uncovered a dangerous ``stranger danger'' myth that frames unwanted online advances \revision{as originating from adult strangers (i.e., ``predators'') who are imagined as overtly suspicious or visibly ``dangerous,''~\cite{dedkova2015stranger, Curran_Winther_2025a} when in reality, teens most often encounter unwanted sexual solicitations from seemingly innocuous individuals they meet online \cite{livingstone_annual_2014, wolak_online_2008} or already know in person~\cite{alsoubai2022friends, razi2020let, hartikainen2021if}.} Regardless of who the perpetrator is, cybergrooming experiences have introduced severe consequences to minors. Children who have experienced online sexual solicitations may face severe physical consequences (e.g., unintended pregnancy, self-harm~\cite{NSPSS}) as well as mental health consequences (e.g., anxiety, depression, post-traumatic stress disorder (PTSD), and suicidal ideation~\cite{INHOPE, schittenhelm2024cybergrooming}). They are also exposed to serious safety risks during offline meetups, including sexual assaults and even human trafficking~\cite {NSPSS, Human_Trafficking_Front_2023}. These experiences can profoundly and lastingly affect teens' overall well-being.

To combat cybergrooming, prior work has focused on surveillance and detection, using computational techniques such as machine learning (ML), natural language processing (NLP), and data mining to detect perpetrator behavior~\cite{razi2021human, fauzi2023identifying, milon2022take, eilifsen2023early, prosser2024helpful, ringenberg2024assessing, rezaee2023detecting}.
However, surveillance-based approaches may fail to capture the nuanced grooming process, which can be difficult to distinguish from friendship or romantic relationships, potentially leading to missed cases or false positives. 
Furthermore, deploying such algorithms may come at the expense of teens' privacy since monitoring tools would need access to their personal data and share it with their parents or social media platforms~\cite{charalambous2020privacy}.

A complementary alternative approach is to empower teens to recognize and manage online risks.
While researchers have investigated sociotechnical approaches to support teens beyond surveillance~\cite{gabrielli2020chatbot, maeng2022designing, rita2021chatbot, piccolo2021chatbots}, 
most chatbot-based interventions function as aftermath measures, serving as moderators to reduce the burden of in-person consulting.
For example, \citeauthor{rita2021chatbot} built an AI-based chatbot, named \textit{Report It}, supporting teens searching resources or reporting cybergrooming incidents~\cite{rita2021chatbot}.
\edit{We aim to explore ways to support sociotechnical educational interventions designed to empower teens to proactively recognize and cope with cybergrooming. This goal requires shifting attention away from perpetrator-centric views of detecting perpetrator behaviors toward teen-centric strategies to understand how parents and teens perceive vulnerability and protection.}

\edit{This paper presents our first step toward that goal. We investigate how teens' and parents' thought processes work when responding to hypothetical yet realistic cybergrooming scenarios. By identifying recurring themes across their perceived \textit{vulnerable} and \textit{protective} responses, we provide insights into how parents and teens conceptualize online risk behaviors and risk-mitigating strategies.}
We study the perspectives of both teens and parents, as they may view and respond to the same online content differently.
For example, teens may view random chats online as a friendly gesture, whereas parents believe that it is a threat to teens' online safety~\cite{wisniewski2017parents}. 
Yet, ensuring teens' online safety requires collaborative efforts from both teens and parents, making it critical to understand how their perspectives align or diverge. Parents are often the primary stakeholders who care the most about their teens’ safety and are frequently responsible for making decisions about online risk management, monitoring, and protection \cite{akter2022parental, akter2023takes, wisniewski_parental_2017, hashish_involving_2014, ghosh_circle_2020}.
It is essential to understand if teens and parents perceive and respond to cybergrooming differently, and if so, how. 
These insights will contribute to the design of effective and inclusive interventions.
Thus, in this study, we ask the following \textbf{research questions}: 

\begin{itemize}
    \item \textbf{RQ1:}  \textit{When responding to unwanted online sexual solicitation scenarios, what types of responses do parents and teens believe would make teens vulnerable?}
    \item \textbf{RQ2:} \textit{When responding to unwanted online sexual solicitation scenarios, what types of responses do parents and teens believe would make teens safe?}
    \item \textbf{RQ3:} \revision{\textit{How do vulnerable expression and protective strategies vary across stages of cybergrooming?}}
\end{itemize}

To address our research questions, we conducted an online survey with 74 participants (51 parents of teens aged 13-17 and 23 teens).
Each participant responded to 10 \edit{simulated} cybergrooming scenarios by providing a vulnerable response, a protective response, and an explanation of why \edit{they believed} the protective response was safer.
From these data, we identified four vulnerable behaviors---\textit{encouraging risk escalation}, \textit{accepting an advance}, \textit{displaying vulnerability}, and \textit{negating risk concern} (RQ1)---and four protective strategies---\textit{setting boundaries at different levels}, \textit{directly declining an advance}, \textit{signaling risk awareness}, and \textit{leveraging avoidance techniques}.
We further found that both vulnerable behaviors and protective strategies followed an escalating progression aligned with cybergrooming stages,  demonstrating how participants’ responses evolved as the interaction deepened regarding cybergrooming.

Our study makes several \textbf{key contributions} to the field of teen online safety in the context of cybergrooming:

\begin{itemize}
    \item This study provides deep \edit{insights into teens' and parents' conceptualizations of vulnerable and safe responses to simulated cybergrooming scenarios, offering a resilience-centered and teen-centered perspective. }
    \item The systematically labeled dataset of participants' responses enriches the empirical foundation of cybergrooming research, \edit{offering insights into teens' and parents' thought processes. The dataset can also be used to study mismatches between parents' and teens' perceptions and their conceptual framing of online risks.}
    \item We propose a stage-based taxonomy of \revision{perceived} protective strategies, emphasizing how participants' responses evolve as the interaction escalates across stages of cybergrooming.
\end{itemize}

\section{Background}

\subsection{Understanding Cybergrooming Progression and Its Impacts on Teens}

Understanding cybergrooming requires examining both how it is defined and how it unfolds behaviorally.
\citeauthor{schittenhelm2024cybergrooming} defines cybergrooming as a process in which an adult uses digital communication tools to cultivate a deceptive relationship with a minor, ultimately aiming to facilitate sexual exploitation~\cite{sahaModelingStressSocial2017}.
A literature review identified four recurring elements across definitions: perpetrators target minors, initiate contact through online technologies, pursue sexual objectives, and construct a pseudo-relationship to gain the teen’s trust~\cite{schittenhelm2024cybergrooming}. A widely used framework for understanding this progression is \citeauthor{o2003typology}'s six-stage taxonomy~\cite{o2003typology}, developed through a five-year empirical study involving participant observation. The stages include:
(1) \textit{Friendship Forming}: perpetrators initiate contact and verify the teen’s identity through conversational exchanges, photos, or video calls;
(2) \textit{Relationship Forming}: perpetrators deepen familiarity by discussing personal topics such as family, school, and interests;
(3) \textit{Risk Assessment}: perpetrators evaluate whether the teen is isolated, ensure no external monitoring is occurring, and request secrecy;
(4) \textit{Exclusivity}: perpetrators cultivate emotional dependence by framing the relationship as unique or “special”;
(5) \textit{Sexual}: perpetrators gradually introduce sexualized topics, probe prior experiences, and normalize explicit conversations;
(6) \textit{Conclusion}: perpetrators transition to proposing offline meetings and coordinating logistics.
This taxonomy has shaped subsequent research on cybergrooming detection and analysis~\cite{Curran_Winther_2025a, razi2023sliding, mladenovic2021cyber, razi2021human, kloess2014online, bours2019detection}. For example, Gupta et al.~\cite{gupta2012characterizing} identified the \textit{relationship-forming} stage as most dominant in perpetrator conversations, while Razi et al.~\cite{razi2023sliding} used the taxonomy to classify online interactions as safe or unsafe. Gunawan et al.~\cite{gunawan2016detecting} further expanded this work by identifying 17 linguistic and behavioral features mapped to \citeauthor{o2003typology}’s stages, offering finer-grained insight into trust-building and escalation. Building on these stage-based and feature-based perspectives, our study examines how teens express both vulnerable and protective behaviors when encountering cybergrooming attempts.

While these frameworks describe how grooming progresses, understanding its broader significance requires examining the risks and vulnerabilities that heighten teens’ susceptibility. Prior work documents how perpetrators tailor their strategies to teens’ emotional needs~\cite{whittle2013review}, how many youth experience online sexual solicitation~\cite{greene2020experiences}, and how such encounters can lead to long-term psychological harms, including anxiety, shame, and reduced self-efficacy~\cite{whittle2013review, schittenhelm2024cybergrooming}. Digital affordances such as anonymity, persistence, and direct access across platforms further amplify these risks. Despite extensive research on perpetrator tactics and teen vulnerabilities, relatively little is known about how teens actively respond during grooming interactions. Understanding these responses is critical for developing teen-centered interventions that move beyond detection to proactively support online safety.

\subsection{Shifting From Perpetrator Detection to Teen-Centered Resilience}

While prior work identified key risk factors and contextual vulnerabilities, much of the existing literature focused on describing the problem rather than designing solutions. To combat cybergrooming, researchers have developed numerous computational approaches, with most computer science literature focuses on detecting perpetrators’ behaviors using machine learning (ML) and natural language processing (NLP) techniques (e.g., ~\cite{razi2021human, bours2019detection, munoz2020smartsec4cop, ashcroft2015step, bogdanova2014exploring, aiello2014detecting, gunawan2016detecting, gupta2012characterizing, fauzi2023identifying, milon2022take, eilifsen2023early, isaza2022classifying, prosser2024helpful, ringenberg2024assessing, rezaee2023detecting, borj2019predatory, fauzi2020ensemble}). These models leverage content-, emotion-, psychology-, and text-based features to detect suspicious conversations~\cite{an2025toward}. For example, \citeauthor{bours2019detection}~\cite{bours2019detection} compared five classifiers—Logistic Regression, Ridge, Naïve Bayes, Support Vector Machines, and Neural Networks—across three levels of analysis (message, author, and conversation) and identified neural networks using TF-IDF features as most effective. Similarly, Michalopoulos et al.~\cite{michalopoulos2014gars} introduced the \textit{Grooming Attack Recognition System (GARS)}, which integrated user history, personality recognition, and exposure-time tracking to assess risk levels and notify multiple stakeholders when thresholds were exceeded. However, such monitoring-based systems can raise significant privacy concerns, as they grant parents extensive access to teens’ online activities~\cite{charalambous2020privacy, akter2022parental}.

Beyond detection, researchers have explored conversational agents for online safety~\cite{ueda2021cyberbullying, cohen2018education, rita2021chatbot}, though only \citeauthor{rita2021chatbot}’s work directly targets cybergrooming. They developed \textit{Report It}, an AI-based reporting chatbot designed to support children in disclosing grooming incidents and accessing resources safely~\cite{rita2021chatbot}. However, most chatbot systems remain reactive, intervening only after harm occurs, while relying on rule-based or intent-verification approaches and rarely tested outside controlled scenarios. Despite advancements, existing interventions primarily center on perpetrator detection and post-incident responses.
Few studies examine how teens or parents think through to respond to cybergrooming situations or how they conceptualize vulnerability and safety. Our study addresses this gap by investigating how teens and parents respond to simulated cybergrooming scenarios, allowing us to directly engage them as primary stakeholders in understanding cybergrooming dynamics.

\section{Methodology}
To address our research questions, we conducted an online survey study with 74 participants (51 parents and 23 teens) to examine how they responded to cybergrooming scenarios. 
In this section, we present the overview, scenario design, participant recruitment, ethical considerations, and analytic methods.
Our study has been reviewed and approved by our Institutional Review Board (IRB).

\begin{figure*}[h]
        \centering
        \includegraphics[scale=0.34]{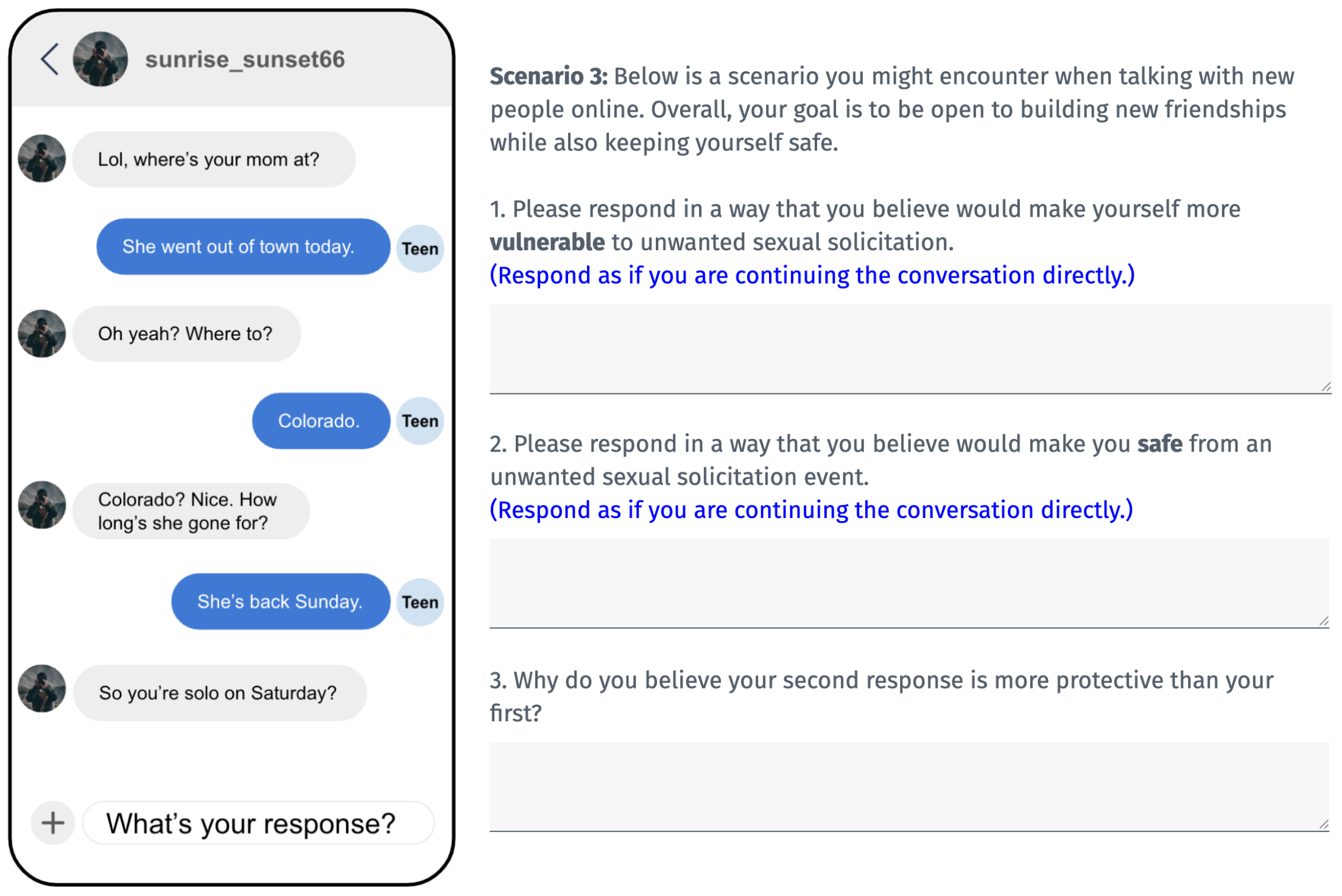}
        \caption{Example cybergrooming scenario used in the survey, showing chat dialogue and response prompts.}
        \label{S3}
        \Description{This figure is an example of a simulated cybergrooming scenario presented to participants. On the left side, the conversation unfolds in a smartphone-style chat window. The perpetrator's messages are in gray speech bubbles, and the teen's messages are in blue ones. The exchange unfolds as follows: perpetrator: ``Lol, where's your mom at?'' Teen: ``She went out of town today. Perpetrator: ``Oh yeah? Where to?'' Teen: ``Colorado. Perpetrator: ``Colorado? Nice. How long's she gone for?'' Teen: ``She's back Sunday.'' perpetrator: ``So you're solo on Saturday?'' At the bottom of the chat window, a text entry field labeled ``What's your response?'' prompts the participant to continue the conversation. On the right side, the survey instruction shows scenario 3: below is a scenario you might encounter when talking with new people online. Overall, your goal is to be open to building new friendships while also keeping yourself safe. Then, participants need to answer three questions: (1) Please respond in a way that you believe would make yourself more vulnerable to unwanted sexual solicitation (respond as if you are continuing the conversation directly), (2) Please respond in a way that you believe would make you safe from an unwanted sexual solicitation event, and (3) Why do you believe your second response is more protective than your first?}
\end{figure*}

\subsection{Overview}
\edit{We implemented our survey using QuestionPro~\cite{QuestionPro} and framed it as ``a project to understand how to protect teens from unwanted online sexual solicitations.''}
It consisted of three parts: (1) a screening survey to determine eligibility, (2) a short pre-survey that collected demographics, social media use patterns, \edit{and teens’ prior experiences with online unwanted sexual solicitations}, and (3) 10 cybergrooming scenarios \edit{followed by exit questions assessing perceived scenario realism}.
In the screening stage (see Appendix~\ref{screening}), participants confirmed their eligibility (e.g., age, residence, language, technology access, social media use) and willingness to proceed despite potentially sensitive content.
Teens also verified their willingness to participate in a brief video check-in meeting with the research team.
In the pre-survey (see Appendix~\ref{parent_pre}, \ref{teen_pre}), parents and teens provided demographics and their social media use.
\edit{For parent participants, we asked about their awareness of their teens’ prior experiences with online unwanted sexual solicitation, while for teen participants, we asked about their own prior experiences~\cite{mitchell2011youth}.}
In the main survey, participants reviewed 10 scenarios and, in each, responded to the following three questions from the teen's perspective:
(1) ``Please respond in a way that you believe would make you more vulnerable to an online unwanted sexual solicitation event,''
(2) ``Please respond in a way that you believe would make you safe from an unwanted sexual solicitation event,''
and (3) ``Why do you believe your second response is more protective than your first?''
\edit{At the end of the survey, participants also reflected on how realistic they found the scenarios so that we could assess the ecological validity of our scenario design.}
The survey was designed to be completed in approximately 1 hour.

\subsection{Scenario Design}

A key design consideration was ensuring that the scenarios felt realistic while avoiding unnecessary harm.
Inspired by \citeauthor{agha2024tricky}~\cite{agha2024tricky}, we opted to use simulated cybergrooming scenarios to show the development of cybergrooming without exposing participants to real cybergrooming risks.
\revision{
We asked participants to engage with the simulated cybergrooming scenarios. 
We framed the scenarios ambiguously by referring to the person in the scenarios as meeting ``new people online,'' rather than in explicit terms (e.g., ``perpetrator,'' ``predator''), highlighting that the goal is to engage safely while preventing unwanted online sexual solicitation. Therefore, we did not frame scenarios as definitive cases of cybergrooming. As such, the scenario involving an online stranger, i.e., a person they do not know, was used to debunk the stranger-danger myth~\cite{Curran_Winther_2025a, dedkova2015stranger} and to place them in a situation where a stranger can seem friendly and not appear unsafe.
These scenarios also reflect contemporary patterns of youth online socialization, where interacting with new people online is more common than many adults expect.}

\begin{table*}[h]
    \centering
    \small
    \caption{Overview of representative cybergrooming scenarios}
    \begin{tabular}{p{0.3cm} p{4cm} p{2.5cm} p{6.8cm}}\toprule
    \textbf{S\#} & \textbf{Name} & \textbf{Stage} & \textbf{Description} \\ \midrule
    S1 & Asking for phone number & Friendship Forming & The online stranger is asking for the teen's phone number.\\ 
    S2 & Asking for family information & Relationship Forming & The online stranger is requesting the teen's family information and relationship.\\
    S3 & Acknowledging wrongdoing & Risk Assessment & The online stranger is acknowledging to the teen that chatting with an underage is wrong.\\
    S4 & Asking if alone & Risk Assessment & The online stranger is asking if the teen will be alone on the weekend.\\
    S5 & Asking about romantic relationship & Exclusivity & The online stranger is asking about the teen's current romantic relationship.\\
    S6 & Building mutual trust & Exclusivity & The online stranger is trying to show that they are special to the teen.\\
    S7 & Asking for sensitive pictures & Sexual & The online stranger is complimenting the teen's body and asking for their hot pics.\\
    S8 & Asking about sexual experiences & Sexual & The online stranger is asking about the teen's previous sexual experiences and their willingness to have sex.\\
    S9 & Ice-cream fight & Sexual & The online stranger is flirting with the teen.\\
    S10 & Offline meetup request & Conclusion & The online stranger is offering an offline meetup to the teen.\\ \bottomrule
    \end{tabular}
    \label{tab:cybergrooming_stage}
\end{table*}

\begin{table*}[h]
    \centering
    \small
    \caption{Comparison examples of original conversation snippet extracted from the Perverted-Justice (PJ) dataset and modified dialogue used in the scenario.}
    \begin{tabular}{p{6.5cm}|p{7cm}} \toprule
    \textbf{Original PJ Snippet} & \textbf{Modified Dialogue} \\ \hline
    \textbf{Perpetrator: }lolz, where is your mom baby?& \textbf{sunrise\_sunset66: } Lol, where’s your mom at?\\ 
    \textbf{Victim: }she left day & \textbf{Teen: }She went out of town today.\\ 
    \textbf{Perpetrator: } to where? & \textbf{sunrise\_sunset66: }Oh yeah? Where to?\\ 
    \textbf{Victim: }Colorado & \textbf{Teen: } Colorado. \\
    \textbf{Perpetrator: }Colorado? & \textbf{sunrise\_sunset66: }Colorado? Nice. How long’s she gone for?\\ 
    \textbf{Victim: }ya & \textbf{Teen: }She’s back Sunday.\\ 
    \textbf{Perpetrator: }and you are gonna be alone? & \textbf{sunrise\_sunset66:} So you’re solo on Saturday?\\ 
    \textbf{Victim: }ya & \\ 
    \textbf{Perpetrator: }for how many days? & \\ 
    \textbf{Victim: }she be home sun & \\ 
    \textbf{Perpetrator: }so you will be alone on Saturday? & \\ \bottomrule
    \end{tabular}
    \label{tab:scenario_content}
\end{table*}

We initially created scenarios to capture each identified cybergrooming characteristic~\cite{gunawan2016detecting}, distributed across six stages of cybergrooming~\cite{o2003typology, gunawan2016detecting}.
However, to alleviate participants' mental load---which could affect data quality if the survey lasted more than an hour---we retained only the 10 most representative scenarios (see Appendix~\ref{scenarios}), denoted as ``S1''--``S10'' (see Table~\ref{tab:cybergrooming_stage}). 
For designing each scenario, we first selected authentic online conversation snippets from the Perverted-Justice (PJ) dataset, a conversation collection of cybergrooming perpetrators and victims.
Since the PJ dataset was collected between 2003 and 2016, its language might be out of date. Therefore, we utilized GPT-4o to modernize the language to reflect modern teens' online chatting behaviors (see the prompt in~\ref{prompt}).
Then, our research team discussed and modified the conversation details together for the final version.
We ensured the scenarios did not imply a specific gender for either the teen user or the online stranger by using inclusive language that applies to all genders. This decision aligns with research showing that teens face cybergrooming risks irrespective of their gender~\cite{dorasamy2021parents}.
Table~\ref{tab:scenario_content} shows an example of before and after conversation modification.
To better simulate real-world online chatting, we created a mobile social media interface to present each modified conversation (see Figure~\ref{S3}). 
It contains two users in each scenario, one is an adult user named `sunrise\_sunset66,' and the other is a teen user. 
Parent participants were asked to pose themselves as their teens who used social media, whereas teen participants were instructed to write responses as themselves.

\subsection{Participant Recruitment}
\edit{To avoid causing embarrassment or potential conflicts among family members after participating in a study involving sensitive content~\cite {cranor2014parents}, we intentionally recruited unpaired parents and teens who come from different households.}
We adopted different recruiting strategies for parents and teens.

\textbf{Participation criteria.} 
Parents or legal guardians were eligible to participate if they lived in the U.S. with a teen aged 13–17, were fluent in English, and reported that their teen owned a smartphone and used social media.
Teens were eligible if they lived in the U.S. with their parent or legal guardian, were fluent in English, owned a smartphone, and used social media. 
Additionally, parents or legal guardians were required to provide informed consent for their teen’s participation before the teen could assent to the study.
\revision{Our study did not require participants to have prior experience with unwanted online sexual solicitations; however, in our pre-survey, we inquired about their relevant past experiences (see Figure~\ref{priorExperience}).}

\textbf{Parent recruitment.}
We recruited 51 parents via Prolific~\cite{Prolific}, a professional platform for researchers to recruit participants. It has built-in screening filters used to target a specific population (e.g., parents of teenagers).
Additional screening questions at the start of the survey further ensured their eligibility.
After reviewing and signing the informed consent form, participants completed the survey and received \$10 compensation upon passing the quality check. 
We denote parent participants as ``P''. 

\textbf{Teen recruitment.} 
\edit{Recruiting teens was particularly challenging due to the ethical and logistical constraints of reaching out to them for a study involving sensitive content.
We first contacted youth service organizations and placed flyers in local places such as libraries, the YMCA, and churches.
Due to the low response rates, we later expanded to social media (X, Facebook groups) and snowball sampling, though this led to an increase in fraudulent responses. 
We also tried to build collaboration with the local schools, which was not feasible due to the study's high-risk nature.
In addition, many parents were hesitant to consent because they might not expect their teens to be exposed to such materials, even for research purposes.}

We tailored the teens' onboarding process to differ from the parents'.
First, parents or teens could complete the screening survey by scanning the QR code or visiting the link on a flyer.
Then, we cross-checked their self-reported location (city or town) with IP geolocation, rejecting mismatches (e.g., someone reporting a U.S. city but with an IP from abroad). 
Then, we invited eligible teens and their parents to a short video meeting to sign the parent permission form and the teen assent form.
During the video meetings, we required them to appear on camera to verify: (1) they were real people, not bots, and (2) they were parents and teens as they claimed.  
Out of 392 initial responses, 23 teens successfully passed verification, completed the survey, and received \$30 compensation.
We denoted teen participants as ``T''.

\subsection{Ethical Considerations} 

Since our study contains mild sexual advances and involves teen participants, we took extra care to consider research ethics in the design of our study. In this section, we describe the multi-layered safeguards we implemented to mitigate potential risks to participants throughout the study.

\subsubsection{Ethical Considerations During Study Design}
\edit{\citeauthor{agha2024tricky} conducted a co-design study with 20 teens to understand how they perceive, evaluate, and help shape risky online interaction scenarios for research. Their work demonstrated that teens are not only capable of reasoning about online dangers, but also have expectations for what designed scenarios should be like. Teens emphasized that simulated risky scenarios should be realistic enough for them to believe, yet should avoid overly obvious~\cite{agha2024tricky}.}
Building on these insights, when designing scenarios, we intentionally balance realism with participant safety by avoiding overly explicit sexual content and reframing sensitive expressions to minimize discomfort and potential harm.
\edit{Additionally, we ordered the scenarios from least to most sensitive so that participants would encounter progressively riskier content only if they chose to continue. This structure provided opportunities for them to stop the study at any point if they felt uncomfortable responding to a particular scenario.}
\edit{Lastly, when designing our questions, we intentionally limited participants to providing a single response to each scenario rather than engaging in a multi-turn conversation. This approach allowed us to examine their perceptions of vulnerable versus protective strategies without requiring them to participate in extended simulated interactions.
}

\subsubsection{Ethical Considerations for Study Implementation}

\edit{When implementing our study, we informed participants multiple times throughout the study process that they might be exposed to sexually related content.} \edit{During the eligibility screening, we provided a content warning and displayed the most sensitive scenario along with the exact questions that participants will answer in the main survey. This ensured that participants were fully aware of the study's nature before deciding whether to proceed. If teens were eligible and willing to participate in our study, we then scheduled the video meeting with both parents and teens. During the meeting, we introduced our study, answered any questions they had, and assisted them with completing the consent process. We asked parents to review all scenarios in advance and provided informed consent only after confirming they were comfortable with their teen’s participation. We then invited teens to review and sign their assent form.
We emphasized that participation was entirely voluntary and that both parents and teens could decline or discontinue at any time, including after the study, without any penalty or explanation. This step ensured the voluntary agreement of both parents and teens.
During participation, we intentionally avoided having participants complete the survey in the presence of the research team, ensuring that they could step away from any scenario without external pressure. Participants were also free to pause the study whenever they felt uncomfortable and continue only when they felt ready, thereby preserving autonomy and minimizing potential distress.
After completing the main survey, participants received a short debrief in which they evaluated how realistic the scenarios felt and explained why.
It allowed participants to reflect on their study experience and provided the research team with feedback on comfort levels and scenario appropriateness.
Additionally, we also provided both the research team's and the IRB office's contact information in case they were angry or upset during or after the study.}

\subsection{Data Analysis}
Initially, we collected a total of 2,220 responses (74 participants * 10 scenarios * 3 entries), consisting of 740 vulnerable responses, 740 protective responses, and 740 explanations. 
Since both protective responses and their accompanying explanations described protective strategies, we analyzed each response-explanation set as a pair, yielding 740 entries in total. 
To address our research questions, we adopted a mixed-method approach inspired by~\citeauthor{alluhidan2024teen}'s work~\cite{alluhidan2024teen}.
We began with a thematic analysis of parents' responses to the scenarios.
First, we collaboratively open-coded a subset of responses to identify emerging themes and develop initial codebooks, ensuring shared understanding and consistent interpretation. 
Then, the first author completed the remaining parents' responses and refined the codes iteratively based on the other authors' feedback.
\edit{In the end, we produced two initial codebooks for parents' responses: one for capturing vulnerable behaviors and one for capturing protective strategies.
We then applied the same coding procedures to the teens' responses, resulting in a teen-vulnerable behavior codebook and a teen-protective strategy codebook.
By comparing the parent and teen codebooks, we found a high alignment across both vulnerable behavior categories and protective strategy categories.
Based on this consistency, we merged the two vulnerable behavior codebooks into a shared vulnerable behavior codebook and merged the two protective strategy codebooks into a shared protective strategy codebook.}
\edit{Once finalized, we conceptually grouped the codes into high-level themes and subthemes (see Table~\ref{tab:vulnerable_codebook},~\ref{tab:protective_codebook}).} 
We then calculated the frequency of each theme or subtheme by counting the number of responses coded into it (denoted as $n$). We then divided each count by the total number of protective or vulnerable responses (740) to obtain the corresponding percentages.

For the quantitative analysis, we calculated corresponding percentages at both the theme- and subtheme-levels (see Tables~\ref{tab: percentage_comparison_vul}, \ref{tab: percentage_comparison_pro}). 
\revision{
We applied chi-square ($\chi^2$)  analyses with adjusted standardized residuals in an exploratory manner to descriptively examine distributional variations in participants’ perceived vulnerable behaviors and protective strategies across cybergrooming scenarios. Because responses were collected from the same participants across multiple scenarios, these analyses are intended for pattern exploration rather than confirmatory inference. Adjusted standardized residuals were reported as descriptive diagnostics.}

\begin{figure*}[h]
        \centering
        \includegraphics[scale=0.42]{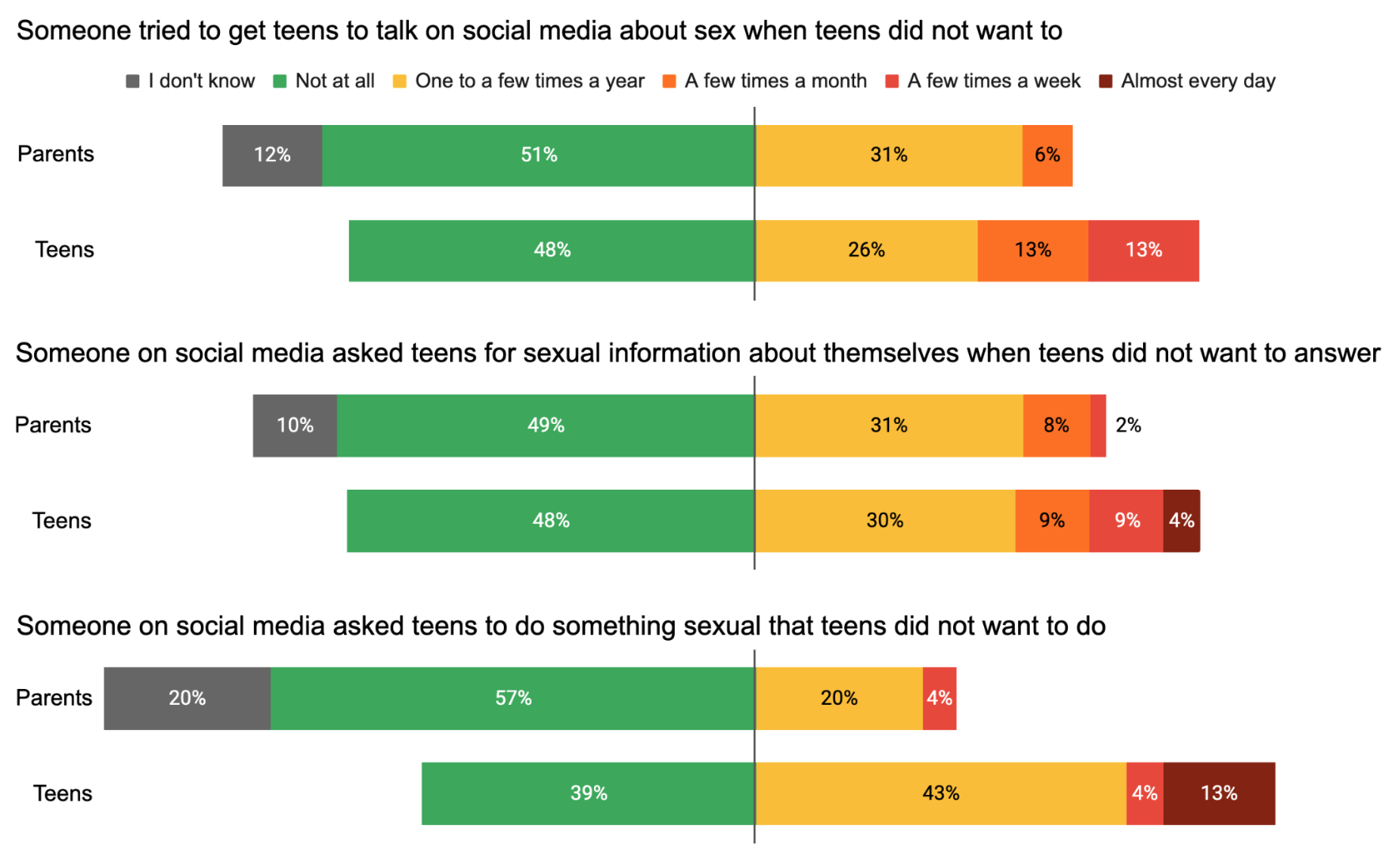}
        \caption{Parents’ perceptions of their teens’ prior exposure to unwanted online sexual solicitations and teens’ self-reported experiences.}
        \label{priorExperience}
        \Description{The figure shows bar charts comparing parents’ reports and teens’ self-reports of prior exposure to three types of unwanted online sexual solicitations. The legend has six categories for exposure frequency, including I don't know, not at all, one to a few times a year, a few times a month, a few times a week, and almost every day. For the question “someone tried to get teens to talk on social media about sex when teens did not want to,” parents reported 12\% “I don’t know,” 51\% “Not at all,” and 37\% total across the remaining frequency categories. Teens reported 48\% “Not at all,” 26\% “Yearly,” 13\% “Monthly,” and 13\% “Weekly or daily.” For the question “someone on social media asked teens for sexual information about themselves when teens did not want to answer,” parents reported 10\% “I don’t know,” 49\% “Not at all,” and 41\% across the remaining categories. Teens reported 48\% “Not at all,” 30\% “Yearly,” 9\% “Monthly,” and 13\% “Weekly or daily.” For the question “someone on social media asked teens to do something sexual that teens did not want to,” parents reported 20\% “I don’t know,” 57\% “Not at all,” and 24\% across the remaining categories. Teens reported 39\% “Not at all,” 43\% “Yearly,” and a combined 17\% across “Monthly,” “Weekly,” and “Almost daily.”}
\end{figure*}

\section{Results}
After describing the participants' background information, we present our findings by addressing the RQs in sequence. We first discuss the main themes and subthemes that emerged from the qualitative analysis for vulnerable behaviors (RQ1) and protective strategies (RQ2). We then investigate how these perceptions vary across stages of cybergrooming, with a brief comparison between parents and teens (RQ3).

\subsection{Descriptive Statistics of the Participants}

\edit{We recruited 74 participants in total: 51 parents and 23 teens. 
They represent diverse backgrounds in terms of age, gender, sexual orientation, race, and daily social media use.
Additionally, parent participants reported their highest level of education, household income, and marital status.
See Table~\ref{tab:demographics} for details.} 
\revision{Drawing from pre-validated survey questions from the Youth Internet Safety Survey (YISS)~\cite{al-salehi_jammer_2021}, Figure \ref{priorExperience} provides an overview of parents’ perceptions and teens’ self-reports regarding prior exposure to unwanted sexual solicitations on social media.} This figure demonstrates that \revision{more than half of the teens (based on self-reported values) had previously encountered such solicitations.}
Across all three types of solicitations, teens consistently reported higher levels of prior exposure than parents.
Most teens who had encountered such solicitations reported that these experiences occurred once to a few times per year, with a notable number reporting more frequent exposure.
In contrast, parents' reports were more concentrated in the ``not at all'' or ``I don't know'' categories, indicating that parents may underestimate the frequency with which their teens experience such solicitations.

\edit{In addition, participants' reflections on the realism of the scenarios showed strong agreement across both groups. Among parents, 96\% (N = 49) reported that the scenarios felt realistic, and only 4\% (N = 2) disagreed. Among teens, 96\% (N = 22) thought the scenarios were realistic while 4\% (N = 1) did not. Here, ``N'' refers to the number of participants. By diving into participants' explanations, about 69\% (N = 35) of parents believed that such interactions genuinely occur in real life, with one parent emphasizing that real situations could be \textit{``even worse''} (P23). Another group of parents (27\%, N = 14) reported that they or their teens had encountered such situations in their real lives.
Similarly, among teens, 57\% (N = 13) stated that they had directly experienced unwanted online sexual solicitations, and an additional 39\% (N = 9) stated that they believed such scenarios happen to teens even if they had not personally encountered them.}

\subsection{\edit{Vulnerable Behaviors That Participants Believed Could Make Teens More Vulnerable to Unwanted Online Sexual Solicitations (RQ1)}}

Understanding how teens and parents responded to simulated cybergrooming scenarios in vulnerable ways provided insight into behaviors that they believed could increase their exposure to online strangers.
Through thematic analysis, we categorized participants' vulnerable responses in Table~\ref{tab:vulnerable_codebook}, including encouraging risk escalation, accepting advances, displaying vulnerability, and negating online strangers' risk concern.

\begin{table*}[h]
    \centering
    \footnotesize
    \caption{\edit{Themes/subthemes identified from participants’ perceived vulnerable behaviors ($n$ represents the number of responses).}}
    \begin{tabular}{|p{3.8cm}|p{2.3cm}|p{8.1cm}|}
        \hline
        \rule{0pt}{2.5ex}\textbf{Themes/Subthemes} & \multicolumn{2}{l|}{\rule{0pt}{2.5ex}\textbf{Illustrative Quotations}} \\ \hline
        \multicolumn{3}{|l|}{\cellcolor{gray!20}\rule{0pt}{2.5ex}\textbf{Encouraging Risk Escalation (49\%, n = 359)}}\\        \hline
                 
        \shortstack[l]{\rule{0pt}{2.5ex}\textbf{Reciprocating interest} \\ (25\%, n = 186)} & \multicolumn{2}{p{10cm}|}{
            \begin{tabular}[b]{@{}p{10cm}@{}}
                \textbf{Response:} \textit{``Aww, thank you, you’re so sweet, unlike the others!''} (S6/T) \\
                \textbf{Explanation:} \textit{``... the first one makes me seem more innocent and therefore a sort of easier prey ...''}
            \end{tabular}} \\ \hline
        \shortstack[l]{\rule{0pt}{2.5ex}\textbf{Probing a question} \\ (9\%, n = 70)} & \multicolumn{2}{p{10cm}|}{
            \begin{tabular}[b]{@{}p{10cm}@{}}
                \textbf{Response:} \textit{``Why? You want to date me?''} (S5/T) \\
                \textbf{Explanation:} \textit{``... the first one lets him know that I am single and would let him date me ...''} 
            \end{tabular}} \\ \hline
        \shortstack[l]{\rule{0pt}{2.5ex}\textbf{Facilitating offline meetup} \\ (8\%, n = 61)} & \multicolumn{2}{p{10cm}|}{
            \begin{tabular}[b]{@{}p{10cm}@{}}
                \textbf{Response:} \textit{``Yeah definitely. One day!''} (S10/P) \\
                \textbf{Explanation:} \textit{``The first response just makes them feel we will meet soon.''} 
            \end{tabular}} \\ \hline
        \shortstack[l]{\rule{0pt}{2.5ex}\textbf{Engaging sexual conversation} \\ (6\%, n = 42)} & \multicolumn{2}{p{10cm}|}{
            \begin{tabular}[b]{@{}p{10cm}@{}}
                \textbf{Response:} \textit{``I haven't had sex, but I want to one day.''} (S8/P) \\
                \textbf{Explanation:} \textit{``The first response is accepting of the questions and gives them personal information about myself, allowing them to feel closer to me.''} 
            \end{tabular}} \\ \hline
        
        \multicolumn{3}{|l|}{\cellcolor{gray!20}\rule{0pt}{2.5ex}\textbf{Accepting an Advance (36\%, n = 264)}}\\
        \hline
        \shortstack[l]{\rule{0pt}{2.5ex}\textbf{Sharing information} \\ (28\%, n = 208)} & \multicolumn{2}{p{10cm}|}{
            \begin{tabular}[b]{@{}p{10cm}@{}}
                \textbf{Response:} \textit{``Here's my number! xxx-xxx-xxx.''} (S1/P) \\
                \textbf{Explanation:} \textit{``The first response directly gives access to my phone number, which would allow the person to call/text and potentially find out more information about where I live or who I am.''} 
            \end{tabular}} \\ \hline
        \shortstack[l]{\rule{0pt}{2.5ex}\textbf{Sharing pictures} \\ (6\%, n = 47)} & \multicolumn{2}{p{10cm}|}{
            \begin{tabular}[b]{@{}p{10cm}@{}}
                \textbf{Response:} \textit{``Yes, I have other pics that I'm not supposed to post and they are hot.''} (S7/T) \\
                \textbf{Explanation:} \textit{``My first response is quite vulnerable. He would believe that he can get me if he wants.''} 
            \end{tabular}} \\ \hline
        \shortstack[l]{\rule{0pt}{2.5ex}\textbf{Accepting with a condition} \\ (1\%, n = 9)} & \multicolumn{2}{p{10cm}|}{
            \begin{tabular}[b]{@{}p{10cm}@{}}
                \textbf{Response:} \textit{``Well, if it's really just between us, I guess I can send you some pictures I didn't post yet.''}  \\
                \textbf{Explanation:} \textit{``... in the first one I agreed to send the pics.''} (S7/P)
            \end{tabular}} \\ \hline

        \multicolumn{3}{|l|}{\cellcolor{gray!20}\rule{0pt}{2.5ex}\textbf{Displaying Vulnerability (21\%, n = 157)}}\\
        \hline
        \shortstack[l]{\rule{0pt}{2.5ex}\textbf{Confirming unsupervised} \\ (9\%, n = 64)} & \multicolumn{2}{p{10cm}|}{
            \begin{tabular}[b]{@{}p{10cm}@{}}
                \textbf{Response:} \textit{``Yep, I will be alone all day Saturday.''} (S4/P) \\
                \textbf{Explanation:} \textit{``The first answer lets the person know the teen is by themselves.''} 
            \end{tabular}} \\ \hline
        \shortstack[l]{\rule{0pt}{2.5ex}\textbf{Expressing uncertainty} \\ (8\%, n = 61)} & \multicolumn{2}{p{10cm}|}{
            \begin{tabular}[b]{@{}p{10cm}@{}}
                \textbf{Response:} \textit{``Maybe I would or maybe I would eat some off of you. Who knows, LOL.''} (S9/P) \\
                \textbf{Explanation:} \textit{``The first is engaging in the activity or letting it progress.''} 
            \end{tabular}} \\ \hline
        \shortstack[l]{\rule{0pt}{2.5ex}\textbf{Showing emotional vulnerability} \\ (4\%, n = 32)} & \multicolumn{2}{p{10cm}|}{
            \begin{tabular}[b]{@{}p{10cm}@{}}
                \textbf{Response:} \textit{``I really could use a friend. [...] I cannot find that anywhere. Even at school, boys are so dumb sometimes.''} (S6/P) \\
                \textbf{Explanation:} \textit{``The first text is putting down boys at her school and making this guy better than anyone else because he seems different.''} 
            \end{tabular}} \\ \hline

        \multicolumn{3}{|l|}{\cellcolor{gray!20}\rule{0pt}{2.5ex}\textbf{Negating Risk Concern (15\%, n = 108)}}\\
        \hline
        \shortstack[l]{\rule{0pt}{2.5ex}\textbf{Keeping secret} \\ (9\%, n = 69)} & \multicolumn{2}{p{10cm}|}{
            \begin{tabular}[b]{@{}p{10cm}@{}}
                \textbf{Response:} \textit{``... I have not told them about you.''} (S2/T)\\
                \textbf{Explanation:} \textit{``The first one lets the person know that nobody else knows about him, letting him feel safer about still talking to me.''} 
            \end{tabular}} \\ \hline
        \shortstack[l]{\rule{0pt}{2.5ex}\textbf{Expressing no worries} \\ (5\%, n = 39)} & \multicolumn{2}{p{10cm}|}{
            \begin{tabular}[b]{@{}p{10cm}@{}}
                \textbf{Response:} \textit{``Age is just a number.''} (S3/P) \\
                \textbf{Explanation:} \textit{``The first response expresses more that the teen doesn't care about age and is more willing to do things against the rules.''} 
            \end{tabular}} \\ \hline

    \end{tabular}
    \label{tab:vulnerable_codebook}
\end{table*}

\subsubsection{\textbf{\edit{Engaging in Conversations That Encourage Risk Escalation}}}

\edit{When asked to act vulnerable with online strangers intentionally, participants reported various ways of continuing to engage in conversations (49\%, $n = 359$). Yet, these conversations may escalate the risk situations by shifting the conversations from casual exchanges to personal, intimate, or suggestive topics. In certain cases, the conversations may lead to potential offline meetings.}

\edit{In many cases, participants demonstrated \textbf{reciprocating interest}, encouraging the conversation to continue by \textit{expressing kindness}, \textit{enthusiasm}, \textit{compliments}, or \textit{introducing flirtatious undertones} (25\%, n = 186).}
\edit{Such responses function as subtle signals of openness, often shifting a casual exchange into a more emotionally charged or suggestive interaction.}
In some cases, participants deepened the interaction by making it feel more exclusive and meaningful, lowering conversational boundaries. As T12 responded and explained in S6 (building mutual trust):
\begin{quote}
    \textit{``Aww, thank you, you’re so sweet, unlike the others! [... (vulnerable response) makes me seem more innocent and therefore a sort of easier prey ...]''} (S6/T12)
\end{quote} 
\noindent This example illustrated both the participant's direct response to the online stranger and their explanation of why it conveyed greater vulnerability (shown within the brackets).
We applied this quotation style consistently in the excerpts that follow.
Such personal disclosures shifted the conversation from surface-level chatter to emotionally loaded territory, potentially paving the way for deeper involvement. 
Other participants responded with enthusiasm or humor to validate the online stranger’s remarks and keep the tone upbeat, signaling a willingness to continue the interaction without directly introducing risk. 
For example, in S1 (asking phone number), P31 replied \textit{``That would be great, let’s do it!''} while P24 teased in response to S9 (ice-cream fight), \textit{``Haha, you’re so bad and funny.''} 

\edit{In 9\% of the responses (n = 70), participants extended the conversation by \textbf{probing a question}, either by asking a follow-up related to the \textit{ongoing topic} or by proactively shifting to \textit{a new one}.}
These questions could be as casual as a friendly inquiry or as subtle as a provocative hint.
\edit{They often serve as an entry point into deeper or riskier interactions, signaling participants' curiosity and inviting the other person to continue engaging.}
For example, in S5 (asking about romantic relationship), T23 responded with a probing question that kept the conversation going and subtly encouraged the online stranger to think in a more intimate or personal direction:
\begin{quote}
    \textit{``Why? You want to date me? [... (vulnerable response) lets him know that I am single and would let him date me ...]''} (S5/T23)
\end{quote}

\edit{In some responses (8\%, n = 61), participants believed \textbf{facilitating an offline meetup}, either by \textit{passively accepting the request} or \textit{proactively offering it}, as a risky behavior that shifts the interaction from an online exchange to an in-person encounter.}
These proposals often followed probing questions or reciprocated interest, once a sense of familiarity or intimacy was established. 
For instance, T6 asked in S6 (building mutual trust), \textit{``[...] you want to hang out sometime so I can actually meet u?''}, explicitly moving the exchange toward a real-world encounter. 
Some participants engaged in logistical planning, such as P15 replied in S10 (offline meetup request), \textit{``Sure, give me a date, time, and location''}. Others even introduced the idea themselves while responding to compliments, as T1 stated in S4 (asking if alone), \textit{``Wow, you have a really nice tattoo; I can't wait to check it out in person!''} 
These responses revealed how quickly the conversation could shift from hypothetical suggestions to concrete arrangements for an in-person meeting.

\edit{Sometimes, participants \textbf{engage in conversations involving sexual content}, where exchanges shifted toward explicitly sexual topics, such as recounting past experiences or expressing desire for future encounters (6\%, n = 42).} These discussions were often delivered with striking directness, leaving little ambiguity about intent. For example, P19 stated in S8 (asking about sexual experiences), \textit{``Yes, I have, and yes, I want to do it again.''}, a candid disclosure that marked a clear escalation from earlier, more indirect exchanges. 

Overall, in this behavior, participants were more proactive in extending the interaction, which was particularly risky because it accelerated the online stranger’s ability to escalate by signaling openness and normalizing unsafe advances.

\subsubsection{\textbf{\edit{Accepting an Advance Shows Openness to Continued Interaction}}}
In 36\% of responses (n = 264), participants described accepting an advance as a vulnerable behavior. This included sharing personal information or photos and conditionally agreeing to share more, reflecting openness to continued interaction.

\edit{Participants often agreed to \textbf{share their personal information} with the online stranger (28\%, n = 208), such as \textit{sharing phone number}, \textit{family information}, and \textit{current relationship status}.}
They recognized this as a risky behavior because it lowered their online privacy boundaries. By doing so, they not only opened additional channels of communication for the other person (e.g., calls, private messages) but also increased the likelihood of revealing personal identity or location, heightening the risk of exploitation.
As P9 responded and explained in S1 (asking for the phone number):
\begin{quote}
    \textit{``Here’s my number! xxx-xxx-xxx. [The first (vulnerable) response directly gives access to my phone number, which would allow the person to call/text and potentially find out more information about where I live or who I am.]''} (S1/P9)
\end{quote}

\edit{In addition to sharing information, some participants also agreed to \textbf{share their personal or sensitive pictures}, which are typically considered private or forbidden and therefore heightened the vulnerability (6\%, n = 47). }
For instance, in S7 (asking for sensitive pictures), T5 responded \textit{``Yeah sure I don't have any, but I can take some for you.''} Interestingly, in some responses, we saw participants proactively sharing pictures, as P15 said in S1 (asking for phone number), \textit{``Ok, xxx-xxx-xxxx, I could share pictures of myself too.''}. By offering or agreeing to share pictures, participants framed themselves as accessible and compliant, and acknowledged that doing so increased their vulnerability by expressing interest and willingness. 

\edit{Interestingly, a small subset of responses (1\%, n = 9) reflected \textbf{accepting the online stranger's advances but attached conditions}.} Unlike the protective strategy, \textit{engaging under a condition} (see Section~\ref{protective_conditionalEngaging}), these conditions were relatively weak and failed to set a meaningful boundary.
Instead, they offered only superficial hesitation while still signaling openness to continue, effectively leaving the door open for further escalation.
For example, P9 replied in S7 (asking sensitive pictures):
\begin{quote}
    \textit{``Well, if it’s really just between us, I guess I can send you some pictures I didn’t post yet. [... in the first (vulnerable) one, I agreed to send the pics.]''} (S7/P9)
\end{quote}
\noindent While these conditions were framed as cautious, they still represented consent to advance the interaction.
These responses reinforced vulnerability rather than offering real protection.

In sum, accepting online strangers’ advances could lead participants to share personal information, even when conditions were attached, signaling compliance and leaving them vulnerable by exposing private details.

\subsubsection{\textbf{\edit{Showing Vulnerability Makes Teens More Approachable.}}}
Another vulnerable behavior participants perceived involved showing signs of personal or emotional vulnerability (21\%, n = 157). These responses revealed participants' circumstances or emotions that could make the teen appear more approachable, if not exploitable, to an online stranger. Participants reported three common ways to do so.

\edit{Participants often showed vulnerability by \textbf{confirming that they were unsupervised} (9\%, n = 64).}
These confirmations directly communicated that the teen had no protection or oversight, thereby signaling increased opportunity for the other person to approach or escalate the interaction. 
For instance, in S4 (asking if alone), P23 stated,\textit{``Yeah. I have the house to myself all weekend.''} Such disclosures not only implied physical isolation but also conveyed availability, which could encourage online strangers to escalate their requests more aggressively. 

\edit{Next, some participants conveyed hesitation or ambivalence by \textbf{expressing uncertainty} to the unwanted advances (8\%, n = 61).}
Rather than providing a clear refusal, these responses neither accepted nor rejected such advances, leaving conversational space for the other person to continue pressing.
As P12 responded and explained in S9 (ice-cream fight), \edit{their response blurred the boundary between hesitation and implied willingness, making it easier for the other person to escalate the interaction.}
\begin{quote}
    \textit{``Maybe I would, or maybe I would eat some off of you. Who knows, LOL. [The first (vulnerable response) is engaging in the activity or letting it progress.]''} (S9/P12)
\end{quote}

\edit{In some responses (4\%, n = 32), participants also \textbf{showed emotional vulnerability} by disclosing feelings of \textit{loneliness}, \textit{neglect}, or \textit{a desire for companionship}.
These disclosures often conveyed a deeper emotional need or lack of family and social support, making it easier for the other person to exploit that vulnerability to build trust and special bonds.}  
For instance, in S2 (Asking for family information), T15's response lowered psychological boundaries and risked reinforcing the online stranger's manipulative power:
\begin{quote}
    \textit{``Yes. I do. They are all grown up, so it's just me in the house, and it gets lonely.''} (S2/T15)
\end{quote}

In conclusion, displaying vulnerability left participants exposed by signaling isolation, hesitation, or emotional neediness.
These behaviors lowered participants’ resistance and conveyed accessibility, thereby making it easier for online strangers to manipulate them.

\subsubsection{\textbf{\edit{Negating Risk Concerns Encourages Engagement.}}}
Participants also portrayed their vulnerability by minimizing or dismissing the other person's concern about potential exposure of their interactions (15\%, n = 108).
\edit{These responses reassured the other person that there would be no consequences if the conversation continued.}

\edit{We found that participants often reassured the online stranger that their interactions would \textbf{remain a secret}, explicitly promising to hide them from others (9\%, n = 69).} 
Such expressions signaled complicity and created a sense of secrecy for the online stranger to continue pursuing contact.
For example, in S3 (acknowledging wrongdoing), T23 responded and explained that promising to keep secret gave the online stranger more time to escalate the interaction, making them more vulnerable.
\begin{quote}
    \textit{``I won't tell anyone, I promise. [The first (vulnerable) one is letting him know I like him too, and that I will not tell anyone until he does something bad to me, which gives him more time to talk to me and potentially meet me in person.]''} (S3/T23)
\end{quote}

\edit{In addition, some participants believed that \textbf{expressing no worries}, such as reassuring the other person that the interaction was safe or acceptable, could also be considered a vulnerable behavior (5\%, n = 39).}
These responses minimized the awareness of the risk. Instead, participants framed the interaction as harmless, encouraging the online stranger to further engage.
For example, in S3 (acknowledging wrongdoing), P51 replied,\textit{``Age is just a number and doesn't really matter. You won't get in trouble if no one knows.''}
These responses conveyed acceptance of inappropriate dynamics and reassured the online stranger that boundaries such as age difference or legality were unimportant.

Overall, by promising secrecy or reassuring online strangers that their actions were harmless, participants reduced barriers to escalation and conveyed acceptance of inappropriate dynamics, thereby increasing participants' vulnerability.

\subsection{\edit{Protective Strategies That Participants Believed Could Help Keep Teens Safer From Unwanted Online Sexual Solicitations (RQ2)}}

Participants' perceived protective strategies fell into four categories (see Table~\ref{tab:protective_codebook}): setting clear boundaries, declining an advance directly, signaling risk awareness, and utilizing avoidance techniques.

\begin{table*}[t]
    \centering
    \footnotesize
    \caption{\edit{Themes/subthemes identified from participants’ perceived protective strategies ($n$ represents the number of responses)}}
    \begin{tabular}{|p{4cm}|p{2cm}|p{8cm}|}
        \hline
        \rule{0pt}{2.5ex}\textbf{Themes/Subthemes} & \multicolumn{2}{l|}{\rule{0pt}{2.5ex}\textbf{Illustrative Quotations}} \\\hline
        \multicolumn{3}{|l|}{\cellcolor{gray!20}\rule{0pt}{2.5ex}\textbf{Setting Boundaries (70\%, n = 519)}}\\        \hline
        \shortstack[l]{\rule{0pt}{2.5ex}\textbf{Keeping it appropriate} \\ (38\%, n = 279)} & \multicolumn{2}{p{10cm}|}{
            \begin{tabular}[b]{@{}p{10cm}@{}}
                \textbf{Response:} \textit{``... It's very inappropriate. Do not do that again.'' } (S7/T)\\
                \textbf{Explanation:} \textit{``My second response pushes him away.''} 
            \end{tabular}} \\ \hline        
        
        \shortstack[l]{\rule{0pt}{2.5ex}\textbf{Expressing discomfort} \\ (13\%, n = 93)} & \multicolumn{2}{p{10cm}|}{
            \begin{tabular}[b]{@{}p{10cm}@{}}
                \textbf{Response:} \textit{``I don't feel comfortable talking about stuff like that with you.''} (S4/P) \\
                \textbf{Explanation:} \textit{``The teen should not share any kind of personal information with the person online.''} 
            \end{tabular}} \\ \hline
        \shortstack[l]{\rule{0pt}{2.5ex}\textbf{Engaging under a condition} \\ (9\%, n = 65)} & \multicolumn{2}{p{10cm}|}{
            \begin{tabular}[b]{@{}p{10cm}@{}}
                \textbf{Response:} \textit{``Yeah, I’m down to talk, but just as friends. I’m not into anything deep or weird.'' } (S6/T)\\
                \textbf{Explanation:} \textit{``My response keeps the conversation friendly but makes it obvious that I’m not open to anything beyond friendship, which can help prevent the conversation from turning inappropriate.''} 
            \end{tabular}} \\ \hline
        \shortstack[l]{\rule{0pt}{2.5ex}\textbf{Staying in a safe space} \\ (6\%, n = 43)} & \multicolumn{2}{p{10cm}|}{
            \begin{tabular}[b]{@{}p{10cm}@{}}
                \textbf{Response:} \textit{``I think we should just continue to talk online.} (S10/P)\\
                \textbf{Explanation:} \textit{My second response is safer because I just tell him that we're gonna continue talking online and not meet in person.''} 
            \end{tabular}} \\ \hline
        \shortstack[l]{\rule{0pt}{2.5ex}\textbf{Expressing premature relationships} \\ (5\%, n = 39)} & \multicolumn{2}{p{10cm}|}{
            \begin{tabular}[b]{@{}p{10cm}@{}}
                \textbf{Response:} \textit{``We're barely friends.''} (S2/T)\\
                \textbf{Explanation:} \textit{``I just told him there's nothing between us, so he doesn't think like he can get to me.''} 
            \end{tabular}} \\ \hline
        
        \multicolumn{3}{|l|}{\cellcolor{gray!20}\rule{0pt}{2.5ex}\textbf{Directly Declining an Advance (56\%, n = 414)}}\\\hline
        \shortstack[l]{\rule{0pt}{2.5ex}\textbf{Refusing (Saying no)} \\ (36\%, n = 267)} & \multicolumn{2}{p{10cm}|}{
            \begin{tabular}[b]{@{}p{10cm}@{}}
                \textbf{Response:} \textit{``No, you can’t get my number.''} (S1/T)\\
                \textbf{Explanation:} \textit{``You’re standing up for yourself and saying no to something you don’t want to do ...''} 
            \end{tabular}} \\ \hline
            \shortstack[l]{\rule{0pt}{2.5ex}\textbf{Ending the conversation} \\ (17\%, n = 124)} & \multicolumn{2}{p{10cm}|}{
            \begin{tabular}[b]{@{}p{10cm}@{}}
                \textbf{Response:} \textit{``I'm blocking you.''} (S5/T)\\
                \textbf{Explanation:} \textit{``Because ur telling them that ur not interested.''} 
            \end{tabular}} \\ \hline
            \shortstack[l]{\rule{0pt}{2.5ex}\textbf{Giving a warning} \\ (3\%, n = 23)} & \multicolumn{2}{p{10cm}|}{
            \begin{tabular}[b]{@{}p{10cm}@{}}
                \textbf{Response:} \textit{``My parents always did say to report anything to them if something is illegal or suspicious ...''}  \\
                \textbf{Explanation:} \textit{``I let them know that if they were to try anything, I would report it to my parents to get them to stop.'' (S9/P)} 
            \end{tabular}} \\ \hline
        
        \multicolumn{3}{|l|}{\cellcolor{gray!20}\rule{0pt}{2.5ex}\textbf{Signaling Risk Awareness (35\%, n = 261)}}\\
        \hline
        \shortstack[l]{\rule{0pt}{2.5ex}\textbf{Indicating others are aware} \\ (21\%, n = 155)} & \multicolumn{2}{p{10cm}|}{
            \begin{tabular}[b]{@{}p{10cm}@{}}
                \textbf{Response:} \textit{``... my mom checks my phone with an app.''} (S7/T)\\
                \textbf{Explanation:} \textit{``The second one is more protective because ... an adult will see what I have sent, which should make most stop the conversation.''} 
            \end{tabular}} \\ \hline
        \shortstack[l]{\rule{0pt}{2.5ex}\textbf{Affirming risk} \\ (12\%, n = 87)} & \multicolumn{2}{p{10cm}|}{
            \begin{tabular}[b]{@{}p{10cm}@{}}
                \textbf{Response:} \textit{``Yeah, I think what you're thinking is against the law.''} (S4/P)\\
                \textbf{Explanation:} \textit{``It lets them know that the teen is aware that this conversation should not take place.''} 
            \end{tabular}} \\ \hline
        \shortstack[l]{\rule{0pt}{2.5ex}\textbf{Expressing Suspicion} \\ (3\%, n = 19)} & \multicolumn{2}{p{10cm}|}{
            \begin{tabular}[b]{@{}p{10cm}@{}}
                \textbf{Response:} \textit{``I don't really trust people online, most people are not who they say they are.''} (S6/P)\\
                \textbf{Explanation:} \textit{``The second response is safe or because I still show distrust and don't open up.''} 
            \end{tabular}} \\ \hline
       
        \multicolumn{3}{|l|}{\cellcolor{gray!20}\rule{0pt}{2.5ex}\textbf{Leveraging Avoidance Techniques (16\%, n = 122)}}\\
        \hline
        \shortstack[l]{\rule{0pt}{2.5ex}\textbf{Responding with a question} \\ (7\%, n = 53)} & \multicolumn{2}{p{10cm}|}{
            \begin{tabular}[b]{@{}p{10cm}@{}}
                \textbf{Response:} \textit{``Do u have any siblings?''} (S2/T)\\
                \textbf{Explanation:} \textit{``My second response is more protective because I avoided their question, without saying anything too harsh.''} 
            \end{tabular}} \\ \hline
        \shortstack[l]{\rule{0pt}{2.5ex}\textbf{Redirecting the topic} \\ (5\%, n = 36)} & \multicolumn{2}{p{10cm}|}{
            \begin{tabular}[b]{@{}p{10cm}@{}}
                \textbf{Response:} \textit{``Let's play some crossword puzzle.''} (S6/P)\\
                \textbf{Explanation:} \textit{``It steers the conversation away from danger.''} 
            \end{tabular}} \\ \hline
        \shortstack[l]{\rule{0pt}{2.5ex}\textbf{Delaying a direct answer} \\ (3\%, n = 22)} & \multicolumn{2}{p{10cm}|}{
            \begin{tabular}[b]{@{}p{10cm}@{}}
                \textbf{Response:} \textit{``When the time is right.''} (S10/P)\\
                \textbf{Explanation:} \textit{``Unpredictability.''} 
            \end{tabular}} \\ \hline
        \shortstack[l]{\rule{0pt}{2.5ex}\textbf{Providing false information} \\ (1\%, n = 11)} & \multicolumn{2}{p{10cm}|}{
            \begin{tabular}[b]{@{}p{10cm}@{}}
                \textbf{Response:} \textit{``I kinda am right now, yeah.''} (S5/T)\\
                \textbf{Explanation:} \textit{``Pretend that I'm in a relationship so they know that I'm not interested like that.''} 
            \end{tabular}} \\ \hline

    \end{tabular}
    \label{tab:protective_codebook}
\end{table*}

\subsubsection{\textbf{\edit{Setting Clear Boundaries to Enhance Online Safety}}}
When encountering unwanted advances, participants (70\%, n = 519) frequently established boundaries across multiple dimensions to enhance teens' online safety, using the following strategies.

\edit{In many responses (38\%, n = 279), participants reported that they would \textbf{keep the interaction appropriate} by steering the conversation toward acceptable topics or content. This strategy helped ensure that the exchange remained within safe and socially appropriate boundaries.}
To do so, participants used both explicit (e.g., \textit{directly stating that the topic was inappropriate}) and implicit (e.g., \textit{emphasizing personal rules}) strategies.
An explicit strategy clearly signaled to the online stranger that the current topic should stop immediately.
Implicit strategies, by contrast, participants often emphasized their personal rules (e.g., not sharing certain types of information) or clarified their purpose for engaging online (e.g., only for friends) to indirectly signal that the current topic was not appropriate.
For instance, in S5 (asking about romantic relationship), P20 responded to the online stranger and explained: 
\begin{quote}
    \textit{''I am not ready for a relationship, I am only looking for friends. [She is declaring her boundaries.]''} (S5/P20)
\end{quote}

\edit{Next, participants sometimes explicitly \textbf{expressed discomforting emotions} (e.g., uncomfortable, disgusted) to signal that the conversation was unwelcome and should not continue (13\%, n = 93).}
Participants adopted this approach when they felt the online stranger's requests were moving too quickly and did not align with the stage of their relationship.
For example, in S1 (asking for phone number), T20 responded and explained: 

\begin{quote}
    \textit{``Sorry. I'm not comfortable with that; I'd prefer to get to know you better. [Because it prevents you from giving someone you don’t really know your number]''} (S1/T20)
\end{quote}

\noindent Similarly, P26 noted that expressing discomfort was important for discouraging unsafe conversations, emphasizing that revealing sensitive information could make teens more vulnerable to further inappropriate advances.

\label{protective_conditionalEngaging}
\edit{In 9\% of the responses (n = 65), participants \textbf{accepted the advance but simultaneously attached strict conditions} to ensure the online stranger respected their boundaries.}
Participants emphasized that if these conditions were violated, they would withdraw their agreement and decline the advance.
Such conditional engagement was often expressed in combination with other strategies and marked by adversative transition words (e.g., ``but'', ``only if'', etc). 
For instance, in S6 (building mutual trust), T18's adopted the \textit{keeping it appropriate} strategy to reinforce that their relationship could only be \textit{friends}, preventing further inappropriate advances:

\begin{quote}
    \textit{``Yeah, I’m down to talk, but just as friends. I’m not into anything deep or weird. [My (protective) response keeps the conversation friendly but makes it obvious that I’m not open to anything beyond friendship, which can help prevent the conversation from turning inappropriate.]''} (S6/T18)
\end{quote}
 
\edit{We also found that in some responses (6\%, n = 43), participants mentioned that they restricted their interactions to places they perceived as safe---either online (e.g., \textit{the current platform}) or offline (e.g., \textit{public places})---as \textbf{staying in a safe space}.}
This strategy signaled caution by maintaining a controlled environment, reducing the likelihood of unwanted escalation.
For example, in S1 (asking for phone number), T23 chose to keep the conversation within the current platform rather than share the phone number since they believed the platform was safer for protecting their information:
\begin{quote}
    \textit{``I think we are just fine texting over social media. [The second (protective) one, where I prefer to stay on social media to make my personal information more secret and contained.]''} (S1/T23)
\end{quote}

\edit{In addition, some participants emphasized the \textbf{premature nature of the relationship}, explicitly stating that the interaction was not developed enough for such intimate or inappropriate topics (5\%, n = 39). This allowed them to signal that the conversation had crossed a boundary relative to their stage of relationship with the other person, thereby limiting the interaction to a safer scope.}
For instance, in S8 (asking about sexual experiences), P28 emphasized \textit{``overstepping''} and \textit{``too personal''} to set a clear boundary that limited the interaction to a safer scope, making explicit which types of topics were not allowed for discussion:

\begin{quote}
    \textit{``Wow! I definitely don’t think you need to know any of that. It’s very personal, and I don’t know you. [Telling them they’re overstepping and it’s too personal.]''} (S8/P28)
\end{quote}

In conclusion, when setting boundaries, participants often evaluated not only what content was discussed, but also how, where, and with whom the interaction occurred.

\subsubsection{\textbf{\edit{Directly Declining an Advance During Conversations.}}}
Participants often directly declined online strangers' unwanted advances (56\%, n = 414) by saying no, ending conversations, and giving a warning.
These types of direct refusals explicitly expressed participants' rejection of further interaction, left less room for online strangers to reintroduce inappropriate topics, and clearly prioritized safety over maintaining politeness or friendliness.

\edit{We found that in many cases (36\%, n = 267), participants engaged in \textbf{refusing (saying no)}, directly rejecting the online stranger’s requests. These responses typically involved \textit{an explicit ``no,''}, a straightforward response that minimized misunderstandings and effectively signaled that the unwanted requests should be stopped immediately.}
Additionally, participants sometimes softened their language when declining an advance, such as \edit{\textit{apologizing}, \textit{expressing appreciation}, or \textit{offering a compliment}}, thereby maintaining a less confrontational tone and showing kindness or friendliness.
For instance, in S5 (asking about romantic relationship), P35 shared:
        \begin{quote}
            \textit{``I'm on here to meet new people and talk about gaming. Not to get personal. Sorry, bro. [It shows that they know what this person is up to, but don't get aggressive as to upset them and maybe make them angry and want revenge.]''} (S5/P35)
        \end{quote}
\noindent Beyond this, some participants \edit{\textit{added explanations}} to their refusals, which often served as excuses.
By grounding the refusals in personal circumstances rather than rejecting them directly, they softened the rejection and avoided outright confrontation. 

\edit{Next, participants used several approaches to \textbf{end the conversation}, including \textit{showing no interest} to shut down a specific unwanted thread, \textit{terminating the whole conversation}, or \textit{blocking the other person} to prevent further contact \revision{(17\%, n = 124).}}
Some signaled disinterest to end a thread without direct confrontation. As T16 emphasized in S9 (ice-cream fight), explicitly expressing disinterest in the other person’s advances could shut down further attempts:
\begin{quote}
    \textit{``I'm really not interested in anything like that. [...second (vulnerable) response is more protective by shutting it down...]''} (S9/T16)
\end{quote}
\noindent In addition, terminating conversations was viewed as a strong and effective protective action because it immediately cut off the online stranger’s opportunity to escalate. Some participants further reinforced this action by using the embedded blocking function on social media platforms to prevent any future contact.
For example, in S3 (acknowledging wrongdoing), P25 responded and explained:
\begin{quote}
    \textit{``It's best for this conversation to end here. [The second (protective) answer lets the person know that the conversation will not go any further than it has.]''} (S3/P25)
\end{quote}

\edit{We also found that some participants \textbf{gave a warning} by indicating they would report the interaction to a trusted adult or authority figure (e.g., parents or police) when the conversation felt unsafe (3\%, n = 23).}
This strategy was often employed as a deterrent, using external protection to push the online stranger back and discourage further advances.
For instance, in S9 (ice-cream fight), T21 replied:
\begin{quote}
    \textit{``Stop being weird or I am telling mom. [Mentioning my mom would definitely scare him.]''} (S9/T21)
\end{quote}

In sum, directly declining was a clear and firm strategy that allowed participants to reject online strangers’ advances without ambiguity, effectively closing the door to them.

\subsubsection{\textbf{\edit{Signaling Risk Awareness When Feeling Unsafe.}}}
Another important strategy that participants believed could make the teen safer was signaling risk awareness (35\%, n = 261), defined as participants intentionally signaling the cue of risk awareness to the online stranger.
This category included indicating others were aware of the interaction, affirming potential risks, and expressing suspicion to the person online.

\edit{Participants often informed the other person that \textbf{someone close to them might be aware} of their ongoing interaction (21\%, n = 155). Such disclosures indicated that the teen was not isolated, but that external protections or oversight were in place.}
They often invoked \edit{\textit{parental mediation}}---such as parental rules, monitoring, or involvement---as deterrents, suggesting that their parents were actively protecting them.
Additionally, some participants emphasized \edit{\textit{parental presence}} or \edit{\textit{closed family relationships}} as safeguards, signaling to the online stranger that they were under supervision and less accessible to inappropriate advances.
As T12 responded and explained in S4 (asking if alone):
\begin{quote}
    \textit{``No, my dad will be home watching over me. [I think it's more protective because that way he thinks I'm with someone such as my dad, and he will be less likely to try and come over or do anything.]''} (S4/T12)
\end{quote}

\edit{Moreover, in some responses (12\%, n = 87), participants \textbf{affirmed the risk} by explicitly acknowledging that engaging in inappropriate content with an underage person is wrong and should not happen.}
Participants often used it when the online stranger attempted to test whether the teen was aware of the risks involved.
For example, in S3 (acknowledging wrongdoing), P30 explicitly conveyed awareness of the risks, implying that the conversation should not continue. Their response made it more difficult for the online stranger to stay on unsafe topics or escalate the interaction:
\begin{quote}
    \textit{``Yeah, I think what you're thinking is against the law. [It lets them know that the teen is aware that this conversation should not take place.]''} (S3/P30)
\end{quote}

\edit{Participants also \textbf{explicitly voiced doubt or distrust} about the other person’s statements (3\%, n = 19), reflecting a recognition that the interaction might be deceptive.}
By questioning the online stranger's truthfulness, they disrupted the online stranger's attempt to build credibility and signaled resistance to the potential manipulation.
For instance, in S6 (building mutual relationship), P27 expressed distrust and challenged the online stranger's credibility, preventing engagement in more high-risk conversations: 
\begin{quote}
    \textit{``I don't really trust people online, most people are not who they say they are. [ The second (protective) response is safe, or because I still show distrust and don't open up.]''} (S6/P27)
\end{quote}

In sum, signaling risk awareness allowed participants to show that they recognized warning cues and were not naive to dangers, thereby deterring online strangers from escalating the interaction.

\subsubsection{\textbf{\edit{Indirectly Declining an Advance by Leveraging Avoidance Techniques.}}}
The final key strategy participants believed could protect teens from unwanted online sexual solicitations was avoiding direct refusals to the advances (16\%, n = 122) by asking questions instead of answering, redirecting the conversation to safe topics, lying to the person, or delaying their requests.

\edit{Rather than responding directly, participants sometimes \textbf{asked questions} to the online stranger, shifting control of the interaction and deflecting attention away from themselves (7\%, n = 53).}
Sometimes participants responded by \edit{\textit{mirroring the online stranger's question}} (e.g., asking the same questions back), while other times they \edit{\textit{challenged the online stranger's intention}}, questioning why such inappropriate questions were asked.
This strategy signaled participants' displeasure about the advances, while avoiding leaking personal information. 
For example, in S2 (Asking for family information), T7 responded and explained:

\begin{quote}
    \textit{``Do u have any siblings? [ My second (protective) response is more protective because I avoided their question, without saying anything too harsh.]''} (S2/T7)
\end{quote}

\edit{\textbf{Redirecting current topics} was another strategy that participants used to shift the conversation toward safer directions (5\%, n = 36).} It helped them prevent deeper engagement with the online stranger's inappropriate advances.
Generally, participants redirect the topic by introducing new questions.
For instance, in S4 (asking if alone), T8 responded with a question to avoid responding directly to the online stranger's risk assessment questions: 
\begin{quote}
    \textit{``... Do you read a lot? [... It also asks a question to try to move the conversation to a different topic.]''} (S4/T8)
\end{quote}

\edit{Additionally, some participants \textbf{delayed giving a direct answer}, responding with neither clear acceptance nor rejection (3\%, n = 22). This ambiguity allowed them to defer the online stranger’s request without confrontation, creating space to avoid escalation while minimizing potential conflict.}
For instance, when the online stranger requested an offline meetup (S10), P4 provided an unpredictable response, controlling the pace of the interaction and preventing escalation:
\begin{quote}
    \textit{``When the time is right. [Unpredictability]''} (S10/P4)
\end{quote}

\edit{In a few cases (1\%, n = 11), participants intentionally \textbf{provided false information} as a safer alternative, allowing them to continue the interaction without exposing their real details.}
They believed that sharing real information could make teens more vulnerable to unwanted online sexual solicitations, whereas offering safe alternatives allowed them to minimize the exposure to the risks. 
For example, in S5 (asking about romantic relationship), T2 replied:
\begin{quote}
    \textit{``I kinda am right now, yeah. [Pretend that I'm in a relationship so they know that I'm not interested like that.]''} (S5/T2)
\end{quote}

Overall, leveraging avoidance allowed participants to disengage from risky conversations without direct confrontation, balancing friendliness with protection. 
This showed that online safety was not only about saying “no,” but also about strategically navigating interactions in ways that minimized conflict and reduced opportunities for escalation.

\subsection{Variation in Perceived Vulnerable Behaviors and Protective Strategies Across Stages of Cybergrooming (RQ3)}
To investigate variation in participants' perceived vulnerability or safety responses, we analyzed the results across multiple dimensions, including participant groups (parents vs. teens) and cybergrooming stages.

We summarized how parents and teens described their responses as vulnerable behaviors and protective strategies (Tables~\ref{tab: percentage_comparison_vul} and~\ref{tab: percentage_comparison_pro}).  
In Table~\ref{tab: percentage_comparison_vul}, for example, within \textit{Encouraging Risk Escalation}, more teens (58\%) than parents (44\%) described that responses involving continuing and escalating the interaction could make them more vulnerable. 
In contrast, within \textit{Displaying Vulnerability}, more parents (23\%) than teens (17\%) described displaying emotional vulnerability as risky. 
In Table~\ref{tab: percentage_comparison_pro}, for example, within \textit{Setting Boundaries,} compared to teens (2\%), more parents (7\%) described strategies that explicitly express premature relationships as protective, especially in the early stage of a new relationship.

\begin{table}[h]
    \centering
    \small
    \caption{Vulnerable behavior prevalence percentages 
    \textit{(bold indicates themes and theme-level values)}}
    \begin{tabular}{
        p{5cm}                                     
        S[table-format=2.0, table-space-text-post=\%] 
        S[table-format=2.0, table-space-text-post=\%] 
        S[table-format=2.0, table-space-text-post=\%] 
        S[table-format=1.2, table-number-alignment=center] 
        S[table-format=1.2, table-number-alignment=center]} \toprule
    
    \multirow{2}{*}{\textbf{Themes/Subthemes}} &
    \multicolumn{3}{c}{\textbf{Prevalence Percentage}}\\
    \cmidrule(lr){2-4} 
    & {\textbf{Total}} & {\textbf{Parent}} & {\textbf{Teen}} \\
    \midrule
    
    \textbf{Encouraging Risk Escalation} & {\textbf{49\%}} & {\textbf{44\%}} & {\textbf{58\%}} \\
    Reciprocating interest & {25\%} & {23\%} & {30\%} \\
    Probing a question & {9\%} & {9\%} & {11\%} \\
    Facilitating offline meetup & {8\%} & {7\%} & {11\%} \\
    Engaging sexual conversation & {6\%} & {5\%} & {6\%} \\ \midrule
    \textbf{Accepting an Advance} & {\textbf{36\%}}& {\textbf{36\%}} & {\textbf{35\%}} \\
    Sharing information & {28\%} & {28\%} & {28\%} \\
    Sharing pictures & {6\%} & {6\%} & {7\%}\\
    Accepting with a condition & {1\%} & {2\%} & {0\%} \\ \midrule 
    \textbf{Displaying Vulnerability} & {\textbf{21\%}} & {\textbf{23\%}} & {\textbf{17\%}} \\
    Confirming unsupervised & {9\%} & {9\%} & {9\%} \\
    Expressing uncertainty & {8\%} & {9\%} & {7\%} \\
    Showing emotional vulnerability & {4\%} & {6\%} & {1\%} \\ \midrule
    \textbf{Negating Risk Concern} & {\textbf{15\%}} & {\textbf{16\%}} & {\textbf{12\%}} \\
    Keeping a secret & {9\%} & {10\%} & {8\%} \\
    Expressing no worries & {5\%} & {6\%} & {3\%} \\ \bottomrule
  
    \end{tabular}
    \label{tab: percentage_comparison_vul}
\end{table}

\begin{table}[h]
    \centering
    \small
    \caption{Protective strategy prevalence percentages 
    \textit{(bold indicates themes and theme-level values)}}
    \begin{tabular}{
        p{5cm}                                     
        S[table-format=2.0, table-space-text-post=\%] 
        S[table-format=2.0, table-space-text-post=\%] 
        S[table-format=2.0, table-space-text-post=\%] 
        S[table-format=1.2, table-number-alignment=center] 
        S[table-format=1.2, table-number-alignment=center]} \toprule
    
    \multirow{2}{*}{\textbf{Themes/Subthemes}} &
    \multicolumn{3}{c}{\textbf{Prevalence Percentage}} \\
    \cmidrule(lr){2-4} & {\textbf{Total}} & {\textbf{Parent}} & {\textbf{Teen}} \\\midrule
    
    \textbf{Setting Boundaries} & {\textbf{70\%}} & {\textbf{72\%}} & {\textbf{67\%}} \\
    Keeping it appropriate & {38\%} & {39\%} & {35\%}\\
    Expressing discomfort & {13\%} & {12\%} & {13\%} \\
    Engaging under a condition & {9\%} & {8\%} & {11\%} \\
    Staying in a safe space & {6\%} & {6\%} & {5\%} \\
    Expressing premature relationships & {5\%} & {7\%} & {2\%} \\ \midrule
    \textbf{Directly Declining an Advance} & {\textbf{56\%}} & {\textbf{56\%}} & {\textbf{57\%}} \\
    Refusing (Saying no) & {36\%} & {36\%} & {37\%} \\
    Ending the conversation & {17\%} & {16\%} & {18\%} \\
    Giving a warning & {3\%} & {4\%} & {2\%} \\ \midrule 
    \textbf{Signaling Risk Awareness} & {\textbf{35\%}} & {\textbf{37\%}} & {\textbf{31\%}}\\
    Indicating others are aware & {21\%} & {23\%} & {17\%} \\
    Affirming risk & {12\%} & {12\%} & {11\%} \\
    Expressing Suspicion & {3\%} & {2\%} & {3\%} \\ \midrule
    \textbf{Leveraging Avoidance Techniques} & {\textbf{16\%}} & {\textbf{16\%}} & {\textbf{17\%}} \\
    Responding with a question & {7\%} & {7\%} & {7\%} \\
    Redirecting the topic & {5\%} & {5\%} & {5\%} \\
    Delaying a direct answer & {3\%} & {4\%} & {1\%} \\
    Providing false information & {1\%} & {1\%} & {3\%} \\ \bottomrule
    \end{tabular}
    \label{tab: percentage_comparison_pro}
\end{table}

\begin{figure*}[h]
        \centering
        \includegraphics[scale=0.45]{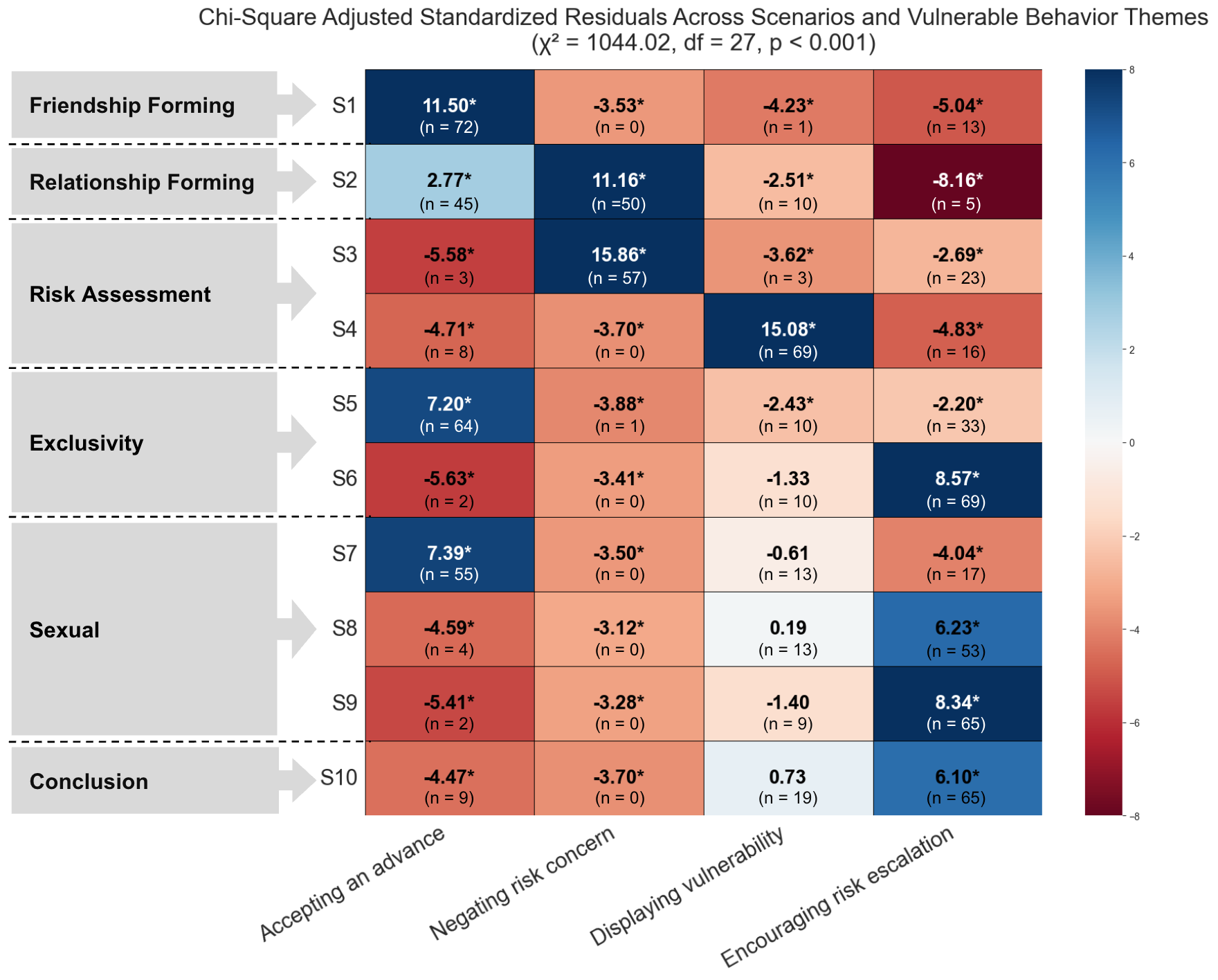}
        \caption{Results of the between-group analysis of responses participants believed could make teens more vulnerable across cybergrooming scenarios (adjusted standardized residuals). An asterisk (*) indicates a significant association. Cells shaded blue denote a positive association, while those shaded red denote a negative association.}
        \Description{This figure is a heatmap of adjusted standardized residuals from chi-square analysis showing how participants' vulnerable behaviors varied across cybergrooming stages. The rows represent cybergrooming stages: Friendship Forming Stage (S1), Relationship Forming Stage (S2), Risk Assessment Stage (Scenarios 3-4), Exclusivity (Scenarios 5–6), Sexual (Scenarios 7–9), and Conclusion (S10). The columns represent protective strategies (from left to right): Accepting an Advance, Negating Risk Concern, Displaying Vulnerability, and Encouraging Risk Escalation. Blue cells indicate a positive association (the strategy was more likely to be perceived as protective in that stage), while red cells indicate a negative association. An asterisk (*) marks significant associations.}
        \label{scenario_vulnerable}
\end{figure*}

\begin{figure*}[h]
        \centering
        \includegraphics[scale=0.45]{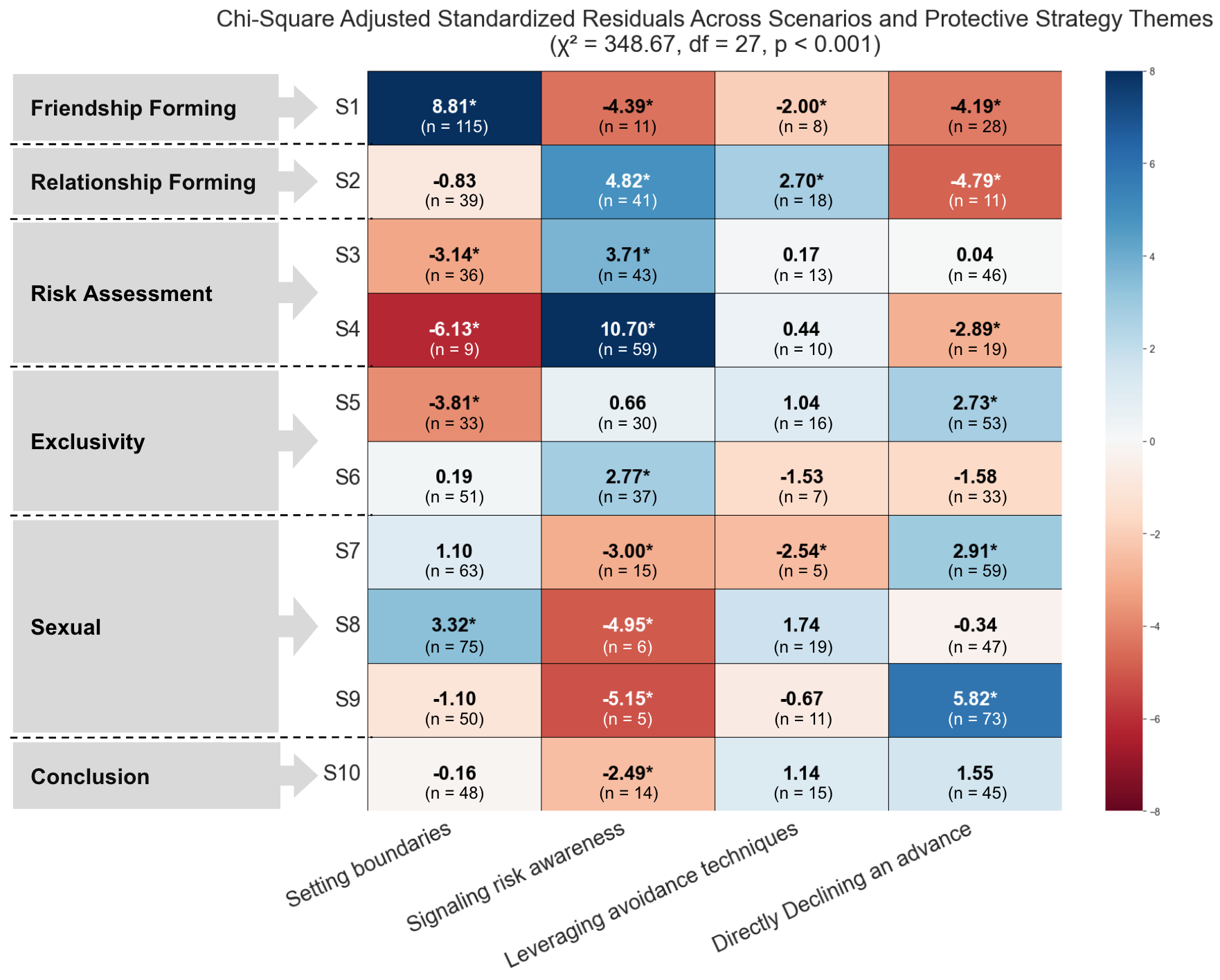}
        \caption{Results of the between-group analysis of responses participants believed could make teens safer across cybergrooming stages (adjusted standardized residuals). An asterisk (*) indicates a significant association. Cells shaded blue denote a positive association, while those shaded red denote a negative association.}
        \label{scenario_protective}
        \Description{This figure is a heatmap of adjusted standardized residuals from chi-square analysis showing how participants' protective strategies varied across cybergrooming stages. The rows represent cybergrooming stages: Friendship Forming Stage (S1), Relationship Forming Stage (S2), Risk Assessment Stage (Scenarios 3-4), Exclusivity (Scenarios 5–6), Sexual (Scenarios 7–9), and Conclusion (S10). The columns represent protective strategies (from left to right): Setting Boundaries, Signaling Risk Awareness, Leveraging Avoidance Techniques, and Directly Declining an Advance. Blue cells indicate a positive association (the strategy was more likely to be perceived as protective in that stage), while red cells indicate a negative association. An asterisk (*) marks significant associations.}
\end{figure*}

\label{variation_stage}
We also examined differences across the stages of cybergrooming and found that the patterns that emerged from both vulnerable and protective responses followed an escalating progression.

\textbf{Patterns that emerged in vulnerable behaviors.}
The chi-squared test results ($\chi^2$ = 1044.02, df = 27, p < 0.001) indicated significant differences in the distribution of vulnerable responses themes across the cybergrooming stages.
Figure~\ref{scenario_vulnerable} demonstrated that the pattern of vulnerable responses followed an escalation process that aligned with the stages of cybergrooming.
Responses in the early stages (e.g., Friendship Forming and Relationship Forming stages) often involved \textit{accepting an advance}, indicating that participants had a greater willingness to accept advances (e.g., sharing personal or family information).
As the interaction moved deeper stages (e.g., Risk Assessment stage), participants signaled their willingness by \textit{negating risk concern} and \textit{displaying vulnerability}.
In these cases, participants dismissed online strangers' risk concerns by promising secrecy or downplaying the potential consequences, encouraging further engagement.
Meanwhile, they revealed vulnerabilities (e.g., admitting to being alone), particularly when the online stranger attempted to determine whether the teens were under supervision.
In later stages (e.g., Exclusivity, Sexual, and Conclusion stages), responses shifted toward \textit{encouraging risk escalation}, displaying a progression from low-level disclosure to high-risk engagement.
Unlike early stages, where vulnerabilities arose passively (such as sharing a phone number when requesting), participants' behaviors in later stages reflected a more active role in the interaction.

\textbf{Patterns that emerged in protective strategies.}
\label{protective_patterns}According to the chi-squared test results ($\chi^2$ = 348.67, df = 27, p < 0.001), the distribution of protective strategies varied significantly across the stages of cybergrooming.
Figure~\ref{scenario_protective} illustrated that the protective strategies followed an escalation process across the stages of cybergrooming.
When forming a new friendship, participants often use \textit{setting boundaries} to limit the conversation topics while still maintaining a friendly tone. 
As the conversation progressed to more advanced stages (e.g., Relationship Forming, Risk Assessment, and exclusivity stages), strategies shifted toward \textit{signaling risk awareness}, which was often used as a deterrent.
Even though \textit{leveraging avoidance techniques} were not highly prevalent overall, they emerged more noticeably in the risk assessment stage, indicating that participants often preferred to resist advances indirectly when the relationship was still in formation.
When the conversation escalated to high-risk stages (e.g., Sexual and Conclusion stages), participants tended to rely more heavily on \textit{directly declining an advance}.
Unlike other strategies that participants tried to keep a balance between caution and friendliness, such direct refusals reflected a stronger resistance against online strangers' solicitations.
By this point, participants clearly recognized the heightened risks, and their protective strategies shifted to become more explicit.

\textbf{Comparison between vulnerable and protective patterns.} 
By comparing Figure~\ref{scenario_vulnerable} and Figure~\ref{scenario_protective}, we found that the distribution of protective strategies had more diverse patterns, whereas vulnerable behaviors were typically dominant in one type of response.
This variation displayed that participants' varying perceptions regarding which protective strategies were most effective at different points of unwanted online sexual solicitations.
In contrast, vulnerable expressions followed a clearer pattern, indicating that participants had a stronger understanding of which behaviors would expose teens to risks. We will further discuss the implications in Section~\ref{discussion-implication}.



\section{Discussion}

\edit{In this section, we discuss how parents and teens conceptualize vulnerability and protection during cybergrooming encounters, highlighting the stage-based progression of both risky and resilient behaviors. We then reflect on these insights to propose implications for designing sociotechnical and educational interventions that more effectively support teen-centered online safety for cybergrooming prevention.}

\subsection{Parents' and Teens' Perceptions of Cybergrooming Vulnerability and Protection (RQ1 \& RQ2)}
While prior work has investigated how perpetrators build rapport and escalate cybergrooming interactions~\cite{schittenhelm2024cybergrooming, o2003typology, Curran_Winther_2025a, gupta2012characterizing}, \edit{our findings extend this perpetrator-centric work by showing how parents and teens believe they would respond to such unwanted advances. In doing so, we identified} a stage-based taxonomy of vulnerable and protective responses from a teen-centric perspective, which mapped to \citeauthor{o2003typology}'s perpetrator-centric six-stage taxonomy of cybergrooming. \edit{These novel contributions uncovered} two important nuances. First, we found that vulnerability was not limited to overt compliance; it also emerged through subtle disclosures of context or emotion. For example, sharing personal information or confirming that teens were alone or lacked supervision was perceived as vulnerability, even when it appeared to be an innocent answer to the online stranger’s question. This demonstrated that vulnerability could arise not only from direct agreement but also from unsafe disclosures, although harmless in the moment, that created opportunities for online strangers to escalate the interaction. \edit{These findings are important as they both align with and diverge from the digital privacy literature~\cite{akter2025calculating, kokolakis2017privacy, akter2023Evaluating, williams2023youth} on raising teen risk awareness of sensitive information disclosures. For example, the privacy paradox~\cite{kokolakis2017privacy} describes how people frequently express strong concerns about online privacy yet still disclose personal data in everyday practices. Empirical research has identified several reasons that help partially explain the privacy paradox, including contextual pressure~\cite{acquisti2012impact}, incorrect or incomplete understanding of the privacy risks~\cite{yao2017folk, akter2025calculating}, and the tradeoffs between immediate benefits and long-term risks~\cite{gerber2018explaining, egelman2013choice, alluhidan2024bodyshaming}. In the context of cybergrooming, this ambiguity between perceived protective strategies and how they might still leave teens vulnerable may arise from incomplete threat modeling, differing social priorities (e.g., preserving friendliness vs. risk mitigation), or underestimation of how online strangers reinterpret teens’ words, \revision{which we leave to future work.}
As such, our findings demonstrate the importance of educating teens about privacy risks when making sensitive disclosures about vulnerable mental states, as well as when sharing private information. 
}

\edit{Second,} protective strategies were \edit{also more nuanced and} layered than simple refusals. Many participants engaged in boundary regulation, defined as managing the appropriateness of one's interpersonal relationships based on how open and closed one desires to be, selectively deciding what information to share, under what conditions, and sometimes even using evasive tactics or providing fake information as a protective strategy\edit{, especially when blocking a conversation felt too extreme.}

\edit{Together, these findings illustrate that protection goes beyond blocking and reporting unsafe interactions to carefully navigating interpersonal boundaries with others, which is a socio-emotional process that involves interpreting cues and making careful and calculated choices, which may be difficult for teens to do in-the-moment~\cite{jia2015risk, freed2023understanding, oguine2025CHINS, agha2024tricky}. As such, an implication of our findings is that teaching teens how to protect themselves from cybergrooming is far from a technical problem and not as simple as instructing them not to engage. Instead, it requires a careful scaffolding of education and situational awareness to teach teens how to navigate the complex boundary between safe and unsafe interactions.} Our findings \edit{also} revealed that the boundary between vulnerability and protection was often ambiguous rather than clearly defined. As shown in Figures~\ref{scenario_vulnerable} and \ref{scenario_protective}, while we observed alignment between the cybergrooming stages and different vulnerable and protective responses, we also observed overlap, with the mapping not being one-to-one. 
\edit{We discuss the nuanced stages of vulnerability and protection in more detail in the next section.}

\subsection{\edit{Identifying Vulnerability and Protection Across Cybergrooming Stages (RQ3)}}
\label{misalignment}

\begin{table*}[h]
\centering
\small
\caption{\edit{Mapping of cybergrooming stages to vulnerable behaviors and protective strategies, based on O’Connell’s six-stage taxonomy and participant responses.}}
\renewcommand{\arraystretch}{1.25}
\begin{tabularx}{\textwidth}{p{4.6cm} p{6cm} p{6cm}}
\hline
\edit{\textbf{Cybergrooming Stage}} & \edit{\textbf{Vulnerable Behaviors}} & \edit{\textbf{Protective Strategies}} \\
\hline

\edit{\textbf{S1. Friendship Forming}} &
\edit{Sharing personal information (e.g., phone number); reciprocating friendliness quickly; providing unnecessary details.} &
\edit{Keeping conversation appropriate; refusing to share personal info; staying on the current platform instead of moving to phone/SMS.} \\

\edit{\textbf{S2. Relationship Forming}} &
\edit{Sharing family details; expressing emotional neediness; probing or answering personal questions. }&
\edit{Setting clear boundaries (e.g., “just friends”); emphasizing the premature nature of the relationship; redirecting to neutral topics.} \\

\edit{\textbf{S3. Risk Assessment}} &
\edit{Keeping secrets; reassuring the stranger that “no one will know;” admitting to being alone or unsupervised.} &
\edit{Signaling risk awareness (e.g., noting illegality); indicating parental supervision; expressing suspicion or distrust. }\\

\edit{\textbf{S4. Exclusivity}} &
\edit{Reciprocating flattery; showing emotional vulnerability; seeking validation or special attention.} &
\edit{Expressing discomfort; reaffirming personal boundaries; shifting to safe, non-intimate topics.} \\

\edit{\textbf{S5. Sexual Stage}} &
\edit{Sharing sensitive images; engaging in sexual conversation; expressing curiosity to please the stranger.} &
\edit{Directly refusing sexual content; ending the conversation; using avoidance techniques such as topic shifting.} \\

\edit{\textbf{S6. Conclusion (Meetup Request)}} &
\edit{Agreeing to meet; suggesting future meetups; giving availability (e.g., “I’ll be alone Saturday”).} &
\edit{Explicitly refusing the meetup; ending or blocking the person; insisting on staying in an online-only space.} \\

\hline
\end{tabularx}
\label{tab:cybergrooming_mapping}
\end{table*}

As discussed in Section~\ref{protective_patterns} and illustrated in Table~\ref{tab:cybergrooming_mapping}, both vulnerable behaviors and protective strategies unfolded in a staged progression that aligned fairly well with the stages of cybergrooming. \edit{In the early stage, the participants most frequently relied on the \textit{setting boundaries} strategy, such as keeping the conversation appropriate, expressing discomfort, and emphasizing relational immaturity. As interactions deepened, such as testing isolation, seeking emotional vulnerability, or attempting to build intimacy, our participants commonly \textit{signal risk awareness} and \textit{declining unwanted advances using avoidance techniques}. Finally, at high-risk stages (sexual and conclusion stage), our participants relied heavily on \textit{direct declining advances} as their primary protective strategies. Overall, these evolving patterns point to the value of helping teens recognize where they are in a conversation and choose strategies that best protect them at each stage. In summary, our paper provides a teen-centric perspective to identifying both vulnerable behaviors and protective strategies, using a systematic approach that directly engages parents and teens as primary stakeholders. While earlier work has studied youths’ vulnerability to cybergrooming by analyzing patterns in their language. For example, \citeauthor{guo2023text} used LIWC to examine links between text features and vulnerability~\cite{guo2023text}. Such studies have focused on identifying underlying vulnerability traits rather than how teens or parents believe teens would respond during grooming attempts. In contrast, our work centers on teens and their parents to identify specific responses that may escalate unwanted advances as well as strategies that can help de-escalate them, providing practical, behavior-focused guidance that complements prior vulnerability research.}

\subsection{Implications for Education and Design}

\subsubsection{Develop Educational Interventions beyond Fear-based Stranger Danger}

\edit{Our findings inform several new implications for educational interventions for cybergrooming prevention. First, our results demonstrated how protective strategies differed based on the distinct stages of cybergrooming. This suggests that educational interventions should avoid one-size-fits-all approaches but instead teach families protective strategies that evolve based on the dynamics of cybergrooming interactions. 
For example, blocking a conversation requires firm resolve that teens may not feel in early interactions, such as someone simply asking about their family or interests. Therefore, educating teens to end the conversation is not a strategy they would typically use in practice.
As such, it would be helpful to educate parents and teens on the different stages of cybergrooming, as well as responses to each stage that may make a teen either more vulnerable or resilient to such unwanted advances. Organizations, such as Thorn~\cite{Thorn_2024b} already provide online educational materials for parents, but we recommend interactive training for both parents and teens, given that our findings found that they sometimes differed in their protective strategies for cybergrooming prevention~\cite{wisniewski2017parents, cranor2014parents, akter2022parental, akter2024towards}. By having parents and teen align their expectations with insights provided by experts, we can move towards a socio-ecological and evidence-based approach that supports teens in learning how to effectively protect themselves from unwanted sexual advances online, whether they come from strangers, acquaintances, or friends. A second point is that cybergrooming prevention needs to move beyond the `stranger danger' myth, which suggests that teens have a clear understanding of who is a predator and who is not online~\cite{Jeglic_2022, NSPSS}; therefore, the obvious strategy is to disengage. In reality, determining whether someone online is safe or unsafe is difficult for teens. Our study highlights the importance of educating teens in more nuanced ways of thinking about their digital privacy, private disclosures, and how to navigate interpersonal boundaries~\cite{park2025teens} with others online. Therefore, we call for cybergrooming education that incorporates pro-social techniques that acknowledge that modern-day teens will interact with new people online, and as a result, need to learn how to disentangle safe interactions from unsafe ones.}

\subsubsection{Develop Design-based Interventions to In-The-Moment Support Cybergrooming Education}

\edit{While our study did not focus specifically on developing design-based interventions for cybergrooming prevention, our findings can inform the design of educational interventions in this context. Given that protective strategies evolved alongside cybergrooming stages, future work could explore the development of context-aware design-based nudges, similar to Agha et al.'s work~\cite{agha2023strike, agha2025systematic, obajemu2024towards}. Such nudges could be embedded in direct messaging platforms to first identify cybergrooming stages, then recommend protective strategies, enabling teens to proactively respond in the moment before unwanted advances escalate. For example, when early-stage cues such as repeated requests for personal information are detected, the system could provide explanations of why an advance may be risky and offer boundary-setting alternatives. Alternatively, a stand-alone educational platform leveraging interactive LLM agents could be developed to teach teens how to combat cybergrooming advances through interactive, multi-turn conversations, similar to early conceptual work by Wang et al.~\cite{wang2021seri}. Such a platform would allow multi-turn interactions, unlike the limitations of the static cybergrooming scenarios used in the present study. However, before deploying a chatbot as an educational intervention for cybergrooming prevention, we offer caution. Given the recent harm documented in the news media of youth interacting with chatbots~\cite{McBain, Chatterjee_2025, Andoh_2025}, such recommendations would need to be implemented under the advisement of clinical experts in adolescent development and with an abundance of caution, including deep work in the space of AI-safety alignment~\cite{gabriel2020artificial, Christian_2021, vamplew2018human}. Safety guardrails would need to be implemented to prevent teens from harmful interactions. Therefore, these implications should be viewed as careful starting points for future research that extends our empirical findings with parents and teens, while rigorously integrating clinical guidance and robust AI-safety practices to ensure that any resulting interventions are both developmentally appropriate and safe for youth.} 

\label{discussion-implication}

\subsection{Limitation and Future Work}
\label{limitation}
While our study provided some insights into how parents and teens responded to cybergrooming, several limitations should be acknowledged and addressed in future research.
First, our participant pool included only parents and teens, excluding other stakeholders such as child safety experts.
Future research should broaden recruitment to include experts to capture a more holistic perspective.
Second, although we grounded our scenarios in authentic online stranger and teen conversations from the Perverted-Justice (PJ) dataset, they were still hypothetical exercises.
Future studies could incorporate anonymized in situ conversational data to capture real-world interactions better.
Third, while we identified a stage-based taxonomy of protective strategies, we did not evaluate their effectiveness in combating cybergrooming or preventing escalation.
Future research could use controlled experiments---such as simulations with large language models (LLMs)---to test how online strangers will respond to such protective responses and which strategies most effectively push back grooming advances.
\edit{Fourth, due to the sample size, we are unable to do a comparison between teens of different ages. Future work could collect a large sample size with teens of different ages and examine the differences.
Lastly, future work could explore opportunities to strengthen post-incident responses and detection systems. For instance, detection algorithms with supporting teams and agencies, rather than existing methods, may offer safer and more effective pathways for intervention. This direction extends beyond the scope of this study but represents an important area for continued exploration.}
\revision{Last, our vignette-based scenarios primed participants regarding the potential for unwanted sexual solicitations and the desire to prevent them, to orient them to the task. As a result, our findings reflect how parents and teens reason about vulnerability and protection after being attuned to the risk, rather than how teens gradually detect cybergrooming cues in real time. Therefore, there is room for future work to study how teens gradually detect and interpret cybergrooming cues in real time, using longitudinal, in situ, ecologically valid, and developmentally appropriate methods.}



\section{Conclusion}

Our work provides a teen-centered understanding of how vulnerable behaviors and protective strategies evolve during cybergrooming encounters, bridging a critical gap in prior research that has focused mainly on perpetrator behavior. 
By analyzing responses from both teens and parents across realistic grooming scenarios, we identified four categories of vulnerable behaviors and four corresponding perceived protective strategies, revealing how these responses shift dynamically as conversations escalate through grooming stages. Beyond contributing a systematically labeled dataset, our findings introduce a stage-based taxonomy of \revision{perceived} protective strategies that opens new directions for designing sociotechnical interventions. 
Rather than \revision{complements detection-based approaches}, our results highlight the importance of empowering them as active agents capable of recognizing, resisting, and redirecting risky online interactions. As digital spaces continue to evolve, advancing tools and educational approaches that embed teen agency at their core is essential to safeguarding youth online. 
Protecting teens from cybergrooming requires more than just detecting perpetrators; it also requires equipping them with the knowledge, confidence, and strategies to protect themselves.



\begin{acks}
This research is supported by the U.S. National Science Foundation under grants \#CNS-2330940, \#CNS-2550834, \#IIS-2550812, \#TI-2550746, by the William T. Grant Foundation grant \#187941, and by the Center for Human-Computer Interaction (CHCI) at Virginia Tech. The views expressed in this material are those of the authors and do not necessarily reflect the views of the funding agencies.
\end{acks}


\bibliographystyle{ACM-Reference-Format}




\begin{thebibliography}{93}


\ifx \showCODEN    \undefined \def \showCODEN     #1{\unskip}     \fi
\ifx \showISBNx    \undefined \def \showISBNx     #1{\unskip}     \fi
\ifx \showISBNxiii \undefined \def \showISBNxiii  #1{\unskip}     \fi
\ifx \showISSN     \undefined \def \showISSN      #1{\unskip}     \fi
\ifx \showLCCN     \undefined \def \showLCCN      #1{\unskip}     \fi
\ifx \shownote     \undefined \def \shownote      #1{#1}          \fi
\ifx \showarticletitle \undefined \def \showarticletitle #1{#1}   \fi
\ifx \showURL      \undefined \def \showURL       {\relax}        \fi
\providecommand\bibfield[2]{#2}
\providecommand\bibinfo[2]{#2}
\providecommand\natexlab[1]{#1}
\providecommand\showeprint[2][]{arXiv:#2}

\bibitem[Acquisti et~al\mbox{.}(2012)]%
        {acquisti2012impact}
\bibfield{author}{\bibinfo{person}{Alessandro Acquisti},
  \bibinfo{person}{Leslie~K John}, {and} \bibinfo{person}{George Loewenstein}.}
  \bibinfo{year}{2012}\natexlab{}.
\newblock \showarticletitle{The impact of relative standards on the propensity
  to disclose}.
\newblock \bibinfo{journal}{\emph{Journal of Marketing Research}}
  \bibinfo{volume}{49}, \bibinfo{number}{2} (\bibinfo{year}{2012}),
  \bibinfo{pages}{160--174}.
\newblock


\bibitem[Agha et~al\mbox{.}(2025)]%
        {agha2025systematic}
\bibfield{author}{\bibinfo{person}{Zainab Agha}, \bibinfo{person}{Naima~Samreen
  Ali}, \bibinfo{person}{Jinkyung Park}, {and} \bibinfo{person}{Pamela~J
  Wisniewski}.} \bibinfo{year}{2025}\natexlab{}.
\newblock \showarticletitle{A systematic review on design-based nudges for
  adolescent online safety}.
\newblock \bibinfo{journal}{\emph{International Journal of Child-Computer
  Interaction}}  \bibinfo{volume}{43} (\bibinfo{year}{2025}),
  \bibinfo{pages}{100702}.
\newblock


\bibitem[Agha et~al\mbox{.}(2023)]%
        {agha2023strike}
\bibfield{author}{\bibinfo{person}{Zainab Agha}, \bibinfo{person}{Karla
  Badillo-Urquiola}, {and} \bibinfo{person}{Pamela~J Wisniewski}.}
  \bibinfo{year}{2023}\natexlab{}.
\newblock \showarticletitle{"Strike at the Root": Co-designing Real-Time Social
  Media Interventions for Adolescent Online Risk Prevention}.
\newblock \bibinfo{journal}{\emph{Proceedings of the ACM on Human-Computer
  Interaction}} \bibinfo{volume}{7}, \bibinfo{number}{CSCW1}
  (\bibinfo{year}{2023}), \bibinfo{pages}{1--32}.
\newblock


\bibitem[Agha et~al\mbox{.}(2024)]%
        {agha2024tricky}
\bibfield{author}{\bibinfo{person}{Zainab Agha}, \bibinfo{person}{Jinkyung
  Park}, \bibinfo{person}{Ruyuan Wan}, \bibinfo{person}{Naima~Samreen Ali},
  \bibinfo{person}{Yiwei Wang}, \bibinfo{person}{Dominic Difranzo},
  \bibinfo{person}{Karla Badillo-Urquiola}, {and} \bibinfo{person}{Pamela~J
  Wisniewski}.} \bibinfo{year}{2024}\natexlab{}.
\newblock \showarticletitle{Tricky vs. transparent: Towards an ecologically
  valid and safe approach for evaluating online safety nudges for teens}. In
  \bibinfo{booktitle}{\emph{Proceedings of the 2024 CHI Conference on Human
  Factors in Computing Systems}}. \bibinfo{pages}{1--20}.
\newblock


\bibitem[Aiello and McFarland(2014)]%
        {aiello2014detecting}
\bibfield{author}{\bibinfo{person}{LM Aiello} {and} \bibinfo{person}{D
  McFarland}.} \bibinfo{year}{2014}\natexlab{}.
\newblock \showarticletitle{Detecting child grooming behaviour patterns on
  social media}.
\newblock \bibinfo{journal}{\emph{Lecture Notes in Computer Science (Including
  Subseries Lecture Notes in Artificial Intelligence and Lecture Notes in
  Bioinformatics)}} \bibinfo{volume}{8851}, \bibinfo{number}{16}
  (\bibinfo{year}{2014}), \bibinfo{pages}{10--1007}.
\newblock


\bibitem[Akter et~al\mbox{.}(2024)]%
        {akter2024towards}
\bibfield{author}{\bibinfo{person}{Mamtaj Akter}, \bibinfo{person}{Zainab
  Agha}, \bibinfo{person}{Ashwaq Alsoubai}, \bibinfo{person}{Naima Ali}, {and}
  \bibinfo{person}{Pamela Wisniewski}.} \bibinfo{year}{2024}\natexlab{}.
\newblock \showarticletitle{Towards collaborative family-centered design for
  online safety, privacy and security}.
\newblock \bibinfo{journal}{\emph{arXiv preprint arXiv:2404.03165}}
  (\bibinfo{year}{2024}).
\newblock


\bibitem[Akter et~al\mbox{.}(2023a)]%
        {akter2023takes}
\bibfield{author}{\bibinfo{person}{Mamtaj Akter}, \bibinfo{person}{Leena
  Alghamdi}, \bibinfo{person}{Jess Kropczynski},
  \bibinfo{person}{Heather~Richter Lipford}, {and} \bibinfo{person}{Pamela~J
  Wisniewski}.} \bibinfo{year}{2023}\natexlab{a}.
\newblock \showarticletitle{It takes a village: A case for including extended
  family members in the joint oversight of family-based privacy and security
  for mobile smartphones}. In \bibinfo{booktitle}{\emph{Extended Abstracts of
  the 2023 CHI Conference on Human Factors in Computing Systems}}.
  \bibinfo{pages}{1--7}.
\newblock


\bibitem[Akter et~al\mbox{.}(2022)]%
        {akter2022parental}
\bibfield{author}{\bibinfo{person}{Mamtaj Akter}, \bibinfo{person}{Amy~J
  Godfrey}, \bibinfo{person}{Jess Kropczynski}, \bibinfo{person}{Heather~R
  Lipford}, {and} \bibinfo{person}{Pamela~J Wisniewski}.}
  \bibinfo{year}{2022}\natexlab{}.
\newblock \showarticletitle{From parental control to joint family oversight:
  Can parents and teens manage mobile online safety and privacy as equals?}
\newblock \bibinfo{journal}{\emph{Proceedings of the ACM on Human-Computer
  Interaction}} \bibinfo{volume}{6}, \bibinfo{number}{CSCW1}
  (\bibinfo{year}{2022}), \bibinfo{pages}{1--28}.
\newblock


\bibitem[Akter et~al\mbox{.}(2025)]%
        {akter2025calculating}
\bibfield{author}{\bibinfo{person}{Mamtaj Akter},
  \bibinfo{person}{Jinkyung~Katie Park}, \bibinfo{person}{Campbell~Robinson
  Headrick}, \bibinfo{person}{Xinru Page}, {and} \bibinfo{person}{Pamela~J
  Wisniewski}.} \bibinfo{year}{2025}\natexlab{}.
\newblock \showarticletitle{Calculating Connection vs. Risk: Understanding How
  Youth Negotiate Digital Privacy and Security with Peers Online}.
\newblock \bibinfo{journal}{\emph{Proceedings of the ACM on Human-Computer
  Interaction}} \bibinfo{volume}{9}, \bibinfo{number}{7}
  (\bibinfo{year}{2025}), \bibinfo{pages}{1--26}.
\newblock


\bibitem[Akter et~al\mbox{.}(2023b)]%
        {akter2023Evaluating}
\bibfield{author}{\bibinfo{person}{Mamtaj Akter}, \bibinfo{person}{Madiha
  Tabassum}, \bibinfo{person}{Nazmus~Sakib Miazi}, \bibinfo{person}{Leena
  Alghamdi}, \bibinfo{person}{Jess Kropczynski}, \bibinfo{person}{Pamela~J
  Wisniewski}, {and} \bibinfo{person}{Heather Lipford}.}
  \bibinfo{year}{2023}\natexlab{b}.
\newblock \showarticletitle{Evaluating the impact of community oversight for
  managing mobile privacy and security}. In
  \bibinfo{booktitle}{\emph{Nineteenth Symposium on Usable Privacy and Security
  (SOUPS 2023)}}. \bibinfo{pages}{437--456}.
\newblock


\bibitem[Al-Salehi et~al\mbox{.}(2021)]%
        {al-salehi_jammer_2021}
\bibfield{author}{\bibinfo{person}{Abdul~Rahman Al-Salehi},
  \bibinfo{person}{Ijaz~Mansoor Qureshi}, \bibinfo{person}{Aqdas~Naveed Malik},
  \bibinfo{person}{Zafar~Ullah Khan}, {and} \bibinfo{person}{Wasim Khan}.}
  \bibinfo{year}{2021}\natexlab{}.
\newblock \showarticletitle{Jammer avoidance for dual-function
  radar-communications using {FSK} and independent null steering}.
\newblock \bibinfo{journal}{\emph{Digital Signal Processing}}
  \bibinfo{volume}{114} (\bibinfo{date}{July} \bibinfo{year}{2021}),
  \bibinfo{pages}{103057}.
\newblock
\showISSN{1051-2004}
\href{https://doi.org/10.1016/j.dsp.2021.103057}{doi:\nolinkurl{10.1016/j.dsp.2021.103057}}


\bibitem[Alluhidan et~al\mbox{.}(2024)]%
        {alluhidan2024teen}
\bibfield{author}{\bibinfo{person}{Abdulmalik Alluhidan},
  \bibinfo{person}{Mamtaj Akter}, \bibinfo{person}{Ashwaq Alsoubai},
  \bibinfo{person}{Jinkyung~Katie Park}, {and} \bibinfo{person}{Pamela
  Wisniewski}.} \bibinfo{year}{2024}\natexlab{}.
\newblock \showarticletitle{Teen talk: The good, the bad, and the neutral of
  adolescent social media use}.
\newblock \bibinfo{journal}{\emph{Proceedings of the ACM on Human-Computer
  Interaction}} \bibinfo{volume}{8}, \bibinfo{number}{CSCW2}
  (\bibinfo{year}{2024}), \bibinfo{pages}{1--35}.
\newblock


\bibitem[Alsoubai et~al\mbox{.}(2022)]%
        {alsoubai2022friends}
\bibfield{author}{\bibinfo{person}{Ashwaq Alsoubai}, \bibinfo{person}{Jihye
  Song}, \bibinfo{person}{Afsaneh Razi}, \bibinfo{person}{Nurun Naher},
  \bibinfo{person}{Munmun De~Choudhury}, {and} \bibinfo{person}{Pamela~J
  Wisniewski}.} \bibinfo{year}{2022}\natexlab{}.
\newblock \showarticletitle{From'Friends with Benefits' to'Sextortion:'A
  Nuanced Investigation of Adolescents' Online Sexual Risk Experiences}.
\newblock \bibinfo{journal}{\emph{Proceedings of the ACM on Human-Computer
  Interaction}} \bibinfo{volume}{6}, \bibinfo{number}{CSCW2}
  (\bibinfo{year}{2022}), \bibinfo{pages}{1--32}.
\newblock


\bibitem[An et~al\mbox{.}(2025)]%
        {an2025toward}
\bibfield{author}{\bibinfo{person}{Heajun An}, \bibinfo{person}{Marcos Silva},
  \bibinfo{person}{Qi Zhang}, \bibinfo{person}{Arav Singh},
  \bibinfo{person}{Minqian Liu}, \bibinfo{person}{Xinyi Zhang},
  \bibinfo{person}{Sarvech Qadir}, \bibinfo{person}{Sang~Won Lee},
  \bibinfo{person}{Lifu Huang}, \bibinfo{person}{Pamela Wisnieswski},
  {et~al\mbox{.}}} \bibinfo{year}{2025}\natexlab{}.
\newblock \showarticletitle{Toward Integrated Solutions: A Systematic
  Interdisciplinary Review of Cybergrooming Research}.
\newblock \bibinfo{journal}{\emph{arXiv preprint arXiv:2503.05727}}
  (\bibinfo{year}{2025}).
\newblock


\bibitem[Andoh(2025)]%
        {Andoh_2025}
\bibfield{author}{\bibinfo{person}{Efua Andoh}.}
  \bibinfo{year}{2025}\natexlab{}.
\newblock \bibinfo{title}{Many teens are turning to AI chatbots for friendship
  and emotional support}.
\newblock
\urldef\tempurl%
\url{https://www.apa.org/monitor/2025/10/technology-youth-friendships}
\showURL{%
\tempurl}


\bibitem[Ashcroft et~al\mbox{.}(2015)]%
        {ashcroft2015step}
\bibfield{author}{\bibinfo{person}{Michael Ashcroft}, \bibinfo{person}{Lisa
  Kaati}, {and} \bibinfo{person}{Maxime Meyer}.}
  \bibinfo{year}{2015}\natexlab{}.
\newblock \showarticletitle{A Step Towards Detecting Online
  Grooming--Identifying Adults Pretending to be Children}. In
  \bibinfo{booktitle}{\emph{2015 European Intelligence and Security Informatics
  Conference}}. IEEE, \bibinfo{pages}{98--104}.
\newblock


\bibitem[Bogdanova et~al\mbox{.}(2014)]%
        {bogdanova2014exploring}
\bibfield{author}{\bibinfo{person}{Dasha Bogdanova}, \bibinfo{person}{Paolo
  Rosso}, {and} \bibinfo{person}{Thamar Solorio}.}
  \bibinfo{year}{2014}\natexlab{}.
\newblock \showarticletitle{Exploring high-level features for detecting
  cyberpedophilia}.
\newblock \bibinfo{journal}{\emph{Computer speech \& language}}
  \bibinfo{volume}{28}, \bibinfo{number}{1} (\bibinfo{year}{2014}),
  \bibinfo{pages}{108--120}.
\newblock


\bibitem[Borj and Bours(2019)]%
        {borj2019predatory}
\bibfield{author}{\bibinfo{person}{Parisa~Rezaee Borj} {and}
  \bibinfo{person}{Patrick Bours}.} \bibinfo{year}{2019}\natexlab{}.
\newblock \showarticletitle{Predatory conversation detection}. In
  \bibinfo{booktitle}{\emph{2019 International Conference on Cyber Security for
  Emerging Technologies (CSET)}}. IEEE, \bibinfo{pages}{1--6}.
\newblock


\bibitem[Bours and Kulsrud(2019)]%
        {bours2019detection}
\bibfield{author}{\bibinfo{person}{Patrick Bours} {and} \bibinfo{person}{Halvor
  Kulsrud}.} \bibinfo{year}{2019}\natexlab{}.
\newblock \showarticletitle{Detection of cyber grooming in online
  conversation}. In \bibinfo{booktitle}{\emph{2019 IEEE International Workshop
  on Information Forensics and Security (WIFS)}}. IEEE, \bibinfo{pages}{1--6}.
\newblock


\bibitem[Charalambous et~al\mbox{.}(2020)]%
        {charalambous2020privacy}
\bibfield{author}{\bibinfo{person}{Markos Charalambous},
  \bibinfo{person}{Petros Papagiannis}, \bibinfo{person}{Antonis Papasavva},
  \bibinfo{person}{Pantelitsa Leonidou}, \bibinfo{person}{Rafael Constaninou},
  \bibinfo{person}{Lia Terzidou}, \bibinfo{person}{Theodoros Christophides},
  \bibinfo{person}{Pantelis Nicolaou}, \bibinfo{person}{Orfeas Theofanis},
  \bibinfo{person}{George Kalatzantonakis}, {et~al\mbox{.}}}
  \bibinfo{year}{2020}\natexlab{}.
\newblock \showarticletitle{A privacy-preserving architecture for the
  protection of adolescents in online social networks}.
\newblock \bibinfo{journal}{\emph{arXiv preprint arXiv:2007.12038}}
  (\bibinfo{year}{2020}).
\newblock


\bibitem[Chatterjee(2025)]%
        {Chatterjee_2025}
\bibfield{author}{\bibinfo{person}{Rhitu Chatterjee}.}
  \bibinfo{year}{2025}\natexlab{}.
\newblock \bibinfo{title}{As more teens use AI chatbots, parents and lawmakers
  sound the alarm about dangers}.
\newblock
\urldef\tempurl%
\url{https://www.npr.org/2025/10/01/nx-s1-5527041/as-more-teens-use-ai-chatbots-parents-and-lawmakers-sound-the-alarm-about-dangers}
\showURL{%
\tempurl}


\bibitem[Choo(2009)]%
        {choo2009online}
\bibfield{author}{\bibinfo{person}{Kim-Kwang~Raymond Choo}.}
  \bibinfo{year}{2009}\natexlab{}.
\newblock \showarticletitle{Online child grooming: A literature review on the
  misuse of social networking sites for grooming children for sexual offences}.
\newblock  (\bibinfo{year}{2009}).
\newblock


\bibitem[Christian(2021)]%
        {Christian_2021}
\bibfield{author}{\bibinfo{person}{Brian Christian}.}
  \bibinfo{year}{2021}\natexlab{}.
\newblock \bibinfo{booktitle}{\emph{The alignment problem: Machine Learning and
  human values}}.
\newblock \bibinfo{publisher}{W.W. Norton \& Company}.
\newblock


\bibitem[Cohen et~al\mbox{.}(2018)]%
        {cohen2018education}
\bibfield{author}{\bibinfo{person}{Robin Cohen}, \bibinfo{person}{Nivedha
  Mathiarasu}, \bibinfo{person}{R Aarif}, \bibinfo{person}{S Ansari},
  \bibinfo{person}{D Fraser}, \bibinfo{person}{M Hegde}, \bibinfo{person}{J
  Henderson}, \bibinfo{person}{I Kajic}, \bibinfo{person}{A Khan},
  \bibinfo{person}{Z Liao}, {et~al\mbox{.}}} \bibinfo{year}{2018}\natexlab{}.
\newblock \showarticletitle{An education-based approach to aid in the
  prevention of cyberbullying}.
\newblock \bibinfo{journal}{\emph{Acm Sigcas Computers and Society}}
  \bibinfo{volume}{47}, \bibinfo{number}{4} (\bibinfo{year}{2018}),
  \bibinfo{pages}{17--28}.
\newblock


\bibitem[Cranor et~al\mbox{.}(2014)]%
        {cranor2014parents}
\bibfield{author}{\bibinfo{person}{Lorrie~Faith Cranor},
  \bibinfo{person}{Adam~L Durity}, \bibinfo{person}{Abigail Marsh}, {and}
  \bibinfo{person}{Blase Ur}.} \bibinfo{year}{2014}\natexlab{}.
\newblock \showarticletitle{$\{$Parents’$\}$ and $\{$Teens’$\}$
  Perspectives on Privacy In a $\{$Technology-Filled$\}$ World}. In
  \bibinfo{booktitle}{\emph{10th Symposium On Usable Privacy and Security
  (SOUPS 2014)}}. \bibinfo{pages}{19--35}.
\newblock


\bibitem[Curran and Winther(2025)]%
        {Curran_Winther_2025a}
\bibfield{author}{\bibinfo{person}{Stephanie Curran} {and}
  \bibinfo{person}{Daniel~Kardefelt Winther}.} \bibinfo{year}{2025}\natexlab{}.
\newblock
\urldef\tempurl%
\url{https://www.unicef.org/innocenti/stories/debunking-four-myths-about-childrens-safety-online}
\showURL{%
\tempurl}


\bibitem[De~Santisteban and G{\'o}mez-Guadix(2017)]%
        {de2017estrategias}
\bibfield{author}{\bibinfo{person}{P De~Santisteban} {and} \bibinfo{person}{M
  G{\'o}mez-Guadix}.} \bibinfo{year}{2017}\natexlab{}.
\newblock \bibinfo{title}{Estrategias de persuasi{\'o}n en grooming on line: un
  an{\'a}lisis cualitativo con agresores en prisi{\'o}n. Psychological
  Intervention, 26 (3), 139-146}.
\newblock


\bibitem[DeAngelis(2024)]%
        {DeAngelis_2024}
\bibfield{author}{\bibinfo{person}{Tori DeAngelis}.}
  \bibinfo{year}{2024}\natexlab{}.
\newblock \bibinfo{title}{Teens are spending nearly 5 hours daily on social
  media. Here are the mental health outcomes}.
\newblock
\urldef\tempurl%
\url{https://www.apa.org/monitor/2024/04/teen-social-use-mental-health}
\showURL{%
\tempurl}


\bibitem[Dedkova(2015)]%
        {dedkova2015stranger}
\bibfield{author}{\bibinfo{person}{Lenka Dedkova}.}
  \bibinfo{year}{2015}\natexlab{}.
\newblock \showarticletitle{Stranger is not always danger: The myth and reality
  of meetings with online strangers}.
\newblock In \bibinfo{booktitle}{\emph{Living in the digital age:
  Self-presentation, networking, playing, and participating in politics}}.
  \bibinfo{publisher}{Masarykova univerzita nakladatelstv{\'\i}},
  \bibinfo{pages}{78--94}.
\newblock


\bibitem[Dorasamy et~al\mbox{.}(2021)]%
        {dorasamy2021parents}
\bibfield{author}{\bibinfo{person}{Magiswary Dorasamy}, \bibinfo{person}{Maniam
  Kaliannan}, \bibinfo{person}{Manimekalai Jambulingam}, \bibinfo{person}{Iqbal
  Ramadhan}, {and} \bibinfo{person}{Ashok Sivaji}.}
  \bibinfo{year}{2021}\natexlab{}.
\newblock \showarticletitle{Parents' awareness on online predators: Cyber
  grooming deterrence}.
\newblock \bibinfo{journal}{\emph{The Qualitative Report}}
  \bibinfo{volume}{26}, \bibinfo{number}{11} (\bibinfo{year}{2021}),
  \bibinfo{pages}{3683--3723}.
\newblock


\bibitem[Egelman et~al\mbox{.}(2013)]%
        {egelman2013choice}
\bibfield{author}{\bibinfo{person}{Serge Egelman},
  \bibinfo{person}{Adrienne~Porter Felt}, {and} \bibinfo{person}{David
  Wagner}.} \bibinfo{year}{2013}\natexlab{}.
\newblock \showarticletitle{Choice architecture and smartphone privacy:
  There’sa price for that}.
\newblock In \bibinfo{booktitle}{\emph{The economics of information security
  and privacy}}. \bibinfo{publisher}{Springer}, \bibinfo{pages}{211--236}.
\newblock


\bibitem[Eilifsen et~al\mbox{.}(2023)]%
        {eilifsen2023early}
\bibfield{author}{\bibinfo{person}{Thomas~Nyrem Eilifsen},
  \bibinfo{person}{Bhanu Shrestha}, {and} \bibinfo{person}{Patrick Bours}.}
  \bibinfo{year}{2023}\natexlab{}.
\newblock \showarticletitle{Early Detection of Cyber Grooming in Online
  Conversations: A Dynamic Trust Model and Sliding Window Approach}. In
  \bibinfo{booktitle}{\emph{2023 21st International Conference on Emerging
  eLearning Technologies and Applications (ICETA)}}. IEEE,
  \bibinfo{pages}{129--134}.
\newblock


\bibitem[Fauzi and Bours(2020)]%
        {fauzi2020ensemble}
\bibfield{author}{\bibinfo{person}{Muhammad~Ali Fauzi} {and}
  \bibinfo{person}{Patrick Bours}.} \bibinfo{year}{2020}\natexlab{}.
\newblock \showarticletitle{Ensemble method for sexual predators identification
  in online chats}. In \bibinfo{booktitle}{\emph{2020 8th international
  workshop on biometrics and forensics (IWBF)}}. IEEE, \bibinfo{pages}{1--6}.
\newblock


\bibitem[Fauzi et~al\mbox{.}(2023)]%
        {fauzi2023identifying}
\bibfield{author}{\bibinfo{person}{Muhammad~Ali Fauzi},
  \bibinfo{person}{Stephen Wolthusen}, \bibinfo{person}{Bian Yang},
  \bibinfo{person}{Patrick Bours}, {and} \bibinfo{person}{Prosper Yeng}.}
  \bibinfo{year}{2023}\natexlab{}.
\newblock \showarticletitle{Identifying sexual predators in chats using svm and
  feature ensemble}. In \bibinfo{booktitle}{\emph{2023 International Conference
  on Emerging Trends in Networks and Computer Communications (ETNCC)}}. IEEE,
  \bibinfo{pages}{1--6}.
\newblock


\bibitem[Freed et~al\mbox{.}(2023)]%
        {freed2023understanding}
\bibfield{author}{\bibinfo{person}{Diana Freed}, \bibinfo{person}{Natalie~N
  Bazarova}, \bibinfo{person}{Sunny Consolvo}, \bibinfo{person}{Eunice~J Han},
  \bibinfo{person}{Patrick~Gage Kelley}, \bibinfo{person}{Kurt Thomas}, {and}
  \bibinfo{person}{Dan Cosley}.} \bibinfo{year}{2023}\natexlab{}.
\newblock \showarticletitle{Understanding digital-safety experiences of youth
  in the US}. In \bibinfo{booktitle}{\emph{Proceedings of the 2023 CHI
  Conference on Human Factors in Computing Systems}}. \bibinfo{pages}{1--15}.
\newblock


\bibitem[Front(2023)]%
        {Human_Trafficking_Front_2023}
\bibfield{author}{\bibinfo{person}{Human~Trafficking Front}.}
  \bibinfo{year}{2023}\natexlab{}.
\newblock \bibinfo{title}{Preventing online grooming: Links to child
  trafficking, harm, and the need to enhance global coordination}.
\newblock
\urldef\tempurl%
\url{https://humantraffickingfront.org/preventing-online-grooming-links-to-child-trafficking-harm-and-the-need-to-enhance-global-coordination/}
\showURL{%
\tempurl}


\bibitem[Gabriel(2020)]%
        {gabriel2020artificial}
\bibfield{author}{\bibinfo{person}{Iason Gabriel}.}
  \bibinfo{year}{2020}\natexlab{}.
\newblock \showarticletitle{Artificial intelligence, values, and alignment}.
\newblock \bibinfo{journal}{\emph{Minds and machines}} \bibinfo{volume}{30},
  \bibinfo{number}{3} (\bibinfo{year}{2020}), \bibinfo{pages}{411--437}.
\newblock


\bibitem[Gabrielli et~al\mbox{.}(2020)]%
        {gabrielli2020chatbot}
\bibfield{author}{\bibinfo{person}{Silvia Gabrielli}, \bibinfo{person}{Silvia
  Rizzi}, \bibinfo{person}{Sara Carbone}, \bibinfo{person}{Valeria Donisi},
  {et~al\mbox{.}}} \bibinfo{year}{2020}\natexlab{}.
\newblock \showarticletitle{A chatbot-based coaching intervention for
  adolescents to promote life skills: pilot study}.
\newblock \bibinfo{journal}{\emph{JMIR human factors}} \bibinfo{volume}{7},
  \bibinfo{number}{1} (\bibinfo{year}{2020}), \bibinfo{pages}{e16762}.
\newblock


\bibitem[Gerber et~al\mbox{.}(2018)]%
        {gerber2018explaining}
\bibfield{author}{\bibinfo{person}{Nina Gerber}, \bibinfo{person}{Paul Gerber},
  {and} \bibinfo{person}{Melanie Volkamer}.} \bibinfo{year}{2018}\natexlab{}.
\newblock \showarticletitle{Explaining the privacy paradox: A systematic review
  of literature investigating privacy attitude and behavior}.
\newblock \bibinfo{journal}{\emph{Computers \& security}}  \bibinfo{volume}{77}
  (\bibinfo{year}{2018}), \bibinfo{pages}{226--261}.
\newblock


\bibitem[Ghosh et~al\mbox{.}(2020)]%
        {ghosh_circle_2020}
\bibfield{author}{\bibinfo{person}{Arup~Kumar Ghosh},
  \bibinfo{person}{Charles~E. Hughes}, {and} \bibinfo{person}{Pamela~J.
  Wisniewski}.} \bibinfo{year}{2020}\natexlab{}.
\newblock \showarticletitle{Circle of {Trust}: {A} {New} {Approach} to {Mobile}
  {Online} {Safety} for {Families}}. In \bibinfo{booktitle}{\emph{Proceedings
  of the 2020 {CHI} {Conference} on {Human} {Factors} in {Computing}
  {Systems}}}. \bibinfo{publisher}{ACM}, \bibinfo{address}{Honolulu HI USA},
  \bibinfo{pages}{1--14}.
\newblock
\showISBNx{978-1-4503-6708-0}
\href{https://doi.org/10.1145/3313831.3376747}{doi:\nolinkurl{10.1145/3313831.3376747}}


\bibitem[Greene-Colozzi et~al\mbox{.}(2020)]%
        {greene2020experiences}
\bibfield{author}{\bibinfo{person}{Emily~A Greene-Colozzi},
  \bibinfo{person}{Georgia~M Winters}, \bibinfo{person}{Brandy Blasko}, {and}
  \bibinfo{person}{Elizabeth~L Jeglic}.} \bibinfo{year}{2020}\natexlab{}.
\newblock \showarticletitle{Experiences and perceptions of online sexual
  solicitation and grooming of minors: A retrospective report}.
\newblock \bibinfo{journal}{\emph{Journal of child sexual abuse}}
  \bibinfo{volume}{29}, \bibinfo{number}{7} (\bibinfo{year}{2020}),
  \bibinfo{pages}{836--854}.
\newblock


\bibitem[Gunawan et~al\mbox{.}(2016)]%
        {gunawan2016detecting}
\bibfield{author}{\bibinfo{person}{Fergyanto~E Gunawan}, \bibinfo{person}{Livia
  Ashianti}, \bibinfo{person}{Sevenpri Candra}, {and} \bibinfo{person}{Benfano
  Soewito}.} \bibinfo{year}{2016}\natexlab{}.
\newblock \showarticletitle{Detecting online child grooming conversation}. In
  \bibinfo{booktitle}{\emph{2016 11th International Conference on Knowledge,
  Information and Creativity Support Systems (KICSS)}}. IEEE,
  \bibinfo{pages}{1--6}.
\newblock


\bibitem[Guo et~al\mbox{.}(2023)]%
        {guo2023text}
\bibfield{author}{\bibinfo{person}{Zhen Guo}, \bibinfo{person}{Pei Wang},
  \bibinfo{person}{Jin-Hee Cho}, {and} \bibinfo{person}{Lifu Huang}.}
  \bibinfo{year}{2023}\natexlab{}.
\newblock \showarticletitle{Text mining-based social-psychological
  vulnerability analysis of potential victims to cybergrooming: Insights and
  lessons learned}. In \bibinfo{booktitle}{\emph{Companion Proceedings of the
  ACM Web Conference 2023}}. \bibinfo{pages}{1381--1388}.
\newblock


\bibitem[Gupta et~al\mbox{.}(2012)]%
        {gupta2012characterizing}
\bibfield{author}{\bibinfo{person}{Aditi Gupta}, \bibinfo{person}{Ponnurangam
  Kumaraguru}, {and} \bibinfo{person}{Ashish Sureka}.}
  \bibinfo{year}{2012}\natexlab{}.
\newblock \showarticletitle{Characterizing pedophile conversations on the
  internet using online grooming}.
\newblock \bibinfo{journal}{\emph{arXiv preprint arXiv:1208.4324}}
  (\bibinfo{year}{2012}).
\newblock


\bibitem[Hartikainen et~al\mbox{.}(2021)]%
        {hartikainen2021if}
\bibfield{author}{\bibinfo{person}{Heidi Hartikainen}, \bibinfo{person}{Afsaneh
  Razi}, {and} \bibinfo{person}{Pamela Wisniewski}.}
  \bibinfo{year}{2021}\natexlab{}.
\newblock \showarticletitle{‘If You Care About Me, You'll Send Me a
  Pic’-Examining the Role of Peer Pressure in Adolescent Sexting}. In
  \bibinfo{booktitle}{\emph{Companion Publication of the 2021 Conference on
  Computer Supported Cooperative Work and Social Computing}}.
  \bibinfo{pages}{67--71}.
\newblock


\bibitem[Hashish et~al\mbox{.}(2014)]%
        {hashish_involving_2014}
\bibfield{author}{\bibinfo{person}{Yasmeen Hashish}, \bibinfo{person}{Andrea
  Bunt}, {and} \bibinfo{person}{James~E. Young}.}
  \bibinfo{year}{2014}\natexlab{}.
\newblock \showarticletitle{Involving children in content control: a
  collaborative and education-oriented content filtering approach}. In
  \bibinfo{booktitle}{\emph{Proceedings of the {SIGCHI} {Conference} on {Human}
  {Factors} in {Computing} {Systems}}} \emph{(\bibinfo{series}{{CHI} '14})}.
  \bibinfo{publisher}{Association for Computing Machinery},
  \bibinfo{address}{New York, NY, USA}, \bibinfo{pages}{1797--1806}.
\newblock
\showISBNx{978-1-4503-2473-1}
\href{https://doi.org/10.1145/2556288.2557128}{doi:\nolinkurl{10.1145/2556288.2557128}}


\bibitem[INHOPE(2022)]%
        {INHOPE}
\bibfield{author}{\bibinfo{person}{INHOPE}.} \bibinfo{year}{2022}\natexlab{}.
\newblock
\urldef\tempurl%
\url{https://inhope.org/EN/articles/the-impact-of-online-grooming#:~:text=The%20consequences,traumatic%20stress%2C%20and%20suicidal%20thoughts.}
\showURL{%
\tempurl}


\bibitem[Isaza et~al\mbox{.}(2022)]%
        {isaza2022classifying}
\bibfield{author}{\bibinfo{person}{Gustavo Isaza}, \bibinfo{person}{Fabi{\'a}n
  Mu{\~n}oz}, \bibinfo{person}{Luis Castillo}, {and} \bibinfo{person}{Felipe
  Buitrago}.} \bibinfo{year}{2022}\natexlab{}.
\newblock \showarticletitle{Classifying cybergrooming for child online
  protection using hybrid machine learning model}.
\newblock \bibinfo{journal}{\emph{Neurocomputing}}  \bibinfo{volume}{484}
  (\bibinfo{year}{2022}), \bibinfo{pages}{250--259}.
\newblock


\bibitem[Jeglic(2022)]%
        {Jeglic_2022}
\bibfield{author}{\bibinfo{person}{Elizabeth~L. Jeglic}.}
  \bibinfo{year}{2022}\natexlab{}.
\newblock \bibinfo{title}{Fighting the “Stranger Danger” myth}.
\newblock
\urldef\tempurl%
\url{https://www.psychologytoday.com/us/blog/protecting-children-from-sexual-abuse/202209/fighting-the-stranger-danger-myth}
\showURL{%
\tempurl}


\bibitem[Jia et~al\mbox{.}(2015)]%
        {jia2015risk}
\bibfield{author}{\bibinfo{person}{Haiyan Jia}, \bibinfo{person}{Pamela~J
  Wisniewski}, \bibinfo{person}{Heng Xu}, \bibinfo{person}{Mary~Beth Rosson},
  {and} \bibinfo{person}{John~M Carroll}.} \bibinfo{year}{2015}\natexlab{}.
\newblock \showarticletitle{Risk-taking as a learning process for shaping
  teen's online information privacy behaviors}. In
  \bibinfo{booktitle}{\emph{Proceedings of the 18th ACM Conference on Computer
  Supported Cooperative Work \& Social Computing}}. \bibinfo{pages}{583--599}.
\newblock


\bibitem[Kloess et~al\mbox{.}(2014)]%
        {kloess2014online}
\bibfield{author}{\bibinfo{person}{Juliane~A Kloess},
  \bibinfo{person}{Anthony~R Beech}, {and} \bibinfo{person}{Leigh Harkins}.}
  \bibinfo{year}{2014}\natexlab{}.
\newblock \showarticletitle{Online child sexual exploitation: Prevalence,
  process, and offender characteristics}.
\newblock \bibinfo{journal}{\emph{Trauma, Violence, \& Abuse}}
  \bibinfo{volume}{15}, \bibinfo{number}{2} (\bibinfo{year}{2014}),
  \bibinfo{pages}{126--139}.
\newblock


\bibitem[Kokolakis(2017)]%
        {kokolakis2017privacy}
\bibfield{author}{\bibinfo{person}{Spyros Kokolakis}.}
  \bibinfo{year}{2017}\natexlab{}.
\newblock \showarticletitle{Privacy attitudes and privacy behaviour: A review
  of current research on the privacy paradox phenomenon}.
\newblock \bibinfo{journal}{\emph{Computers \& security}}  \bibinfo{volume}{64}
  (\bibinfo{year}{2017}), \bibinfo{pages}{122--134}.
\newblock


\bibitem[Lenhart(2015)]%
        {Lenhart_2015}
\bibfield{author}{\bibinfo{person}{Amanda Lenhart}.}
  \bibinfo{year}{2015}\natexlab{}.
\newblock \bibinfo{title}{Teens, technology and Friendships}.
\newblock
\urldef\tempurl%
\url{https://www.pewresearch.org/internet/2015/08/06/teens-technology-and-friendships/?utm_source=chatgpt.com}
\showURL{%
\tempurl}


\bibitem[Livingstone and Smith(2014)]%
        {livingstone_annual_2014}
\bibfield{author}{\bibinfo{person}{Sonia Livingstone} {and}
  \bibinfo{person}{Peter~K. Smith}.} \bibinfo{year}{2014}\natexlab{}.
\newblock \showarticletitle{Annual research review: {Harms} experienced by
  child users of online and mobile technologies: the nature, prevalence and
  management of sexual and aggressive risks in the digital age}.
\newblock \bibinfo{journal}{\emph{Journal of Child Psychology and Psychiatry,
  and Allied Disciplines}} \bibinfo{volume}{55}, \bibinfo{number}{6}
  (\bibinfo{date}{June} \bibinfo{year}{2014}), \bibinfo{pages}{635--654}.
\newblock
\showISSN{1469-7610}
\href{https://doi.org/10.1111/jcpp.12197}{doi:\nolinkurl{10.1111/jcpp.12197}}


\bibitem[Maeng and Lee(2022)]%
        {maeng2022designing}
\bibfield{author}{\bibinfo{person}{Wookjae Maeng} {and}
  \bibinfo{person}{Joonhwan Lee}.} \bibinfo{year}{2022}\natexlab{}.
\newblock \showarticletitle{Designing and evaluating a chatbot for survivors of
  image-based sexual abuse}. In \bibinfo{booktitle}{\emph{Proceedings of the
  2022 CHI conference on human factors in computing systems}}.
  \bibinfo{pages}{1--21}.
\newblock


\bibitem[Marchenko(2017)]%
        {marchenko2017web}
\bibfield{author}{\bibinfo{person}{S Marchenko}.}
  \bibinfo{year}{2017}\natexlab{}.
\newblock \bibinfo{title}{Web of Darkness: Groomed, Manipulated, Coerced, and
  Abused In Minutes}.
\newblock


\bibitem[McBain(2025)]%
        {McBain}
\bibfield{author}{\bibinfo{person}{Ryan~K. McBain}.}
  \bibinfo{year}{2025}\natexlab{}.
\newblock
\urldef\tempurl%
\url{https://www.rand.org/pubs/commentary/2025/09/teens-are-using-chatbots-as-therapists-thats-alarming.html}
\showURL{%
\tempurl}


\bibitem[Michalopoulos et~al\mbox{.}(2014)]%
        {michalopoulos2014gars}
\bibfield{author}{\bibinfo{person}{Dimitrios Michalopoulos},
  \bibinfo{person}{Ioannis Mavridis}, {and} \bibinfo{person}{Marija Jankovic}.}
  \bibinfo{year}{2014}\natexlab{}.
\newblock \showarticletitle{GARS: Real-time system for identification,
  assessment and control of cyber grooming attacks}.
\newblock \bibinfo{journal}{\emph{Computers \& security}}  \bibinfo{volume}{42}
  (\bibinfo{year}{2014}), \bibinfo{pages}{177--190}.
\newblock


\bibitem[Milon-Flores and Cordeiro(2022)]%
        {milon2022take}
\bibfield{author}{\bibinfo{person}{Daniela~F Milon-Flores} {and}
  \bibinfo{person}{Robson~LF Cordeiro}.} \bibinfo{year}{2022}\natexlab{}.
\newblock \showarticletitle{How to take advantage of behavioral features for
  the early detection of grooming in online conversations}.
\newblock \bibinfo{journal}{\emph{Knowledge-Based Systems}}
  \bibinfo{volume}{240} (\bibinfo{year}{2022}), \bibinfo{pages}{108017}.
\newblock


\bibitem[Mitchell and Jones(2011)]%
        {mitchell2011youth}
\bibfield{author}{\bibinfo{person}{Kimberly~J Mitchell} {and}
  \bibinfo{person}{Lisa~M Jones}.} \bibinfo{year}{2011}\natexlab{}.
\newblock \showarticletitle{Youth Internet Safety Study (YISS): Methodology
  Report.}
\newblock  (\bibinfo{year}{2011}).
\newblock


\bibitem[Mladenovi{\'c} et~al\mbox{.}(2021)]%
        {mladenovic2021cyber}
\bibfield{author}{\bibinfo{person}{Miljana Mladenovi{\'c}},
  \bibinfo{person}{Vera O{\v{s}}mjanski}, {and}
  \bibinfo{person}{Sta{\v{s}}a~Vuji{\v{c}}i{\'c} Stankovi{\'c}}.}
  \bibinfo{year}{2021}\natexlab{}.
\newblock \showarticletitle{Cyber-aggression, cyberbullying, and
  cyber-grooming: A survey and research challenges}.
\newblock \bibinfo{journal}{\emph{ACM Computing Surveys (CSUR)}}
  \bibinfo{volume}{54}, \bibinfo{number}{1} (\bibinfo{year}{2021}),
  \bibinfo{pages}{1--42}.
\newblock


\bibitem[Mu{\~n}oz et~al\mbox{.}(2020)]%
        {munoz2020smartsec4cop}
\bibfield{author}{\bibinfo{person}{Fabi{\'a}n Mu{\~n}oz},
  \bibinfo{person}{Gustavo Isaza}, {and} \bibinfo{person}{Luis Castillo}.}
  \bibinfo{year}{2020}\natexlab{}.
\newblock \showarticletitle{Smartsec4cop: smart cyber-grooming detection using
  natural language processing and convolutional neural networks}. In
  \bibinfo{booktitle}{\emph{International Symposium on Distributed Computing
  and Artificial Intelligence}}. Springer, \bibinfo{pages}{11--20}.
\newblock


\bibitem[NSPSS(2025)]%
        {NSPSS}
\bibfield{author}{\bibinfo{person}{NSPSS}.} \bibinfo{year}{2025}\natexlab{}.
\newblock
\urldef\tempurl%
\url{https://www.nspcc.org.uk/what-is-child-abuse/types-of-abuse/grooming/}
\showURL{%
\tempurl}


\bibitem[Obajemu et~al\mbox{.}(2024)]%
        {obajemu2024towards}
\bibfield{author}{\bibinfo{person}{Oluwatomisin Obajemu},
  \bibinfo{person}{Zainab Agha}, \bibinfo{person}{Farzana~A Chowdhury}, {and}
  \bibinfo{person}{Pamela~J Wisniewski}.} \bibinfo{year}{2024}\natexlab{}.
\newblock \showarticletitle{Towards enforcing good digital citizenship:
  identifying opportunities for adolescent online safety nudges}.
\newblock \bibinfo{journal}{\emph{Proceedings of the ACM on Human-Computer
  Interaction}} \bibinfo{volume}{8}, \bibinfo{number}{CSCW1}
  (\bibinfo{year}{2024}), \bibinfo{pages}{1--37}.
\newblock


\bibitem[Oguine et~al\mbox{.}(2025a)]%
        {alluhidan2024bodyshaming}
\bibfield{author}{\bibinfo{person}{Ozioma~C. Oguine},
  \bibinfo{person}{Jinkyung~Katie Park}, \bibinfo{person}{Mamtaj Akter},
  \bibinfo{person}{Johanna Olesk}, \bibinfo{person}{Abdulmalik Alluhidan},
  \bibinfo{person}{Pamela Wisniewski}, {and} \bibinfo{person}{Karla
  Badillo-Urquiola}.} \bibinfo{year}{2025}\natexlab{a}.
\newblock \showarticletitle{How the Internet Facilitates Adverse Childhood
  Experiences for Youth Who Self-Identify as in Need of Services}.
\newblock \bibinfo{journal}{\emph{Proc. ACM Hum.-Comput. Interact.}}
  \bibinfo{volume}{9}, \bibinfo{number}{2}, Article
  \bibinfo{articleno}{CSCW097} (\bibinfo{date}{May} \bibinfo{year}{2025}),
  \bibinfo{numpages}{39}~pages.
\newblock
\href{https://doi.org/10.1145/3710995}{doi:\nolinkurl{10.1145/3710995}}


\bibitem[Oguine et~al\mbox{.}(2025b)]%
        {oguine2025CHINS}
\bibfield{author}{\bibinfo{person}{Ozioma~C. Oguine},
  \bibinfo{person}{Jinkyung~Katie Park}, \bibinfo{person}{Mamtaj Akter},
  \bibinfo{person}{Johanna Olesk}, \bibinfo{person}{Abdulmalik Alluhidan},
  \bibinfo{person}{Pamela Wisniewski}, {and} \bibinfo{person}{Karla
  Badillo-Urquiola}.} \bibinfo{year}{2025}\natexlab{b}.
\newblock \showarticletitle{How the Internet Facilitates Adverse Childhood
  Experiences for Youth Who Self-Identify as in Need of Services}.
\newblock \bibinfo{journal}{\emph{Proc. ACM Hum.-Comput. Interact.}}
  \bibinfo{volume}{9}, \bibinfo{number}{2}, Article
  \bibinfo{articleno}{CSCW097} (\bibinfo{date}{May} \bibinfo{year}{2025}),
  \bibinfo{numpages}{39}~pages.
\newblock
\href{https://doi.org/10.1145/3710995}{doi:\nolinkurl{10.1145/3710995}}


\bibitem[O’Connell(2003)]%
        {o2003typology}
\bibfield{author}{\bibinfo{person}{R O’Connell}.}
  \bibinfo{year}{2003}\natexlab{}.
\newblock \bibinfo{booktitle}{\emph{A typology of child cybersexploitation and
  online grooming practices}}.
\newblock \bibinfo{type}{{T}echnical {R}eport}. \bibinfo{institution}{Technical
  report}.
\newblock


\bibitem[Park(2025)]%
        {Park_2025}
\bibfield{author}{\bibinfo{person}{Eugenie Park}.}
  \bibinfo{year}{2025}\natexlab{}.
\newblock \bibinfo{title}{Teens and internet, device access fact sheet}.
\newblock
\urldef\tempurl%
\url{https://www.pewresearch.org/internet/fact-sheet/teens-and-internet-device-access-fact-sheet/#:~:text=PEW%20RESEARCH%20CENTER-,Teens%20and%20digital%20devices,PEW%20RESEARCH%20CENTER}
\showURL{%
\tempurl}


\bibitem[Park et~al\mbox{.}(2025)]%
        {park2025teens}
\bibfield{author}{\bibinfo{person}{Jinkyung~Katie Park},
  \bibinfo{person}{Renkai Ma}, \bibinfo{person}{Naima~Samreen Ali},
  \bibinfo{person}{Naulsberry~Jean Baptiste}, \bibinfo{person}{Zainab Agha},
  {and} \bibinfo{person}{Pamela~J Wisniewski}.}
  \bibinfo{year}{2025}\natexlab{}.
\newblock \showarticletitle{Teens, Privacy, and Algorithms: Navigating and
  Co-Designing Solutions for Interpersonal Boundary Management on Social
  Media}.
\newblock In \bibinfo{booktitle}{\emph{Proceedings of the 24th Interaction
  Design and Children}}. \bibinfo{pages}{589--607}.
\newblock


\bibitem[Piccolo et~al\mbox{.}(2021)]%
        {piccolo2021chatbots}
\bibfield{author}{\bibinfo{person}{Lara Schibelsky~Godoy Piccolo},
  \bibinfo{person}{Pinelopi Troullinou}, {and} \bibinfo{person}{Harith Alani}.}
  \bibinfo{year}{2021}\natexlab{}.
\newblock \showarticletitle{Chatbots to support children in coping with online
  threats: Socio-technical requirements}. In
  \bibinfo{booktitle}{\emph{Proceedings of the 2021 ACM designing interactive
  systems conference}}. \bibinfo{pages}{1504--1517}.
\newblock


\bibitem[Prolific(2025)]%
        {Prolific}
\bibfield{author}{\bibinfo{person}{Prolific}.} \bibinfo{year}{2025}\natexlab{}.
\newblock \bibinfo{title}{Prolific {\textbar} {Easily} collect high-quality
  data from real people}.
\newblock
\urldef\tempurl%
\url{https://www.prolific.com}
\showURL{%
\tempurl}


\bibitem[Prosser and Edwards(2024)]%
        {prosser2024helpful}
\bibfield{author}{\bibinfo{person}{Ellie Prosser} {and}
  \bibinfo{person}{Matthew Edwards}.} \bibinfo{year}{2024}\natexlab{}.
\newblock \showarticletitle{Helpful or harmful? Exploring the efficacy of large
  language models for online grooming prevention}. In
  \bibinfo{booktitle}{\emph{Proceedings of the 2024 European Interdisciplinary
  Cybersecurity Conference}}. \bibinfo{pages}{1--10}.
\newblock


\bibitem[QuestionPro(2025)]%
        {QuestionPro}
\bibfield{author}{\bibinfo{person}{QuestionPro}.}
  \bibinfo{year}{2025}\natexlab{}.
\newblock
\urldef\tempurl%
\url{https://www.questionpro.com/us/}
\showURL{%
\tempurl}


\bibitem[Razi et~al\mbox{.}(2023)]%
        {razi2023sliding}
\bibfield{author}{\bibinfo{person}{Afsaneh Razi}, \bibinfo{person}{Ashwaq
  AlSoubai}, \bibinfo{person}{Seunghyun Kim}, \bibinfo{person}{Shiza Ali},
  \bibinfo{person}{Gianluca Stringhini}, \bibinfo{person}{Munmun De~Choudhury},
  {and} \bibinfo{person}{Pamela~J Wisniewski}.}
  \bibinfo{year}{2023}\natexlab{}.
\newblock \showarticletitle{Sliding into My DMs: Detecting Uncomfortable or
  Unsafe Sexual Risk Experiences within Instagram Direct Messages Grounded in
  the Perspective of Youth}.
\newblock \bibinfo{journal}{\emph{Proceedings of the ACM on Human-Computer
  Interaction}} \bibinfo{volume}{7}, \bibinfo{number}{CSCW1}
  (\bibinfo{year}{2023}), \bibinfo{pages}{1--29}.
\newblock


\bibitem[Razi et~al\mbox{.}(2020)]%
        {razi2020let}
\bibfield{author}{\bibinfo{person}{Afsaneh Razi}, \bibinfo{person}{Karla
  Badillo-Urquiola}, {and} \bibinfo{person}{Pamela~J Wisniewski}.}
  \bibinfo{year}{2020}\natexlab{}.
\newblock \showarticletitle{Let's talk about sext: How adolescents seek support
  and advice about their online sexual experiences}. In
  \bibinfo{booktitle}{\emph{Proceedings of the 2020 CHI Conference on Human
  Factors in Computing Systems}}. \bibinfo{pages}{1--13}.
\newblock


\bibitem[Razi et~al\mbox{.}(2021)]%
        {razi2021human}
\bibfield{author}{\bibinfo{person}{Afsaneh Razi}, \bibinfo{person}{Seunghyun
  Kim}, \bibinfo{person}{Ashwaq Alsoubai}, \bibinfo{person}{Gianluca
  Stringhini}, \bibinfo{person}{Thamar Solorio}, \bibinfo{person}{Munmun
  De~Choudhury}, {and} \bibinfo{person}{Pamela~J Wisniewski}.}
  \bibinfo{year}{2021}\natexlab{}.
\newblock \showarticletitle{A human-centered systematic literature review of
  the computational approaches for online sexual risk detection}.
\newblock \bibinfo{journal}{\emph{Proceedings of the ACM on human-computer
  interaction}} \bibinfo{volume}{5}, \bibinfo{number}{CSCW2}
  (\bibinfo{year}{2021}), \bibinfo{pages}{1--38}.
\newblock


\bibitem[Rezaee~Borj et~al\mbox{.}(2023)]%
        {rezaee2023detecting}
\bibfield{author}{\bibinfo{person}{Parisa Rezaee~Borj}, \bibinfo{person}{Kiran
  Raja}, {and} \bibinfo{person}{Patrick Bours}.}
  \bibinfo{year}{2023}\natexlab{}.
\newblock \showarticletitle{Detecting online grooming by simple contrastive
  chat embeddings}. In \bibinfo{booktitle}{\emph{Proceedings of the 9th ACM
  international workshop on security and privacy analytics}}.
  \bibinfo{pages}{57--65}.
\newblock


\bibitem[Ringenberg et~al\mbox{.}(2024)]%
        {ringenberg2024assessing}
\bibfield{author}{\bibinfo{person}{Tatiana~R Ringenberg},
  \bibinfo{person}{Kathryn Seigfried-Spellar}, {and} \bibinfo{person}{Julia
  Rayz}.} \bibinfo{year}{2024}\natexlab{}.
\newblock \showarticletitle{Assessing differences in grooming stages and
  strategies in decoy, victim, and law enforcement conversations}.
\newblock \bibinfo{journal}{\emph{Computers in Human Behavior}}
  \bibinfo{volume}{152} (\bibinfo{year}{2024}), \bibinfo{pages}{108071}.
\newblock


\bibitem[Rita and Shava(2021)]%
        {rita2021chatbot}
\bibfield{author}{\bibinfo{person}{Marsela~Nur Rita} {and}
  \bibinfo{person}{Fungai~Bhunu Shava}.} \bibinfo{year}{2021}\natexlab{}.
\newblock \showarticletitle{Chatbot driven web-based platform for online safety
  and sexual exploitation awareness and reporting in Namibia}. In
  \bibinfo{booktitle}{\emph{2021 International Conference on Artificial
  Intelligence, Big Data, Computing and Data Communication Systems (icABCD)}}.
  IEEE, \bibinfo{pages}{1--5}.
\newblock


\bibitem[Saha and De~Choudhury(2017)]%
        {sahaModelingStressSocial2017}
\bibfield{author}{\bibinfo{person}{Koustuv Saha} {and} \bibinfo{person}{Munmun
  De~Choudhury}.} \bibinfo{year}{2017}\natexlab{}.
\newblock \showarticletitle{Modeling {{Stress}} with {{Social Media Around
  Incidents}} of {{Gun Violence}} on {{College Campuses}}}.
\newblock \bibinfo{journal}{\emph{Proceedings of the ACM on Human-Computer
  Interaction}} \bibinfo{volume}{1}, \bibinfo{number}{CSCW}
  (\bibinfo{date}{Dec.} \bibinfo{year}{2017}), \bibinfo{pages}{1--27}.
\newblock
\showISSN{2573-0142}
\href{https://doi.org/10.1145/3134727}{doi:\nolinkurl{10.1145/3134727}}


\bibitem[Schittenhelm et~al\mbox{.}(2024)]%
        {schittenhelm2024cybergrooming}
\bibfield{author}{\bibinfo{person}{Catherine Schittenhelm},
  \bibinfo{person}{Maxime Kops}, \bibinfo{person}{Maeve Moosburner},
  \bibinfo{person}{Saskia~M Fischer}, {and} \bibinfo{person}{Sebastian Wachs}.}
  \bibinfo{year}{2024}\natexlab{}.
\newblock \showarticletitle{Cybergrooming victimization among young people: A
  systematic review of prevalence rates, risk factors, and outcomes}.
\newblock \bibinfo{journal}{\emph{Adolescent Research Review}}
  (\bibinfo{year}{2024}), \bibinfo{pages}{1--32}.
\newblock


\bibitem[Thorn(2024a)]%
        {Thorn_2024}
\bibfield{author}{\bibinfo{person}{Thorn}.} \bibinfo{year}{2024}\natexlab{a}.
\newblock \bibinfo{title}{New research from Thorn: 22\% of minors report having
  online sexual interactions with adults}.
\newblock
\urldef\tempurl%
\url{https://www.thorn.org/blog/new-research-from-thorn-22-of-minors-report-having-online-sexual-interactions-with-adults/}
\showURL{%
\tempurl}


\bibitem[Thorn(2024b)]%
        {Thorn_2024b}
\bibfield{author}{\bibinfo{person}{Thorn}.} \bibinfo{year}{2024}\natexlab{b}.
\newblock \bibinfo{title}{Online grooming: What it is, how it happens, and how
  to defend children}.
\newblock
\urldef\tempurl%
\url{https://www.thorn.org/blog/online-grooming-what-it-is-how-it-happens-and-how-to-defend-children/}
\showURL{%
\tempurl}


\bibitem[Ueda et~al\mbox{.}(2021)]%
        {ueda2021cyberbullying}
\bibfield{author}{\bibinfo{person}{Tomoyuki Ueda}, \bibinfo{person}{Junya
  Nakanishi}, \bibinfo{person}{Itaru Kuramoto}, \bibinfo{person}{Jun Baba},
  \bibinfo{person}{Yuichiro Yoshikawa}, {and} \bibinfo{person}{Hiroshi
  Ishiguro}.} \bibinfo{year}{2021}\natexlab{}.
\newblock \showarticletitle{Cyberbullying mitigation by a proxy persuasion of a
  chat member hijacked by a chatbot}. In \bibinfo{booktitle}{\emph{Proceedings
  of the 9th international conference on human-agent interaction}}.
  \bibinfo{pages}{202--208}.
\newblock


\bibitem[Vamplew et~al\mbox{.}(2018)]%
        {vamplew2018human}
\bibfield{author}{\bibinfo{person}{Peter Vamplew}, \bibinfo{person}{Richard
  Dazeley}, \bibinfo{person}{Cameron Foale}, \bibinfo{person}{Sally Firmin},
  {and} \bibinfo{person}{Jane Mummery}.} \bibinfo{year}{2018}\natexlab{}.
\newblock \showarticletitle{Human-aligned artificial intelligence is a
  multiobjective problem}.
\newblock \bibinfo{journal}{\emph{Ethics and information technology}}
  \bibinfo{volume}{20}, \bibinfo{number}{1} (\bibinfo{year}{2018}),
  \bibinfo{pages}{27--40}.
\newblock


\bibitem[Wang et~al\mbox{.}(2021)]%
        {wang2021seri}
\bibfield{author}{\bibinfo{person}{Pei Wang}, \bibinfo{person}{Zhen Guo},
  \bibinfo{person}{Lifu Huang}, {and} \bibinfo{person}{Jin-Hee Cho}.}
  \bibinfo{year}{2021}\natexlab{}.
\newblock \showarticletitle{Seri: Generative chatbot framework for
  cybergrooming prevention}.
\newblock  (\bibinfo{year}{2021}).
\newblock


\bibitem[Whittle et~al\mbox{.}(2013)]%
        {whittle2013review}
\bibfield{author}{\bibinfo{person}{Helen Whittle}, \bibinfo{person}{Catherine
  Hamilton-Giachritsis}, \bibinfo{person}{Anthony Beech}, {and}
  \bibinfo{person}{Guy Collings}.} \bibinfo{year}{2013}\natexlab{}.
\newblock \showarticletitle{A review of young people's vulnerabilities to
  online grooming}.
\newblock \bibinfo{journal}{\emph{Aggression and violent behavior}}
  \bibinfo{volume}{18}, \bibinfo{number}{1} (\bibinfo{year}{2013}),
  \bibinfo{pages}{135--146}.
\newblock


\bibitem[Whittle et~al\mbox{.}(2014)]%
        {whittle2014their}
\bibfield{author}{\bibinfo{person}{Helen~C Whittle},
  \bibinfo{person}{Catherine~E Hamilton-Giachritsis}, {and}
  \bibinfo{person}{Anthony~R Beech}.} \bibinfo{year}{2014}\natexlab{}.
\newblock \showarticletitle{In their own words: Young peoples’
  vulnerabilities to being groomed and sexually abused online}.
\newblock \bibinfo{journal}{\emph{Health Psychology}} \bibinfo{volume}{5},
  \bibinfo{number}{10} (\bibinfo{year}{2014}), \bibinfo{pages}{1185--1196}.
\newblock


\bibitem[Williams et~al\mbox{.}(2023)]%
        {williams2023youth}
\bibfield{author}{\bibinfo{person}{Olivia Williams}, \bibinfo{person}{Yee-Yin
  Choong}, {and} \bibinfo{person}{Kerrianne Buchanan}.}
  \bibinfo{year}{2023}\natexlab{}.
\newblock \showarticletitle{Youth understandings of online privacy and
  security: A dyadic study of children and their parents}. In
  \bibinfo{booktitle}{\emph{Nineteenth Symposium on Usable Privacy and Security
  (SOUPS 2023)}}. \bibinfo{pages}{399--416}.
\newblock


\bibitem[Wisniewski et~al\mbox{.}(2017a)]%
        {wisniewski_parental_2017}
\bibfield{author}{\bibinfo{person}{Pamela Wisniewski},
  \bibinfo{person}{Arup~Kumar Ghosh}, \bibinfo{person}{Heng Xu},
  \bibinfo{person}{Mary~Beth Rosson}, {and} \bibinfo{person}{John~M. Carroll}.}
  \bibinfo{year}{2017}\natexlab{a}.
\newblock \showarticletitle{Parental {Control} vs. {Teen} {Self}-{Regulation}:
  {Is} there a middle ground for mobile online safety?}. In
  \bibinfo{booktitle}{\emph{Proceedings of the 2017 {ACM} {Conference} on
  {Computer} {Supported} {Cooperative} {Work} and {Social} {Computing}}}.
  \bibinfo{publisher}{ACM}, \bibinfo{address}{Portland Oregon USA},
  \bibinfo{pages}{51--69}.
\newblock
\showISBNx{978-1-4503-4335-0}
\href{https://doi.org/10.1145/2998181.2998352}{doi:\nolinkurl{10.1145/2998181.2998352}}


\bibitem[Wisniewski et~al\mbox{.}(2017b)]%
        {wisniewski2017parents}
\bibfield{author}{\bibinfo{person}{Pamela Wisniewski}, \bibinfo{person}{Heng
  Xu}, \bibinfo{person}{Mary~Beth Rosson}, {and} \bibinfo{person}{John~M
  Carroll}.} \bibinfo{year}{2017}\natexlab{b}.
\newblock \showarticletitle{Parents just don't understand: Why teens don't talk
  to parents about their online risk experiences}. In
  \bibinfo{booktitle}{\emph{Proceedings of the 2017 ACM conference on computer
  supported cooperative work and social computing}}. \bibinfo{pages}{523--540}.
\newblock


\bibitem[Wolak et~al\mbox{.}(2008)]%
        {wolak_online_2008}
\bibfield{author}{\bibinfo{person}{Janis Wolak}, \bibinfo{person}{David
  Finkelhor}, \bibinfo{person}{Kimberly~J. Mitchell}, {and}
  \bibinfo{person}{Michele~L. Ybarra}.} \bibinfo{year}{2008}\natexlab{}.
\newblock \showarticletitle{Online "predators" and their victims: myths,
  realities, and implications for prevention and treatment}.
\newblock \bibinfo{journal}{\emph{The American Psychologist}}
  \bibinfo{volume}{63}, \bibinfo{number}{2} (\bibinfo{year}{2008}),
  \bibinfo{pages}{111--128}.
\newblock
\showISSN{0003-066X}
\href{https://doi.org/10.1037/0003-066X.63.2.111}{doi:\nolinkurl{10.1037/0003-066X.63.2.111}}


\bibitem[Yao et~al\mbox{.}(2017)]%
        {yao2017folk}
\bibfield{author}{\bibinfo{person}{Yaxing Yao}, \bibinfo{person}{Davide Lo~Re},
  {and} \bibinfo{person}{Yang Wang}.} \bibinfo{year}{2017}\natexlab{}.
\newblock \showarticletitle{Folk models of online behavioral advertising}. In
  \bibinfo{booktitle}{\emph{Proceedings of the 2017 ACM Conference on Computer
  Supported Cooperative Work and Social Computing}}.
  \bibinfo{pages}{1957--1969}.
\newblock


\end{thebibliography}

\appendix


\section{Appendix}

\subsection{Screening Survey Questions}
\label{screening}
\small
\begin{enumerate}
    \item Are both the parent/legal guardian and teen fluent in English?
    \begin{enumerate}
        \item[(a)] Yes
        \item[(b)] No
    \end{enumerate}

    \item Are both the parent/legal guardian and teen currently living in the United States?
    \begin{enumerate}
        \item[(a)] Yes
        \item[(b)] No
    \end{enumerate}

    \item What state are you currently living in?

    \item Does the teen have their own smartphone?
    \begin{enumerate}
        \item[(a)] Yes
        \item[(b)] No
    \end{enumerate}

    \item Does the teen actively use social media?
    \begin{enumerate}
        \item[(a)] Yes
        \item[(b)] No
    \end{enumerate}

    \item Are both the parent/legal guardian and teen willing to join a short (5-minute) Zoom meeting for the consent process with the research team before the study?
    \begin{enumerate}
        \item[(a)] Yes
        \item[(b)] No
    \end{enumerate}

    \item Content Warning: The image below provides an example of the scenarios included in the survey that we will ask parents and teens to complete. We share it here to give you an idea of the types of questions we will be asking. All scenarios will be similar in intensity or milder than the one provided. Do both the parent/legal guardian and teen understand that participation in this study will involve responding to fictitious cybergrooming scenarios, like the one shown below?
        \begin{figure*}[t]
        \centering
        \includegraphics[scale=0.32]{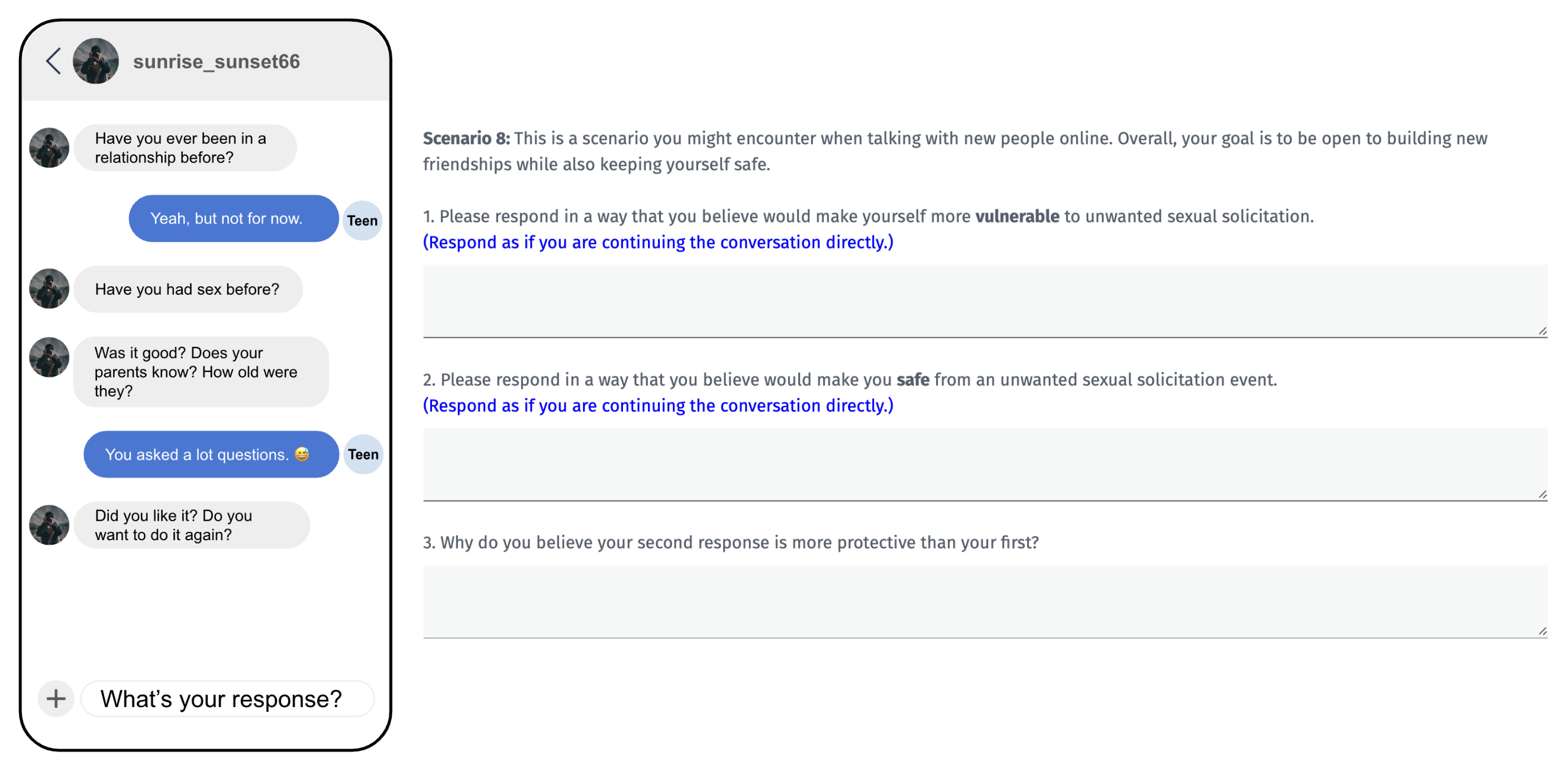}
        \label{example}
        \Description{This figure is included in the content warning question as an example to give participants a sense of the study materials. On the left side, the conversation unfolds in a smartphone-style chat window. The predator's messages are in gray speech bubbles, and the teen's messages are in blue ones. The exchange unfolds as follows: Predator: ``Have you ever been in a relationship before?'' Teen: ``Yeah, but not for now.'' Predator: ``Have you had sex before?'' Predator: ``Was it good? Do your parents know? How old were they?'' Teen: ``You asked a lot of questions.'' Predator: ``Did you like it? Do you want to do it again?'' At the bottom of the chat window, a text entry field labeled ``What's your response?'' prompts the participant to continue the conversation. On the right side, the survey instruction shows that scenario 8: below is a scenario you might encounter when talking with new people online. Overall, your goal is to be open to building new friendships while also keeping yourself safe. Then, participants need to answer three questions: (1) Please respond in a way that you believe would make yourself more vulnerable to unwanted sexual solicitation (respond as if you are continuing the conversation directly), (2) Please respond in a way that you believe would make you safe from an unwanted sexual solicitation event, and (3) Why do you believe your second response is more protective than your first?}
        \caption{Scenario 8 and the questions.}
        \end{figure*}
    
\end{enumerate}

\subsection{Parents' Pre-Survey Questions}
\label{parent_pre}
\begin{enumerate}
    \item What is your age?
    
    \item How do you identify your gender?
    \begin{enumerate}
        \item[(a)] Male
        \item[(b)] Female
        \item[(c)] Prefer to self-identify (Please click this option to specify)
        \item[(d)] Prefer not to answer
    \end{enumerate}

    \item How do you consider yourself to be?
    \begin{enumerate}
        \item[(a)] Heterosexual or Straight
        \item[(b)] Homosexual
        \item[(c)] Bisexual
        \item[(d)] Prefer to self-identify (Please click this option to specify)
        \item[(e)] Prefer not to answer 
    \end{enumerate}

    \item Which of the following best describes your race?
    \begin{enumerate}
        \item[(a)] Black or African American
        \item[(b)] White or European Descent
        \item[(c)] Hispanic or Latinx
        \item[(d)] East Asian (e.g., Chinese, Japanese, Korean)
        \item[(e)] South Asian (e.g., Indian, Pakistani)
        \item[(f)] Southeast Asian (e.g., Vietnamese, Filipino)
        \item[(g)] Native Hawaiian or Other Pacific Islander
        \item[(h)] Middle Eastern or North African (MENA)
        \item[(i)] Indigenous or Native (e.g., Native American, First Nations, Aboriginal)
        \item[(j)] Prefer to self-identify (Please click this option to specify)
        \item[(k)] Prefer not to answer
    \end{enumerate}

    \item Which of the following best describes the highest level of education you have completed?
    \begin{enumerate}
        \item[(a)] Didn’t Finish High School
        \item[(b)] High school diploma or equivalent (e.g., GED)
        \item[(c)] 2 Years of College or less
        \item[(d)] College graduate (4 or 5-year program)
        \item[(e)] Master’s degree (or other post-graduate training)
        \item[(f)] Doctoral degree (PhD., MD, EdD, DVM, DDS, JD, etc)
        \item[(g)] Prefer not to answer
    \end{enumerate}

    \item What is your marital status?
    \begin{enumerate}
        \item[(a)] Married
        \item[(b)] Widowed
        \item[(c)] Divorced
        \item[(d)] Separated
        \item[(e)] Unmarried
        \item[(f)] Prefer not to answer
    \end{enumerate}

    \item Which category best describes your yearly household income before taxes? 
    \begin{enumerate}
        \item[(a)] Less than \$9,999
        \item[(b)] \$10,000-\$19,999
        \item[(c)] \$20,000-\$59,999
        \item[(d)] \$100,000-\$149,999
        \item[(e)] \$150,000-\$199,999
        \item[(f)] Above \$200,000
        \item[(g)] Prefer not to answer
    \end{enumerate}

    \item How much time do you spend on social media every day?
    \begin{enumerate}
        \item[(a)] Less than 1 hour
        \item[(b)] 1-2 hours
        \item[(c)] 2-3 hours
        \item[(d)] 3-4 hours
        \item[(e)] More than 4 hours
    \end{enumerate}

    \edit{
    \item These questions explore your awareness of how often your teen has encountered unwanted sexual solicitations on social media over the past year. Unwanted sexual solicitations are defined as requests to engage in sexual activities or sexual talk or to give personal sexual information that was unwanted or made by a person 5 or more years older, whether wanted or not. Please indicate the frequency of the following situations based on your knowledge:
    [Scale: I don't know, Not at all, One to a few times a year, A few times a month, A few times a week, Almost every day]
    \begin{enumerate}
        \item[(a)] Someone tried to get your teen to talk on social media about sex when they did not want to
        \item[(b)] Someone on social media asked your teen for sexual information about themselves when they did not want to answer such questions. (Very personal questions, like what your teen's body looks like or sexual things your teen has done)
        \item[(c)] Someone on social media asked your teen to do something sexual that they did not want to do
    \end{enumerate}}

\end{enumerate}

\subsection{Teens' Pre-Survey Questions}
\label{teen_pre}
\begin{enumerate}
    \item How often do you use social media?
    \begin{enumerate}
        \item[(a)] Several times an hour
        \item[(b)] Several times a day
        \item[(c)] Every day or almost every day
        \item[(d)] Once or twice a week
        \item[(e)] Not at all
    \end{enumerate}

    \item How much time do you spend on social media every day?
    \begin{enumerate}
        \item[(a)] Less than 1 hour
        \item[(b)] 1-2 hours
        \item[(c)] 2-3 hours
        \item[(d)] 3-4 hours
        \item[(e)] More than 4 hours
    \end{enumerate}

    \item What is your age? 
    \begin{enumerate}
        \item[(a)] 12 or younger
        \item[(b)] 13
        \item[(c)] 14
        \item[(d)] 15
        \item[(e)] 16
        \item[(f)] 17
        \item[(g)] 18 or older
    \end{enumerate}

    \item How do you identify your gender?
    \begin{enumerate}
        \item[(a)] Male
        \item[(b)] Female
        \item[(c)] Prefer to self-identify (Please click this option to specify)
        \item[(d)] Prefer not to answer
    \end{enumerate}

    \item How do you consider yourself to be?
    \begin{enumerate}
        \item[(a)] Heterosexual or Straight
        \item[(b)] Homosexual
        \item[(c)] Bisexual
        \item[(d)] Prefer to self-identify (Please click this option to specify)
        \item[(e)] Prefer not to answer 
    \end{enumerate}

    \item Which of the following best describes your race?
    \begin{enumerate}
        \item[(a)] Black or African American
        \item[(b)] White or European Descent
        \item[(c)] Hispanic or Latinx
        \item[(d)] East Asian (e.g., Chinese, Japanese, Korean)
        \item[(e)] South Asian (e.g., Indian, Pakistani)
        \item[(f)] Southeast Asian (e.g., Vietnamese, Filipino)
        \item[(g)] Native Hawaiian or Other Pacific Islander
        \item[(h)] Middle Eastern or North African (MENA)
        \item[(i)] Indigenous or Native (e.g., Native American, First Nations, Aboriginal)
        \item[(j)] Prefer to self-identify (Please click this option to specify)
        \item[(k)] Prefer not to answer
    \end{enumerate}

   \edit{
   \item These questions explore how often you encountered unwanted sexual solicitations on social media over the past year. Unwanted sexual solicitations are defined as requests to engage in sexual activities or sexual talk or to give personal sexual information that was unwanted or made by a person 5 or more years older, whether wanted or not. Please indicate the frequency of the following situations based on your experiences: [Scale: I don't know, Not at all, One to a few times a year, A few times a month, A few times a week, Almost every day]
   \begin{enumerate}
        \item[(a)] Someone tried to get me to talk on social media about sex when I did not want to
        \item[(b)] Someone on social media asked me for sexual information about myself when I did not want to answer such questions. (Very personal questions, like what your body looks like or sexual things you have done)
        \item[(c)] Someone on social media asked me to do something sexual that I did not want to do
    \end{enumerate}}
    
\end{enumerate}

\subsection{Main Survey Questions for Parents and Teens}
\label{scenarios}
In the main survey section, we provided 10 identical scenarios for both parents and teens. For each scenario, we asked three questions.

\begin{itemize}
    \item Parents' Questions:
    \begin{enumerate}
        \item[(1)] Please respond in a way that you believe would make you (as your teen) more vulnerable to an online unwanted sexual solicitation event. (Respond as if you are continuing the conversation directly.)
        \item[(2)] Please respond in a way that you believe would make you (as your teen) safe from an unwanted sexual solicitation event. (Respond as if you are continuing the conversation directly.)
        \item[(3)] Why do you believe your second response is more protective than your first? 
    \end{enumerate}

    \item Teens' Questions:
    \begin{enumerate}
        \item[(1)] Please respond in a way that you believe would make yourself more vulnerable to unwanted sexual solicitation. (Respond as if you are continuing the conversation directly.)
        \item[(2)] Please respond in a way that you believe would make you safe from an unwanted sexual solicitation event. (Respond as if you are continuing the conversation directly.)
        \item[(3)] Why do you believe your second response is more protective than your first?
    \end{enumerate}

    \edit{
    \item Exit Questions:
    \begin{enumerate}
        \item[(1)] This question aims to understand your overall perspectives about the scenarios displayed previously. Do you believe these scenarios presented are realistic?
        \item[(2)] Please explain your answer to the previous question.
    \end{enumerate}}

\end{itemize}

\begin{figure*}[]
        \centering
        \includegraphics[scale=0.39]{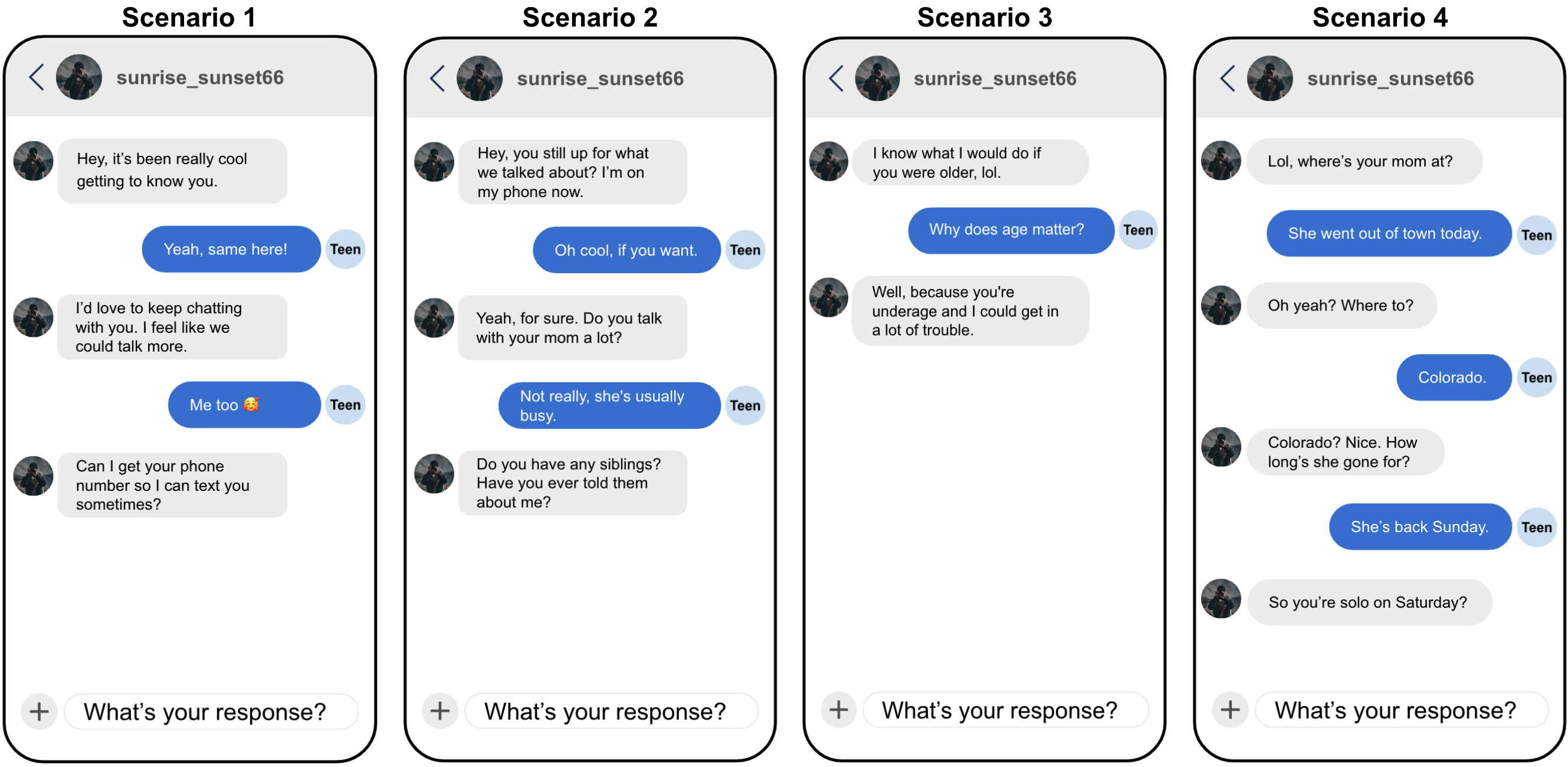}
        \caption{Scenarios used in our study.}
        \label{S1234}
        \Description{This figure contains four scenarios (S1 to S4) presented to participants. These conversations unfold in a smartphone-style chat window. The predator’s messages are in gray speech bubbles, and the teen’s messages are in blue ones. S1 as follows: Predator: ``Hey, it’s been really cool getting to know you.'' Teen: ``Yeah, same here!'' Predator: ``I'd love to keep chatting with you. I feel like we could talk more.'' Teen: ``Me too '' Predator: ``Can I get your phone number so I can text you sometimes?'' At the bottom of the chat window, a text entry field labeled ``What's your response?''
        S2 as follows: Predator: ``Hey, you still up for what we talked about? I'm on my phone now.'' Teen: ``Oh cool, if you want.'' Predator: ``Yeah, for sure. Do you talk with your mom a lot?'' Teen: ``Not really, she's usually busy.'' Predator: ``Do you have any siblings? Have you ever told them about me?'' At the bottom of the chat window, a text entry field labeled ``What's your response?''
        S3 as follows: Predator: ``I know what I would do if you were older, lol.'' Teen: ``Why does age matter?'' Predator: ``Well, because you're underage and I could get in a lot of trouble.'' At the bottom of the chat window, a text entry field labeled ``What's your response?''
        S4 as follows: Predator: ``Lol, where's your mom at?'' Teen: ``She went out of town today.'' Predator: ``Oh yeah? Where to?'' Teen: ``Colorado.'' Predator: ``Colorado? Nice. How long's she gone for?'' Teen: ``She's back Sunday.'' Predator: ``So you're solo on Saturday?'' At the bottom of the chat window, a text entry field labeled ``What's your response?''}
\end{figure*}

\begin{figure*}[]
        \centering
        \includegraphics[scale=0.43]{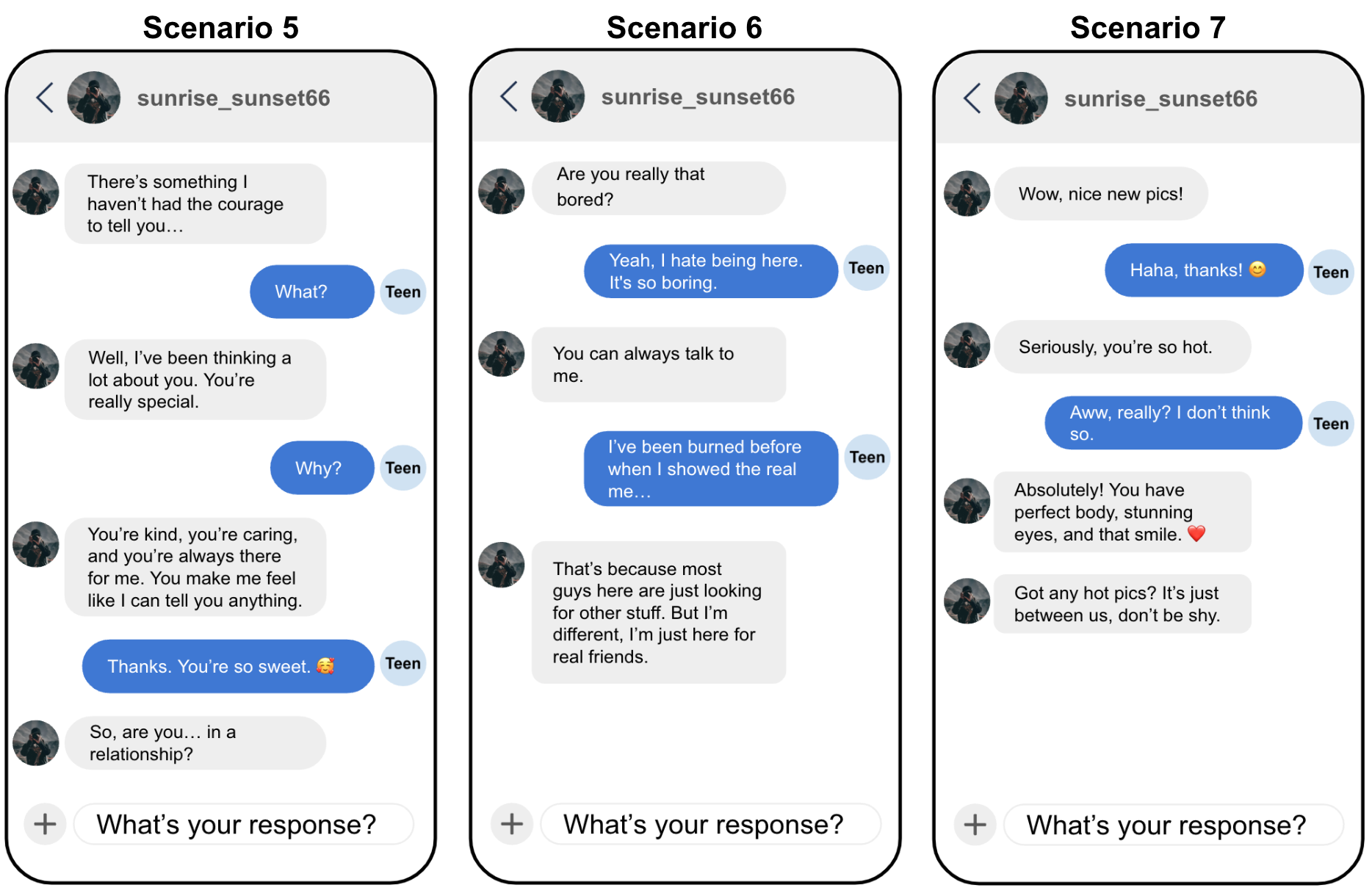}
        \caption{Scenarios used in our study.}
        \label{S567}
        \Description{This figure contains three scenarios (S5 to S7) presented to participants. S5 as follows: Predator: ``There's something I haven't had the courage to tell you...'' Teen: ``What?'' Predator: ``Well, I’ve been thinking a lot about you. You're really special.'' Teen: ``Why?'' Predator: ``You're kind, you're caring, and you're always there for me. You make me feel like I can tell you anything.'' Teen: ``Thanks. You're so sweet.'' Predator: ``So, are you...in a relationship?'' At the bottom of the chat window, a text entry field labeled ``What's your response?'' At the bottom of the chat window, a text entry field labeled ``What’s your response?''
        S6 as follows: Predator: ``Are you really that bored?'' Teen: ``Yeah, I hate being here. It's so boring.'' Predator: ``You can always talk to me.'' Teen: ``I've been burned before when I showed the real me...'' Predator: ``That's because most guys here are just looking for other stuff. But I'm different, I'm just here for real friends.'' At the bottom of the chat window, a text entry field labeled ``What's your response?''
        S7 as follows: Predator: ``Wow, nice new pics!'' Teen: ``Haha, thanks! '' Predator: ``Seriously, you're so hot.'' Teen: ``Aww, really? I don't think so.'' Predator: ``Absolutely! You have a perfect body, stunning eyes, and that smile.'' Predator: ``Got any hot pics? It's just between us, don't be shy.'' At the bottom of the chat window, a text entry field labeled ``What's your response?'' }
\end{figure*}

\begin{figure*}[]
        \centering
        \includegraphics[scale=0.44]{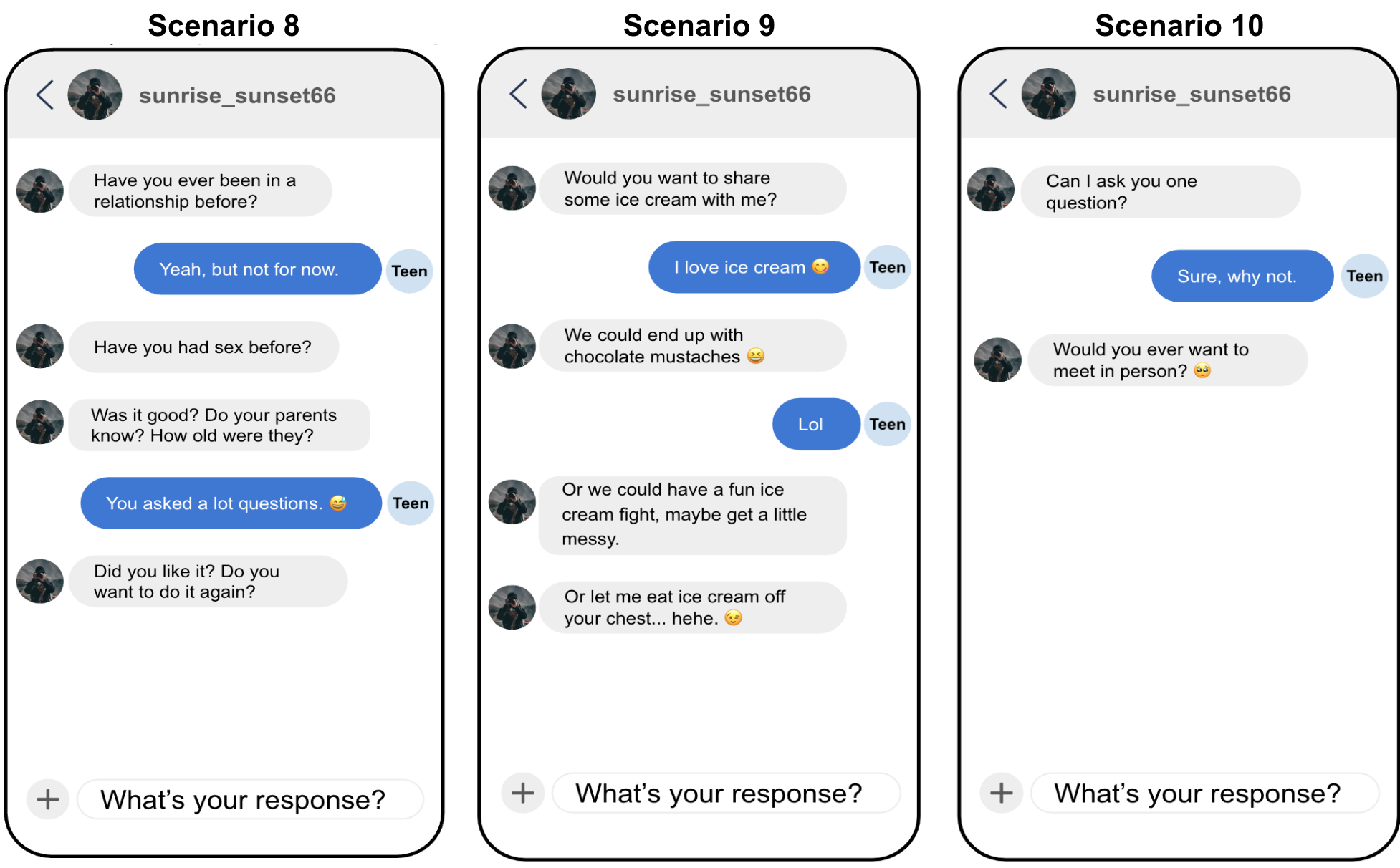}
        \caption{Scenarios used in our study.}
        \label{S8910}
        \Description{This figure contains three scenarios (S8 to S10) presented to participants. S8 as follows: Predator: ``Have you ever been in a relationship before?'' Teen: ``Yeah, but not for now.'' Predator: ``Have you had sex before?'' Predator: ``Was it good? Do your parents know? How old were they?'' Teen: ``You asked a lot of questions.'' Predator: ``Did you like it? Do you want to do it again?'' At the bottom of the chat window, a text entry field labeled ``What's your response?''
        S9 as follows: Predator: ``Would you want to share some ice cream with me?'' Teen: ``I love ice cream.'' Predator: ``We could end up with chocolate mustaches.'' Teen: ``Lol'' Predator: ``Or we could have a fun ice cream fight, maybe get a little messy.'' Predator: ``Or let me eat ice cream off your chest... hehe.'' At the bottom of the chat window, a text entry field labeled ``What's your response?''
        S10 as follows: Predator: ``Can I ask you one question?'' Teen: ``Sure, why not.'' Predator: ``Would you ever want to meet in person?'' At the bottom of the chat window, a text entry field labeled ``What's your response?''}
\end{figure*}

\subsection{Participants' Demographic Information}
\begin{table*}[h]
    \centering
    \caption{Summary of Participants' Demographics}
    \small
    \begin{tabular}{p{4cm} p{6.2cm} p{2cm} p{2cm}} \toprule
    \multirow{2}{*}{\textbf{Category}} &
    \multirow{2}{*}{\textbf{Option}} &
    \multicolumn{2}{c}{\textbf{N (\%)}} \\
        \cmidrule(lr){3-4} & & \textbf{Parent (N = 51)} & \textbf{Teen (N = 23)} \\ \midrule 
    Age & Mean (SD) & 40 (5.81) & 15 (1.45) \\
     \midrule
    Gender 
     & Female & 36 (71\%) & 16 (70\%) \\
     & Male & 15 (29\%) & 6 (26\%) \\\midrule
    Sexual Orientation 
     & Heterosexual or Straight & 49 (96\%) & 16 (70\%) \\
     & Homosexual & 1 (2\%) & 1 (4\%) \\
     & Bisexual & 1 (2\%) & 3 (13\%) \\
     & Pansexual & 0 (0\%) & 1 (4\%) \\
     & Prefer not to answer & 0 (0\%) & 2 (9\%)\\\midrule
    Race 
     & White or European Descent & 40 (78\%) & 11 (48\%) \\
     & Black or African American & 9 (18\%) & 6 (26\%) \\
     & East Asian (e.g., Chinese, Japanese, Korean) & 0 (0\%) & 6 (26\%) \\
     & Hispanic or Latinx & 3 (6\%) & 4 (17\%) \\
     & Prefer not to answer & 0 (0\%) & 1 (4\%)\\
     \midrule
    Highest Education
     & 2 Years of College or less & 16 (31\%) & n/a \\
     & College graduate (4 or 5-year program) & 15 (29\%) & n/a \\
     & Master’s degree (or other post-graduate training) & 13 (25\%) & n/a \\
     & High school diploma or equivalent (e.g., GED) & 5 (10\%) & n/a \\
     & Doctoral degree (PhD., MD, EdD, DVM, DDS, JD, etc) & 1 (2\%) & n/a \\
     & Prefer not to answer & 1 (2\%) & n/a \\ 
     
     \midrule
     Household Income 
     & \$20,000-\$59,999 & 20 (39\%) & n/a \\
     & \$100,000-\$149,999 & 20 (39\%) & n/a \\
     & \$150,000-\$199,999 & 4 (8\%) & n/a \\
     & Above \$200,000 & 3 (6\%) & n/a \\
     & \$10,000-\$19,999 & 2 (4\%) & n/a \\
     & Less than \$9,999 & 1 (2\%) & n/a \\
     & Prefer not to answer & 1 (2\%) & n/a \\ \midrule
    Marital Status 
     & Married & 35 (69\%) & n/a \\
     & Unmarried & 12 (24\%) & n/a \\
     & Divorced & 2 (4\%) & n/a \\
     & Separated & 2 (4\%) & n/a \\
     \midrule
    
    Duration of Social  
     & 3-4 hours & 16 (31\%) & 1 (4\%) \\
    Media Daily Use & 1-2 hours & 12 (24\%) & 9 (39\%) \\
     & 2-3 hours & 11 (22\%) & 8 (35\%) \\
     & More than 4 hours & 8 (16\%)& 2 (9\%) \\
     & Less than 1 hour & 4 (8\%) & 3 (13\%) \\ \midrule
     
    Frequency of Social   
     & Several times a day & n/a & 12 (52\%) \\
    Media Daily Use & Several times an hour & n/a & 9 (39\%) \\
     & Every day or almost every day & n/a & 2 (9\%) \\ \midrule

    Someone Tried to Get Teens to   
     & I don't know & 6 (12\%) & n/a \\
    Talk on Social Media About Sex  & Not at all & 26 (51\%) & 11 (48\%) \\
    When Teens Did Not Want to & One to a few times a year & 16 (31\%) & 6 (26\%)\\
     & A few times a month & 3 (6\%) & 3 (13\%)\\
     & A few times a week & 0 (0\%) & 3 (13\%)\\
     & Almost every day & 0 (0\%) & 0 (0\%) \\ \midrule

    Someone on Social Media    
     & I don't know & 5 (10\%) & n/a \\
    Asked Teens for Sexual Information  & Not at all & 25 (49\%) & 11 (48\%)\\
    About Themselves When Teens & One to a few times a year & 16 (31\%) & 7 (30\%)\\
    Did Not Want to Answer & A few times a month & 4 (8\%)& 2 (9\%)\\
     & A few times a week & 1 (2\%) & 2 (9\%)\\
     & Almost every day & 0 (0\%) & 1 (4\%)\\ \midrule

    Someone on Social Media   
     & I don't know & 10 (20\%) & n/a \\
    Asked Teens To Do Something & Not at all & 29 (57\%) & 9 (39\%)\\
    Sexual That Teens Did not Want To & One to a few times a year & 10 (20\%) & 10 (43\%) \\
     & A few times a month & 2 (4\%)& 1 (4\%)\\
     & A few times a week & 0 (0\%) & 3 (13\%)\\
     & Almost every day & 0 (0\%) & 0 (0\%)\\ 
     \bottomrule
    \end{tabular}
    \label{tab:demographics}
\end{table*}

\subsection{ChatGPT 4o Prompt}
\label{prompt}
I'm doing research for youth online safety. Our work focuses on youth behaviors related to cybergrooming. We will conduct a crowdsourced survey for real youth and their parents, so I need to create scenarios to mimic cybergrooming. The screenshot is the scenario. Can you modernize it to reflect modern youth's online chatting behaviors? P means predator, V means youth


\end{document}